\documentclass[aps,prc,showpacs,floatfix,footinbib]{revtex4}

\usepackage{subfigure}
\usepackage{amsmath}    
\usepackage{graphicx}   
\usepackage{verbatim}  
\usepackage{color}      
\usepackage{hyperref}   
\long\def\/*#1*/{}

\begin{document}

\title{Fluctuations and correlations of conserved charges in an Excluded Volume Hadron Resonance Gas model}
\author{Abhijit Bhattacharyya}
\email{abphy@caluniv.ac.in}
\affiliation{Department of Physics, University of Calcutta,
92, A. P. C. Road, Kolkata - 700009, INDIA}
\author{Supriya Das}
\email{supriya@jcbose.ac.in}
\author{Sanjay K. Ghosh}
\email{sanjay@jcbose.ac.in}
\author{Rajarshi Ray}
\email{rajarshi@jcbose.ac.in}
\author{Subhasis Samanta}
\email{samanta@jcbose.ac.in}
\affiliation{Center for Astroparticle Physics \&
Space Science, Bose Institute, Block-EN, Sector-V, Salt Lake, Kolkata-700091, INDIA 
 \\ \& \\ 
Department of Physics, Bose Institute, \\
93/1, A. P. C Road, Kolkata - 700009, INDIA}
\date{}

\begin{abstract}
We present temperature ($T$) and baryonic chemical potential ($\mu_B$) dependence of higher order fluctuations and correlation 
between conserved charges in Excluded Volume Hadron Resonance Gas (EVHRG) model. Products of moments,
such as ratio of variance to mean ($\sigma^2/M)$, product of skewness and standard deviation ($S\sigma$),
product of kurtosis and variance ($\kappa\sigma^2$), for 
net-proton, net-kaon and net-charge have been evaluated on the phenomenologically determined freeze-out 
curve. Further, products of moments for 
net-proton and net-charge have been compared 
with the experimental data measured by STAR experiment. The dependence of the model result on the hadronic radius parameter has also been discussed.
\end{abstract}


\maketitle
\section{\label{sec:Intro} Introduction}
One of the primary goals of heavy-ion collision experiments at ultra-relativistic energies  is to study the 
thermodynamic aspects of strongly interacting matter at high temperature and density. While at low density 
and high temperature, present experimental as well as lattice data seem to indicate a smooth cross over~\cite{nature05120_Aoki, PRL65_2491_Brown} from 
hadronic to a quark gluon matter, at high density and low temperature the system is expected to have a first order 
transition~\cite{PRD78_074507_Ejiri, PRD29_Pisarski, NPA504_Asakawa, PRD58_096007_Halasz, PRD67_014028_Hatta, PRC79_Bowman}. 
So the first order phase transition at high densities and low temperature should end at a critical end point (CEP) - a second order phase transition point - as one moves towards high temperature and low density region, in the phase diagram of 
strongly interacting matter~\cite{JHEP_Fodor, PTPS_Stephanov, PRD71_114014_Gavai, PRD75_Schaefer}. 

Several experimental programmes have been launched to study the phase transition of strongly interacting matter. 
At present, Relativistic Heavy-Ion Collider (RHIC) at Brookhaven National Laboratory and the Large Hadron Collider (LHC) at CERN are 
providing us plethora of data. In near future the Facility for Anti-proton and Ion Research (FAIR) at GSI will also provide us a large 
amount of information. The low temperature and high $\mu_B$ region will be studied 
in Beam Energy Scan (BES) programme of RHIC and the Compressed Baryonic Matter (CBM) programme at FAIR.
Those experiments would also explore the first order line of the phase diagram along with the location of the CEP.

The unveiling of the nature of phase transition requires a proper understanding of Quantum Chromo Dynamics (QCD), 
the theory of strong interactions. Unfortunately, the non-perturbative nature of the phenomena inhibits the use of 
first principle QCD for the study of 
strongly interacting matter at extreme conditions. In this regard Lattice QCD (LQCD) provides the 
most direct approach to study QCD at high temperature ~\cite{PRD78_114503_Gavai, PRD79_074505_Cheng, 
boyd, engels, fodor1, fodor2, allton1,
allton2,allton3,forcrand, aoki1, nature05120_Aoki, Bazavov, Borsanyi, Borsanyi_JHEP_11_77, PRD77_014511_Cheng}. However LQCD
has its own restrictions due to the discretization of space-time. Furthermore,
at finite chemical potential, LQCD faces the well known sign problem. 

In contrast, effective models~\cite{Ray, Bhattacharyya, PRL102_032301_Stephanov} provide
a simpler alternative for the study of the strongly interacting matter in the non-perturbative domain. Some of these 
models have been quite successful in describing physics of strongly interacting matter. For example, 
Polyakov loop extended Nambu-Jona-Lasinio model (PNJL) has been used to study various aspects of physics of
strongly interacting matter at high temperatures  and found to reproduce the zero density lattice data quite 
successfully~\cite{Ray, Bhattacharyya}. On the other hand, hadron resonance gas (HRG) model
\cite{HRG_Braun-Munzinger} has been very successful in describing the hadron yields in central 
heavy ion collisions from AGS up to RHIC energies~\cite{PLB344_Braun-Munzinge, arXiv:nucl-th/9603004_Cleymans, 
PLB465_Braun-Munzinge, PRC60_054908_Cleymans, PRC73_Becattini, PLB518_Braun-Munzinger, NPA772_Andronic, PLB673_Andronic}.

A reliable way to understand the physics of the phase transition of strongly interacting matter is to study the 
correlations and fluctuations of conserved charges. Susceptibilities 
are related to fluctuations via the fluctuation-dissipation theorem. A measure of the intrinsic
statistical fluctuations in a system close to thermal equilibrium is provided
by the corresponding susceptibilities. At finite temperature and chemical potential fluctuations of conserved charges are
sensitive indicators of the transition from hadronic matter to quark-gluon plasma (QGP). Moreover, the existence of the CEP can be 
signalled by the divergent fluctuations. For the small net baryon number, which       
can be met at different experiments, the transition from hadronic to QGP phase
is continuous and the fluctuations are not expected to lead to any singular behaviour.  
Computations on the lattice have been performed for many of these susceptibilities at zero chemical potentials 
\cite{gottlieb,gavai,bernard1,bernard2}. It has been shown that at vanishing chemical potential the susceptibilities
rise rapidly around the continuous crossover transition region. 

Since the prediction of HRG model for the system at freeze-out is in very good agreement with
the experiments, it would be interesting to study the susceptibilities as well as higher moments 
using this model and its interacting version {\it i.e.} Excluded Volume Hadron resonance gas (EVHRG) 
model~\cite{PLB97_Hagedorn, ZPC51_Rischke, PS48_277_Cleymans, Singh, PRC56_Yen, PRC77_Gorenstein, PLB718_Andronic, PRC85_Fu, 
begunprc88, PLB722_Fu, PRC88_Tawfik}. In fact as the higher moments are 
expected to be more sensitive to the phase transition, any deviation of experimental observation 
from the model results may be taken as an indication of new phenomena. 

HRG model is based on the Dashen, Ma and Bernstein theorem~\cite{dashen} which shows that a dilute system of strongly
interacting matter can be described by a gas of free resonances. The attractive part of the hadron interactions is taken 
care of by these resonances. On the other hand, both the long range attraction as well as the short range repulsion are 
important for the description of strongly interacting matter~\cite{begunprc88}. Moreover, near critical temperature HRG
calculations tends towards Hagedorn divergence which may be due to the
absence of repulsive interaction~\cite{PLB718_Andronic}. This repulsive part is incorporated through the excluded volume
effects in the HRG~\cite{ZPC51_Rischke} and is commonly known as EVHRG model. 
EVHRG equation of states have also been used for the hydrodynamical models of Nucleus-nucleus collision~\cite{hama,werner,satarov}.
In the present work we will be discussing our results of susceptibilities and correlations using EVHRG model.

Recently fluctuations of conserved charges, using HRG model, have been studied 
in Ref.~\cite{Mohanty} where actual experimental acceptances in terms of pseudo-rapidity and transverse momentum are considered.
In Ref.~\cite{PLB722_Fu}, higher moments of net-proton multiplicity have been studied using EVHRG model.

Our aim in the present work is twofold. Firstly we would like to have a physical understanding of the EVHRG model results 
vis-a-vis HRG, lattice and experimental data. Secondly, we would like to study the high density sector to be explored in beam 
energy scan at RHIC and CBM experiment at FAIR.

Here we present the temperature ($T$) and baryonic chemical potential ($\mu_B$) dependence of 
susceptibilities of different conserved 
quantities such as net-baryon number, net-strangeness and net-charge up to order four using EVHRG model.  
Baryon-strangeness and charge-strangeness correlation functions have been evaluated at different $T$ and $\mu_B$. 
Different values of baryon and meson radii have been used to study their effect on susceptibilities and correlations. 
We have discussed experimental observables in the framework of HRG as well as EVHRG models. The product of moments 
of distribution of conserved quantities are related to the ratio of different order of susceptibilities 
\cite{PLB633_275_Ejiri, PLB695_Karsch, PRL105_022302_Aggarwal} 
such as $\sigma^2/M=\chi^{2}/\chi^{1}$, $S \sigma=\chi^{3}/\chi^{2}$ and
$\kappa \sigma^2=\chi^{4}/\chi^{2}$ where $M$ is the mean, $\sigma$ is the standard deviation,
$S$ is the skewness, $\kappa$ is the  kurtosis of the distribution of conserved quantities and 
$\chi^{n}$ are the $n^{th}$ order susceptibilities. In general, higher order susceptibilities are more sensitive to the 
large correlation length and hence the critical point~\cite{PRL102_032301_Stephanov}. This implies that any memory
of the large correlation length retained in the thermal system at freeze-out would be reflected in the behaviour of higher moments. 
We have studied the energy dependence of product of moments for 
net-proton, net-kaon and net-charge and compare our result for 
net-proton and net-charge with experimental data of fluctuation 
along with the transverse 
momentum and pseudo-rapidity acceptance.

The paper is organized as follows. HRG model and its extension to EVHRG are introduced in sec. \ref{sec:HRG}. In sec. \ref{sec:Fluctuation}
we have first discussed the
fluctuations of different conserved charges and correlations among them at zero $\mu_B$ (finite $T$) and compared them with the 
lattice data. Then we have discussed the finite density scenario along with a comparison with the experimental data. Finally we have 
summarized our results in sec. \ref{sec:Conclusion}.

\section{\label{sec:HRG} Hadron resonance gas model}
In this section we are going to discuss HRG and EVHRG model very briefly. The detailed
discussions can be found in Refs.~\cite{HRG_Braun-Munzinger, PLB344_Braun-Munzinge, arXiv:nucl-th/9603004_Cleymans, 
PLB465_Braun-Munzinge, PRC60_054908_Cleymans, PRC73_Becattini, PLB518_Braun-Munzinger, NPA772_Andronic,
PLB673_Andronic, PLB97_Hagedorn, ZPC51_Rischke, PS48_277_Cleymans,  Singh, PRC56_Yen, PLB718_Andronic}.
The grand canonical partition function of a hadron resonance gas~\cite{HRG_Braun-Munzinger, PLB718_Andronic} can be written as 
\begin {equation}
 \ln Z^{id}=\sum_i \ln Z_i^{id},
\end{equation}
where sum is over all the hadrons. $id$ refers to ideal {\it i.e.}, non-interacting HRG.
For particle $i$,
\begin{equation}
 \ln Z_i^{id}=\pm \frac{Vg_i}{2\pi^2}\int_0^\infty p^2\,dp \ln[1\pm\exp(-(E_i-\mu_i)/T)],
\end{equation}
where $V$ is the volume of the system, $g_i$ is the degeneracy factor, $T$ is the temperature, $E_i=\sqrt{{p}^2+m^2_i}$
is the single particle energy, $m_i$ is the mass and $\mu_i=B_i\mu_B+S_i\mu_S+Q_i\mu_Q$ is the
chemical potential. $B_i,S_i,Q_i$ are respectively the baryon number, strangeness
and charge of the particle, $\mu^,s$ being corresponding chemical potentials. The $(+)$ and $(-)$ sign corresponds to fermions and bosons
respectively.
Partition function depends in general on five parameters. However, only three are independent, since $\mu_Q$ and $\mu_S$ 
can be found from conservation of different quantum numbers like baryon number,
charge and strangeness~\cite{HRG_Braun-Munzinger, PLB718_Andronic}.

The partition function is the basic quantity from which one can calculate various thermodynamic quantities of the 
thermal system created in heavy ion collisions. The partial pressure $P_i$, the particle density $n_i$, the energy density 
$\varepsilon_i$, and the entropy density $s_i$ can be calculated using the standard definitions,

\begin{align}\label{eq:p}
 \begin{split}
  P_i^{id}&=\frac{T}{V}\ln Z_i^{id}
  =\pm\frac{g_iT}{2\pi^2}\int_0^\infty p^2\,dp \ln[1\pm\exp(-(E_i-\mu_i)/T)],
 \end{split}
\end{align}

\begin{equation}\label{eq:n}
 n_i^{id}=\frac{T}{V} \left(\frac{\partial \ln Z_i^{id}}{\partial\mu_i}\right)_{V,T}
 =\frac{g_i}{2\pi^2}\int_0^\infty\frac{p^2\,dp}{\exp[(E_i-\mu_i)/T]\pm1},
\end{equation}

\begin{align}
\begin{split}
\varepsilon_i^{id}&=\frac{E_i^{id}}{V}=-\frac{1}{V} \left(\frac{\partial \ln Z_i^{id}}{\partial\frac{1}{T}}\right)_{\frac{\mu}{T}}
=\frac{g_i}{2\pi^2}\int_0^\infty\frac{p^2\,dp}{\exp[(E_i-\mu_i)/T]\pm1}E_i,
\end{split}
\end{align}

\begin{align}\label{eq:s}
 \begin{split}
  s_i^{id}&=\frac{S_i^{id}}{V}=\frac{1}{V}\left(\frac{\partial\left({T \ln Z_i^{id}}\right)}{\partial T}\right)_{V,\mu}
 = \pm\frac{g_i}{2\pi^2}\int_0^\infty p^2\,dp \left[ \ln\left(1\pm\exp(-\frac{(E_i-\mu_i)}{T})\right)\right.
 \left.\pm\frac{(E_i-\mu_i)}{T(\exp((E_i-\mu_i)/T)\pm1)}\right].
 \end{split}
\end{align}

Since Eqs. \ref{eq:p} - \ref{eq:s} determine the thermodynamic properties of the system, those are called equations of state 
(EOS) of the system.

\begin{figure}[]
\centering
 \subfigure {\includegraphics[scale=0.34,angle=-90]{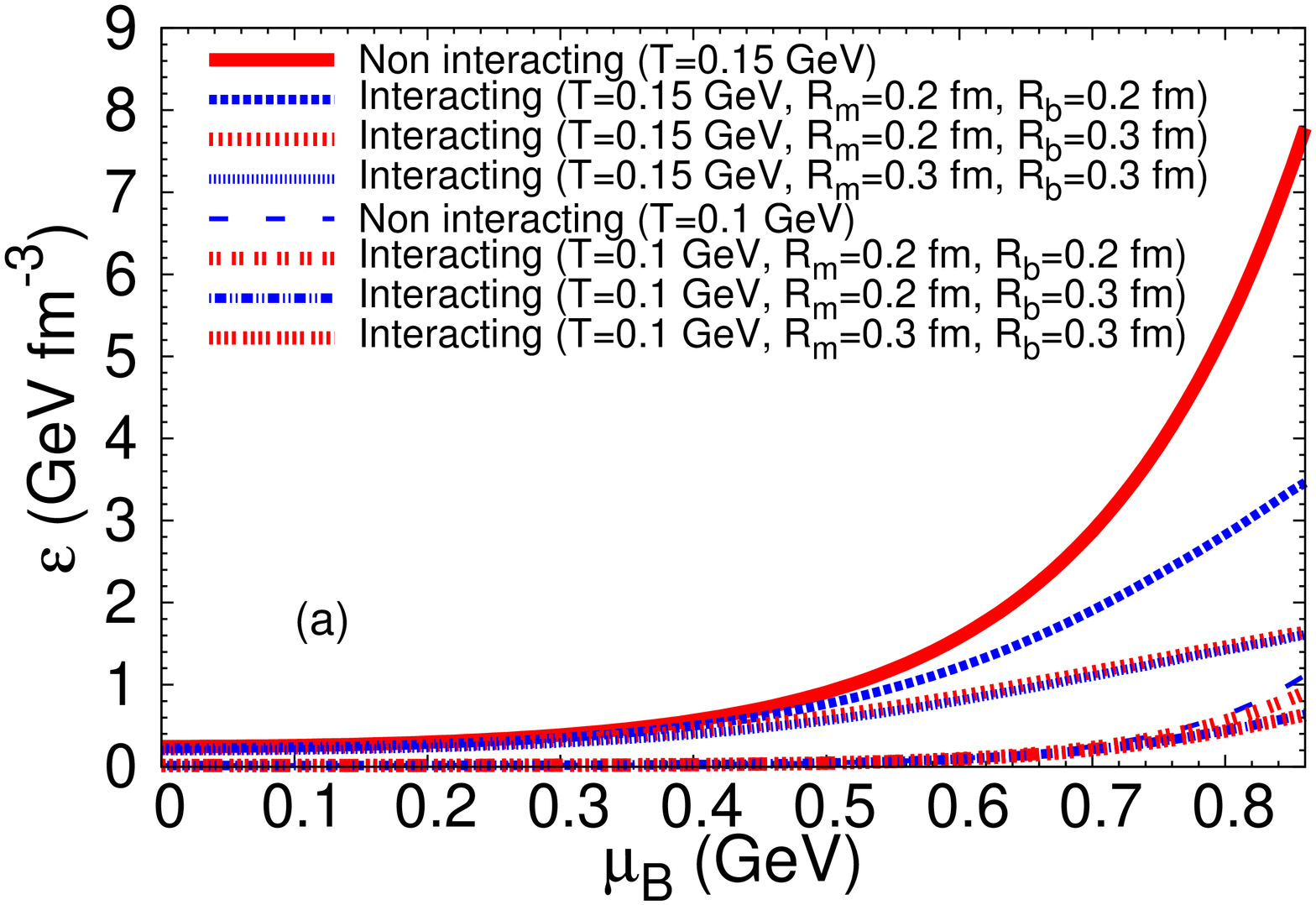}\label {energy_mub}}
 \subfigure {\includegraphics[scale=0.34,angle=-90]{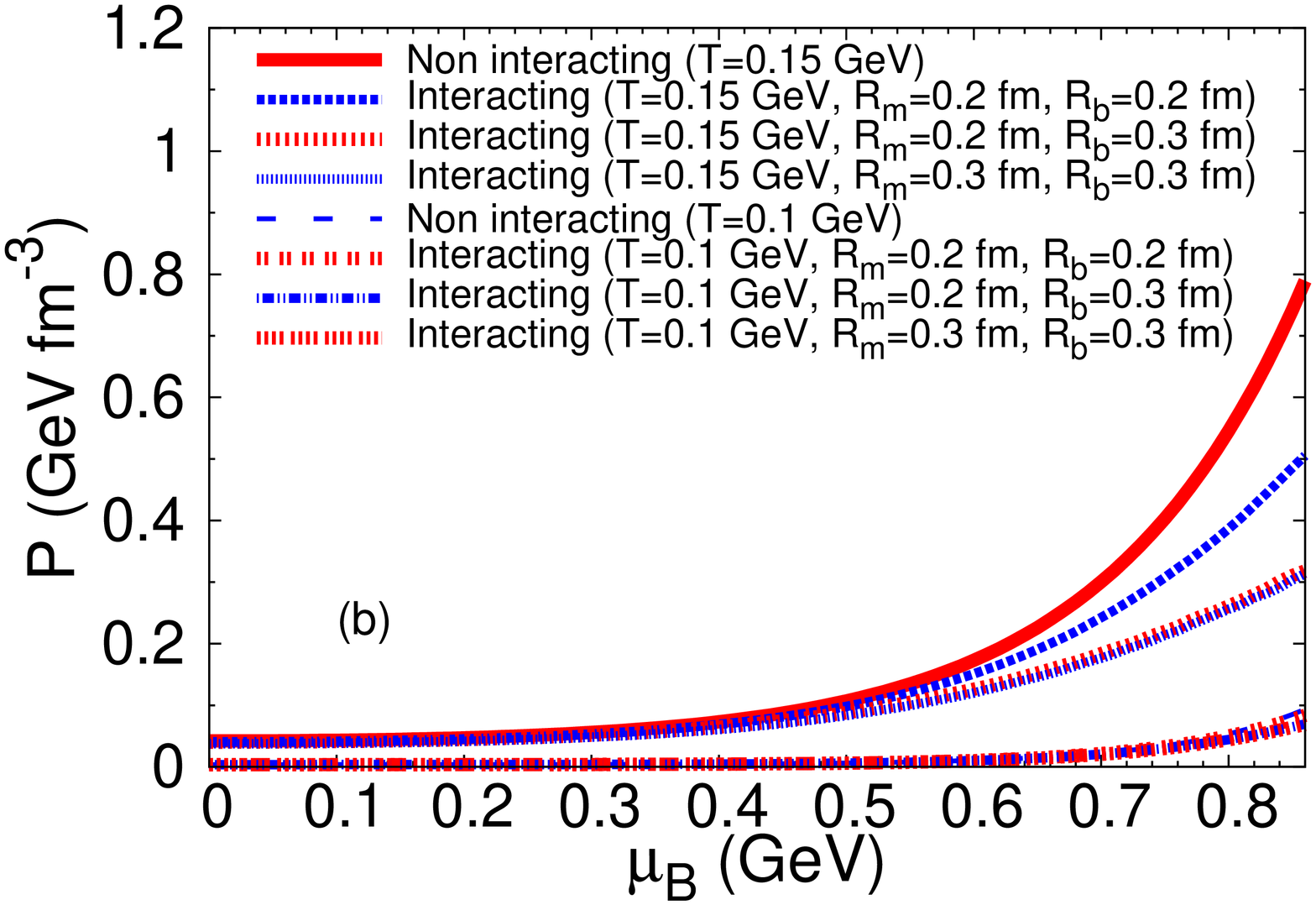}\label {pressure_mub}}
 \subfigure {\includegraphics[scale=0.34,angle=-90]{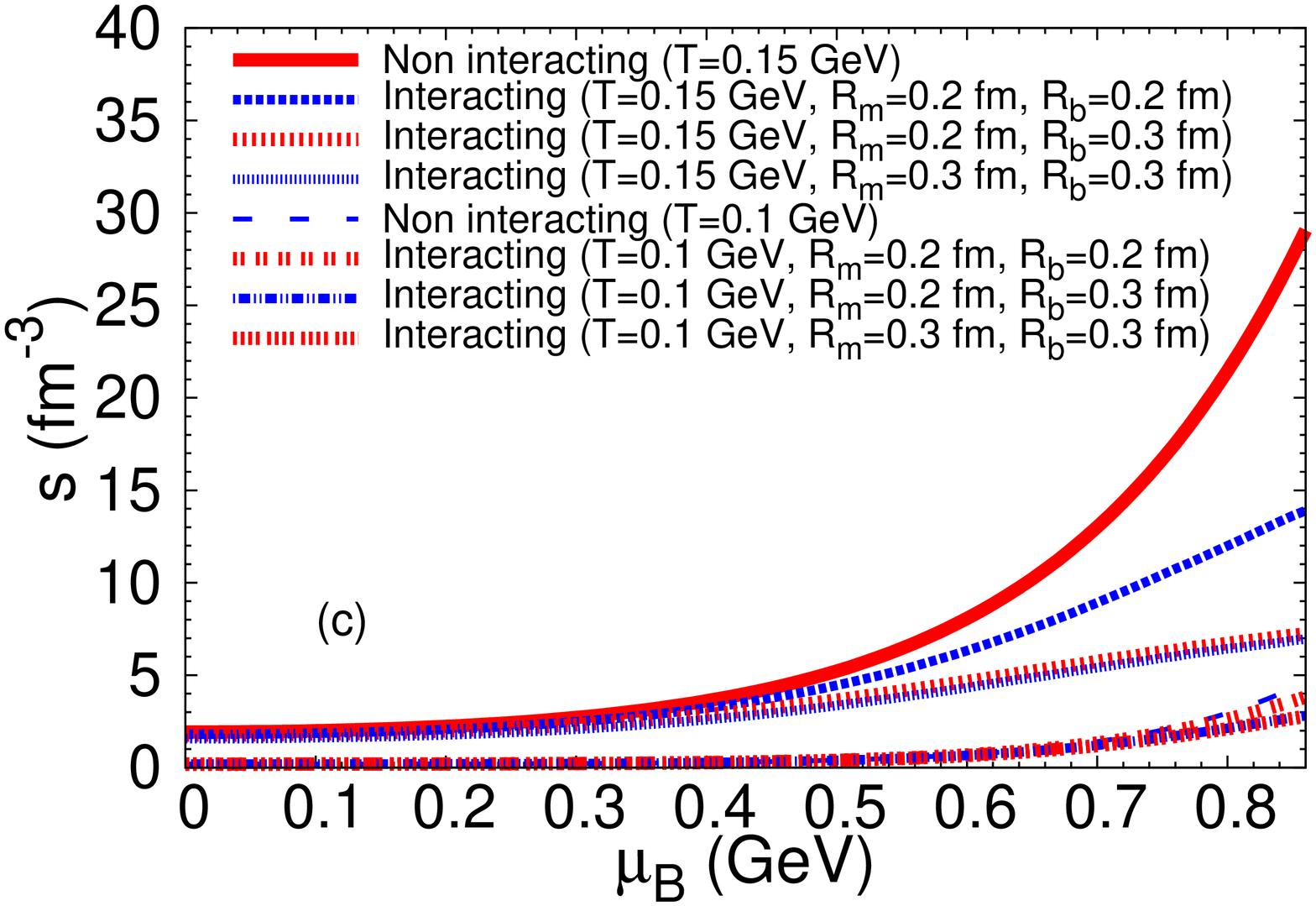}\label {entropy_mub}}
 \subfigure {\includegraphics[scale=0.34,angle=-90]{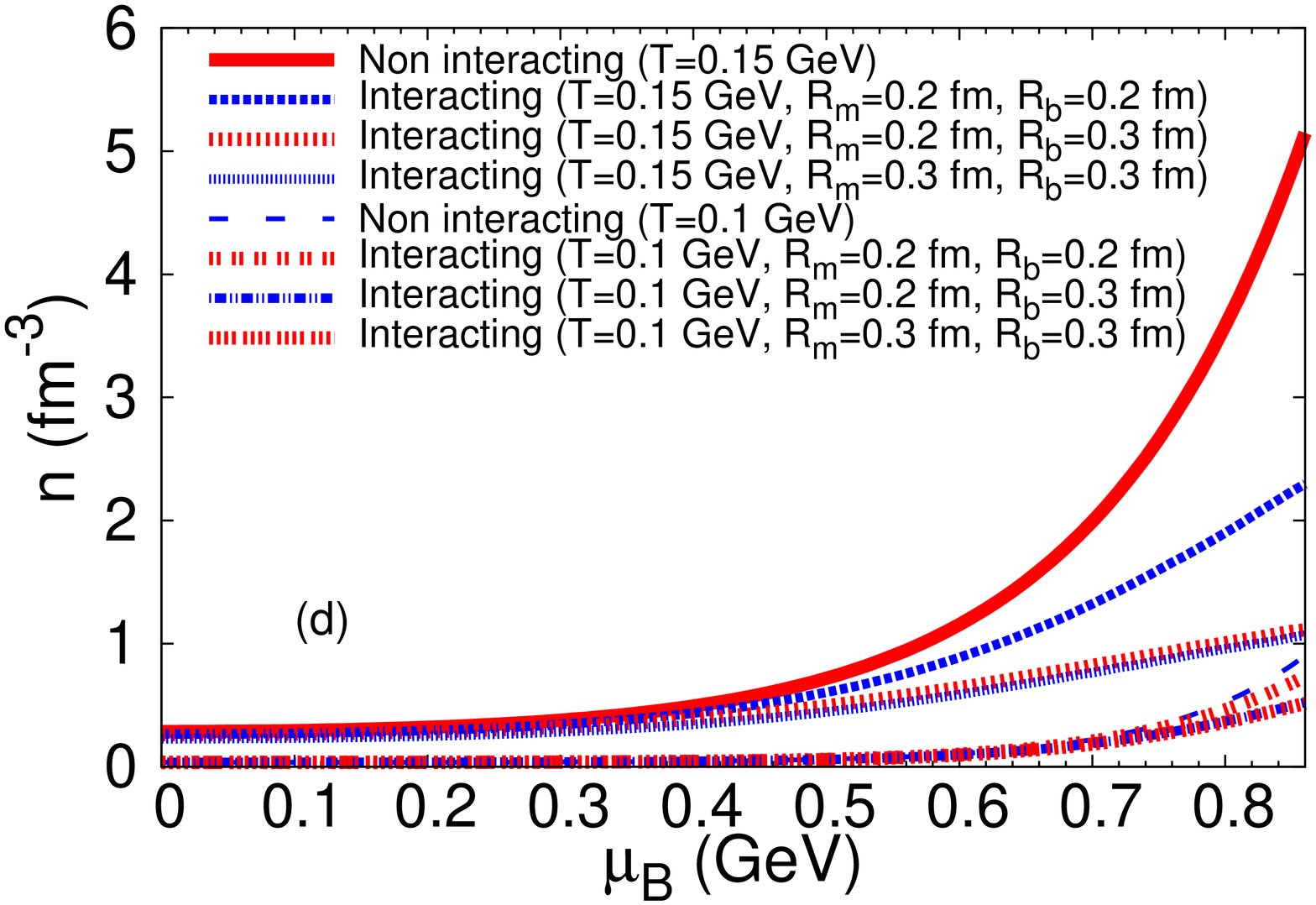}\label {number_mub}}
 \caption{(Color online). \label{fig:EOS_mub}Equations of state  as a function of $\mu_B$ at constant $T$ keeping
 $\mu_S=\mu_Q=0$. Non interacting means non interacting HRG model {\it i.e.} without
 excluded volume correction and interacting refers to EVHRG. $R_m$ refers to radius of mesons and $R_b$ refers to radius of baryons.}
\end{figure}


Let us now discuss the interacting HRG or EVHRG model.
In contrast to the point like hadrons in HRG, in EVHRG model, hadronic phase
is modelled by a gas of interacting hadrons, where the geometrical size of the hadrons are explicitly incorporated
as the excluded volume correction~\cite{ZPC51_Rischke, PS48_277_Cleymans, PRC56_Yen} to approximate the short-range 
repulsive hadron-hadron interaction.

Excluded volume corrections were first introduced in~\cite{PLB97_Hagedorn} but it was thermodynamically inconsistent. A thermodynamically consistent 
excluded volume correction  was first proposed in~\cite{ZPC51_Rischke}.

In EVHRG model pressure can be written as
\begin{equation}\label{eq:p_new}
 P(T,\mu_1,\mu_2,..)=\sum_i P_i^{id}(T,\hat{\mu}_1,\hat{\mu}_2,..),
\end{equation}
where for $i$'th particle the chemical potential is
\begin{equation}\label{eq:mu}
 \hat{\mu_i}=\mu_i-V_{ev,i}P(T,\mu_1,\mu_2,..)
\end{equation}
where
$V_{ev,i}=4 \, \frac{4}{3} \, \pi R_i^3$ is the volume excluded for
 the $i$ th hadron with hard core radius $R_i$.

In an iterative procedure one can get the total pressure. Pressure $P(T,\mu_1,\mu_2,..)$ is suppressed compared to 
the ideal gas pressure $P^{id}$ because of the smaller value of effective chemical potential. The other
thermodynamic quantities like $n_i$, $s$ , $\varepsilon$
can be calculated from Eqs. \ref{eq:p_new} - \ref{eq:mu} as

\begin{align}
 \begin{split}
   n_i=n_i(T,\mu_1,\mu_2,..)&=\frac{\partial P}{\partial\mu_i}
  =\frac{n_i^{id}(T,\hat{\mu_i})}{1+\sum_k V_{ev,k}n_k^{id}(T,\hat{\mu_k})},
 \end{split}
\end{align}

\begin{align}
 \begin{split}
  s=s(T,\mu_1,\mu_2,..)&=\left(\frac{\partial P}{\partial T}\right)_{\mu_1,\mu_2,..}
 =\frac{\sum_i s_i^{id}(T,\hat{\mu_i})}{1+\sum_k V_{ev,k}n_k^{id}(T,\hat{\mu_k})},
 \end{split}
\end{align}

\begin{equation}
 \varepsilon=\varepsilon(T,\mu_1,\mu_2,..)=\frac{\sum_i \varepsilon_i^{id}(T,\hat{\mu_i})}{1+\sum_k V_{ev,k}n_k^{id}(T,\hat{\mu_k})},
\end{equation}
This correction scheme is thermodynamically consistent {\it i.e.} EOS after corrections obey the relation 

\begin{equation}
 s=\frac{\varepsilon+P-\sum_i \mu_i n_i}{T}.
 \end{equation}

In this work we have incorporated all the hadrons listed in the particle data book~\cite{PDG} up to mass of $3$ GeV.
In Fig. \ref{fig:EOS_mub} we have shown $\varepsilon, P, s$ and $n$ as a function of $\mu_B$ at fixed $T$. For 
completeness one should also plot these quantities as a function of $T$ at fixed $\mu_B$. However such relations
are already given in Ref.~\cite{PLB718_Andronic, PRC88_Tawfik} and we do not repeat those here. 
Our results show that there is almost
no effect of interaction till $\mu_B=0.4$ GeV in EOS. Beyond $\mu_B=0.4$ GeV we see quite a substantial change in 
these quantities. The change is more pronounced at higher temperatures. One can see from this figure that at large
$\mu_B$ the energy, entropy and number density in EVHRG model  are suppressed by a factor of $2$, compared to HRG, if we take the radii of all the hadrons to be $0.2$ fm. This is expected as the finite radius acts as repulsive interaction between hadrons. If we increase the size of the baryons further to $0.3$ fm the suppression is even more. However
the thermodynamic quantities are less sensitive to the mesonic radii as can be seen from Fig. \ref{fig:EOS_mub}.
The plot for $R_b= R_m=0.3$ fm is almost same as that for $R_b=0.3$ fm, $R_m=0.2$ fm and is suppressed compared to $R_b=R_m=0.2$ fm. 
This is an expected result as the system is dominated by baryons at high $\mu_B$.

The difference of the EVHRG model, as compared to HRG, is governed by the radius parameter. The electromagnetic charge radii of 
hadrons have been measured by different groups~\cite{amnedolia}.  For example, the radii for $p$, $\Sigma^-$, $\pi^-$ and $K^-$
are around 0.8 fm, 0.9 fm, 0.7 fm and 0.6 fm respectively. One can also define a strong interaction radii~\cite{povh} which comes 
out to be around the same values. In accordance with these results, a value of 0.8 fm for baryons and 0.62 fm for mesons were proposed 
earlier~\cite{PRC56_Yen}. On the other hand, Braun-Munzinger {\it {et al.}}~\cite{PLB465_Braun-Munzinge} argued that a more 
realistic approach is to incorporate repulsive behaviour of the NN potential using hard-core radius ($\sim$ 0.3 fm) as obtained 
from  nucleon-nucleon scattering~\cite{bohr}. The corresponding meson radius should not exceed that of baryons. 
A similar hard-core radius of 0.2-0.3 fm has also been proposed in Ref.~\cite{giltinan} to explain the proton-proton scattering 
data. It has also been shown earlier that to justify a 
hydrodynamic approach to heavy ion collisions within the hadron phase the hard-core hadron radius should be $r \ge 0.2$ fm in
EVHRG model~\cite{PRC77_Gorenstein}. In the present study we have taken an approach similar to~\cite{PLB465_Braun-Munzinge} and 
have used different baryon ($R_b$) and meson ($R_m$) radii between 0.2-0.3 fm.
 
\section{\label{sec:Fluctuation}Fluctuation and Correlation: Lattice vs. EVHRG }

\begin{figure}[]
\centering
  \subfigure {\includegraphics[scale=0.34,angle=-90]{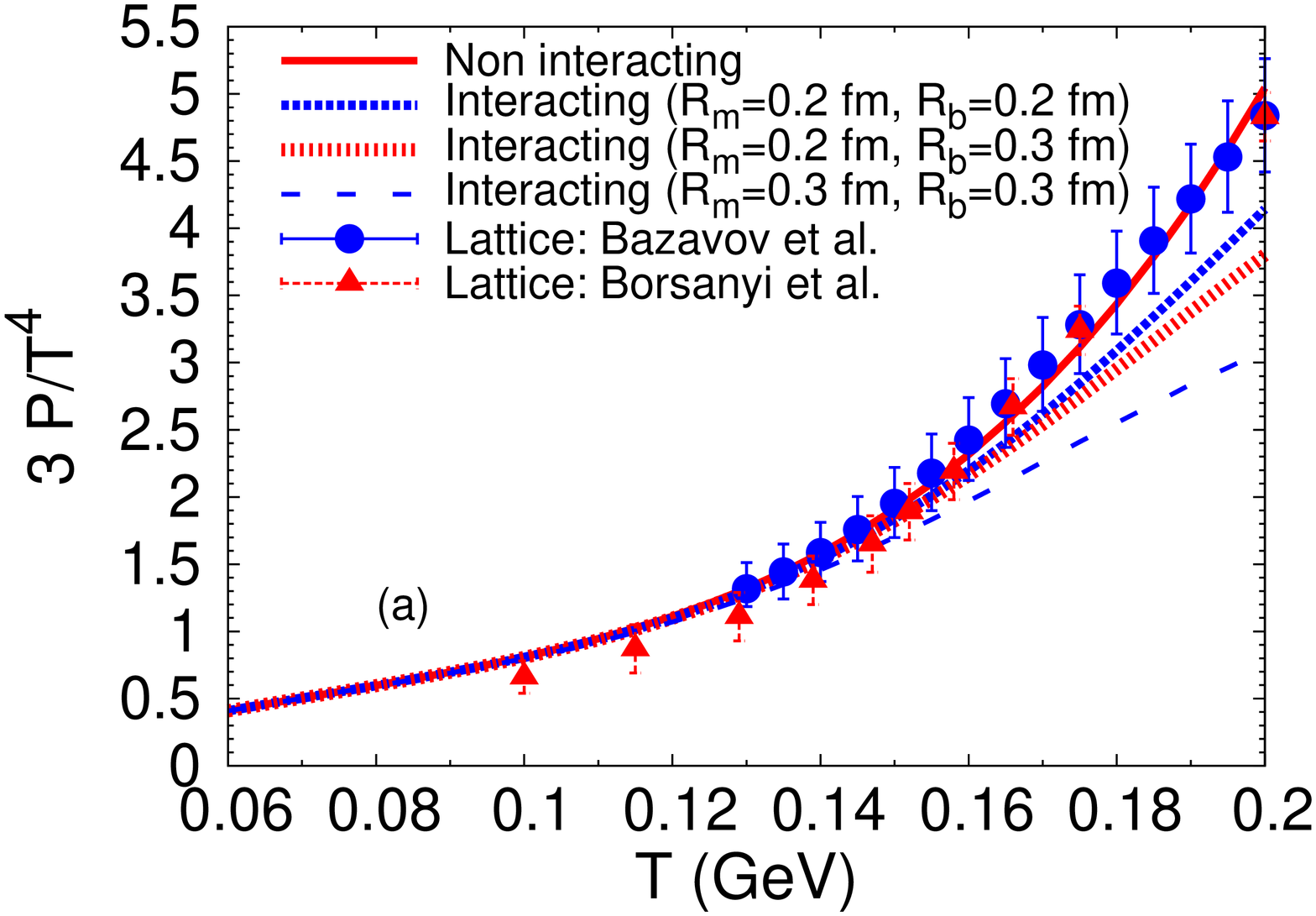}\label {press_temp}}
   \subfigure {\includegraphics[scale=0.34,angle=-90]{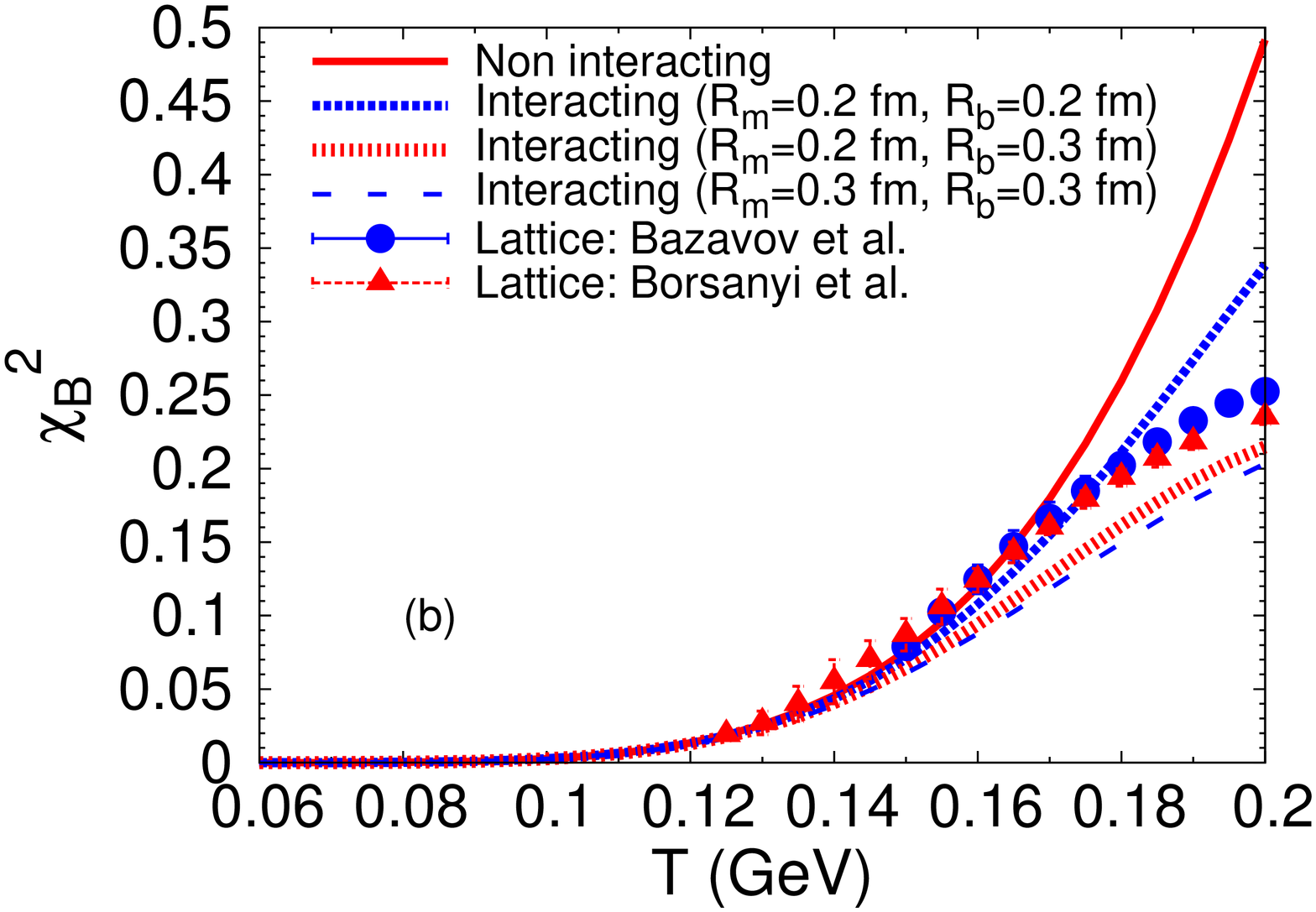}\label {chi_B2_temp}}
    \subfigure {\includegraphics[scale=0.34,angle=-90]{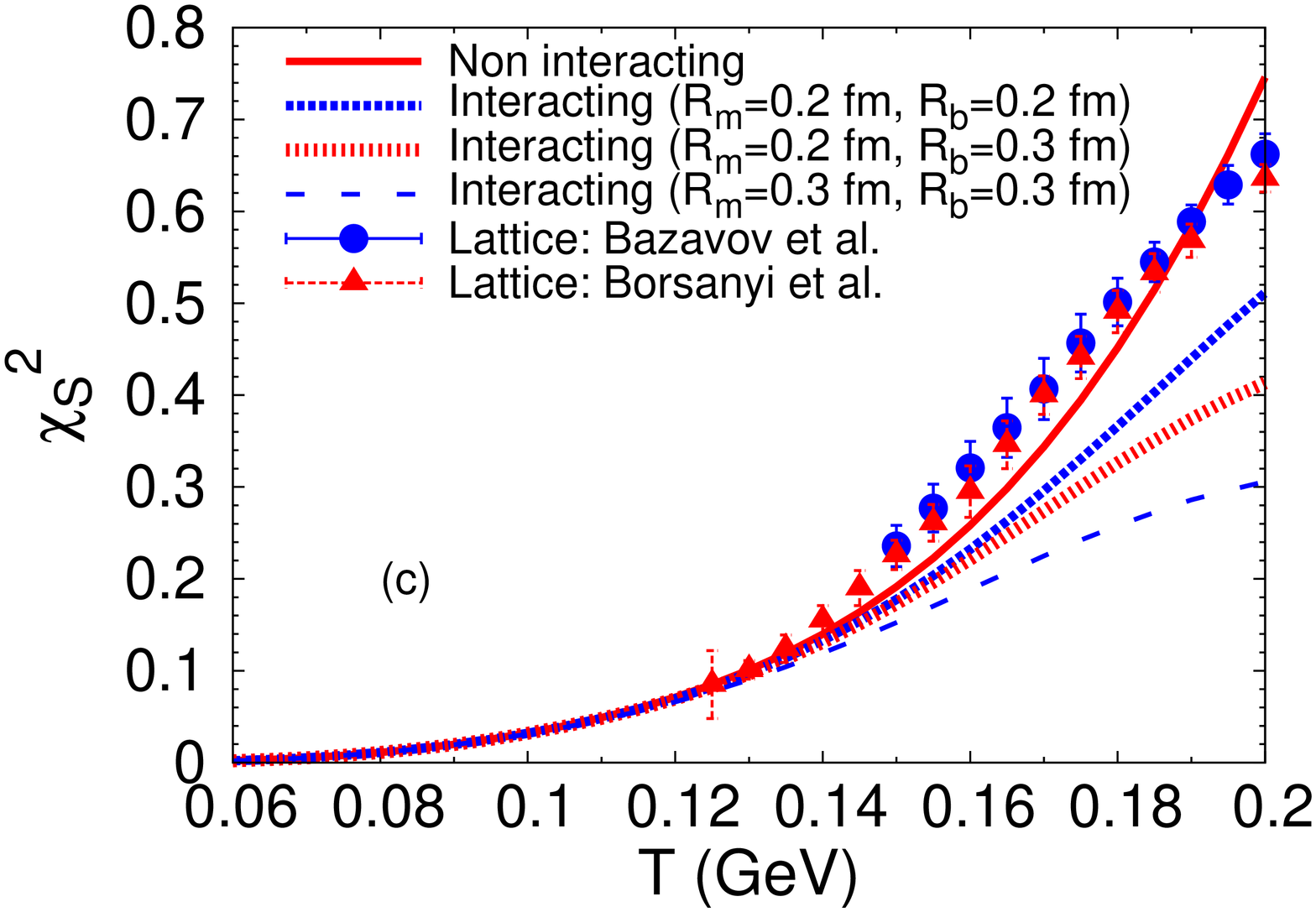}\label {chi_S2_temp}}
     \subfigure {\includegraphics[scale=0.34,angle=-90]{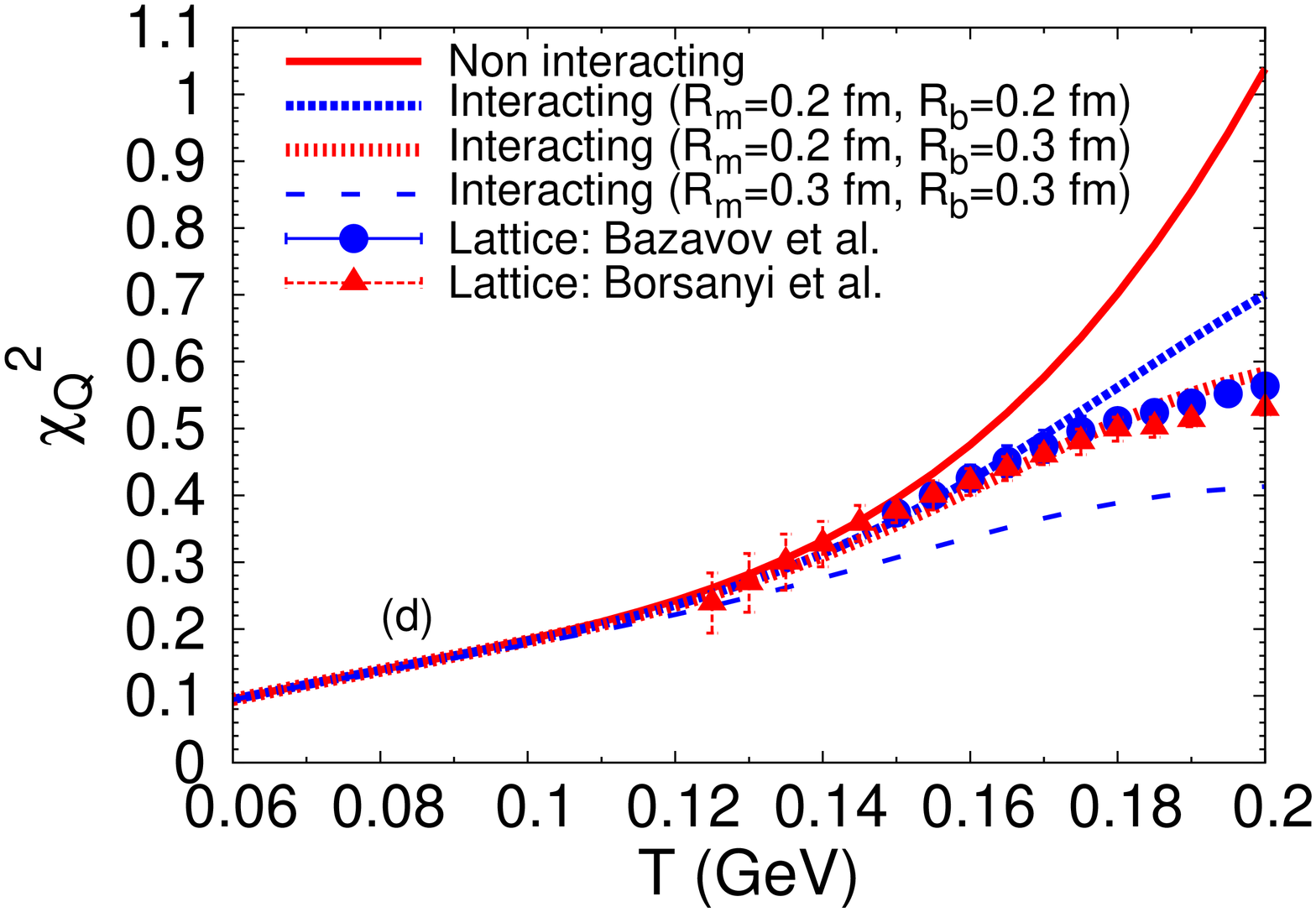}\label {chi_Q2_temp}}
 \caption{(Color online). \label{fig:chi^2_temp}Variation of pressure and second order susceptibilities
 $\chi_B^2$, $\chi_S^2$ and $\chi_Q^2$ 
 with temperature at $\mu=0$. Lattice data for pressure are taken from 
Bazavov {\it et.al}~\cite{Bazavov_eos_contm} and
Bors\'{a}nyi {\it et.al}~\cite{Borsanyi_JHEP_11_77} whereas
those for $\chi^2$ are taken from Bazavov {\it et.al}~\cite{Bazavov} and Bors\'{a}nyi {\it et.al} ~\cite{Borsanyi}}.
 \end{figure}

\begin{figure}
\centering
  \subfigure {\includegraphics[scale=0.34,angle=-90]{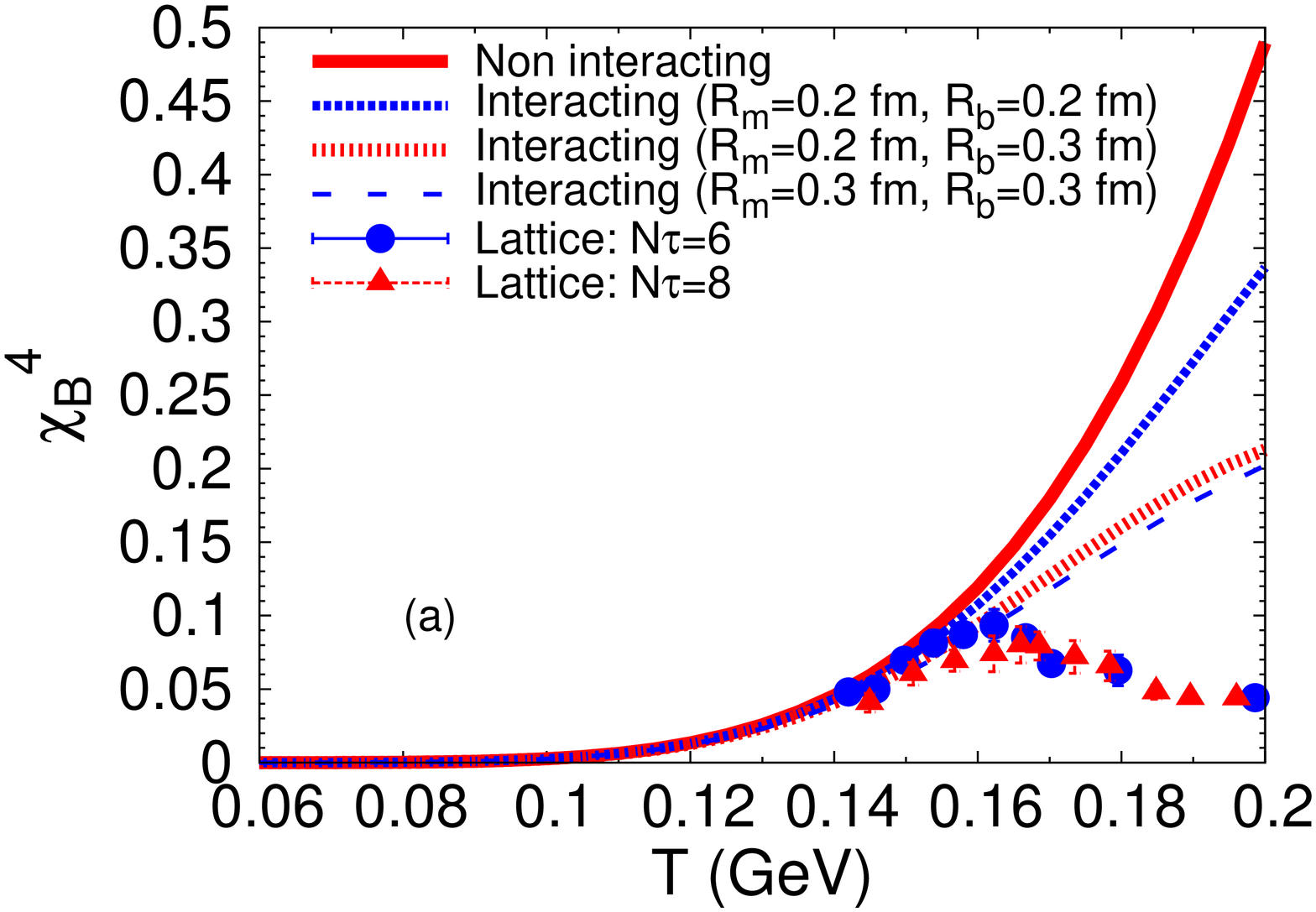}\label {chi_B4_temp}}
   \subfigure {\includegraphics[scale=0.34,angle=-90]{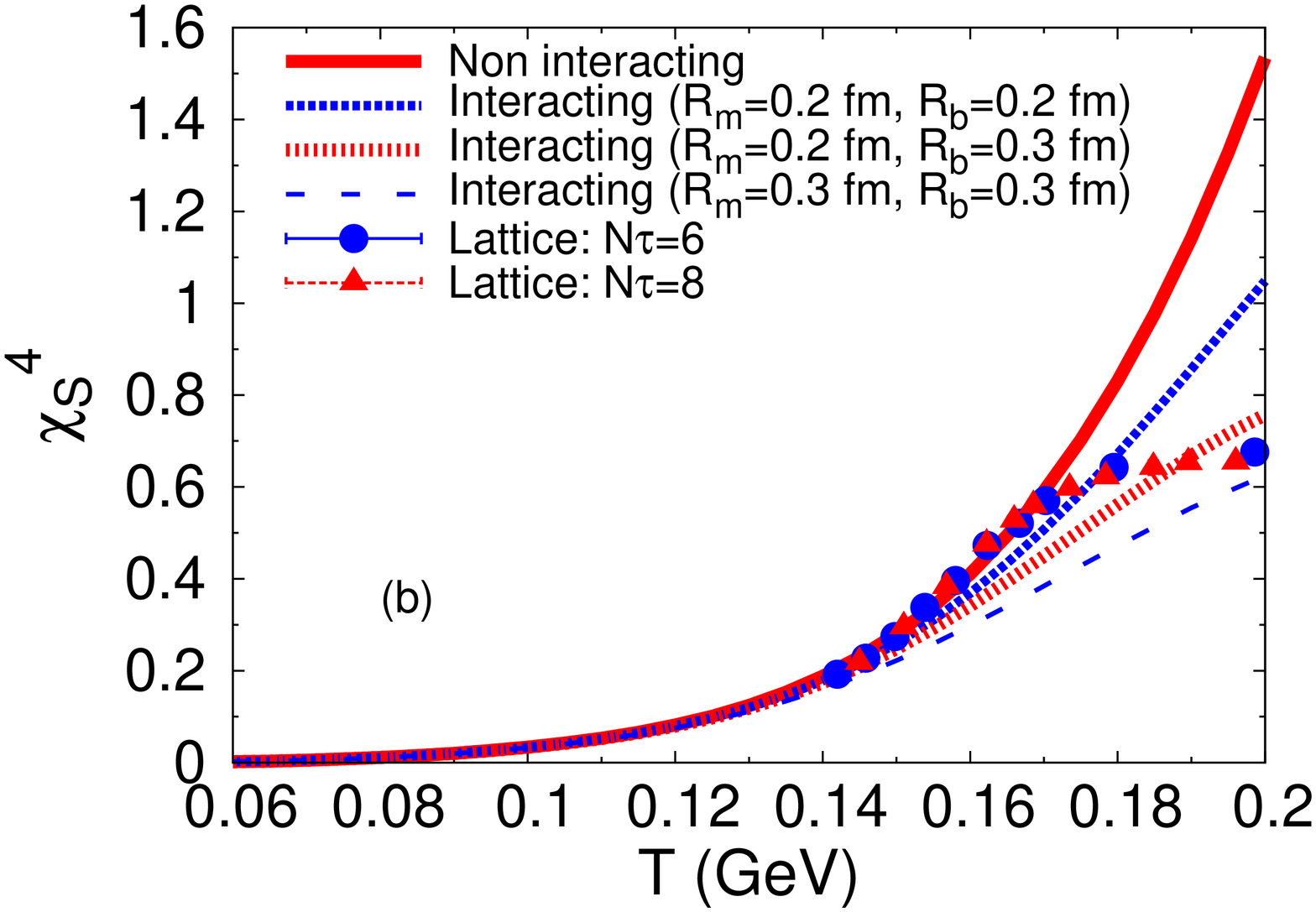}\label {chi_S4_temp}}
    \subfigure {\includegraphics[scale=0.34,angle=-90]{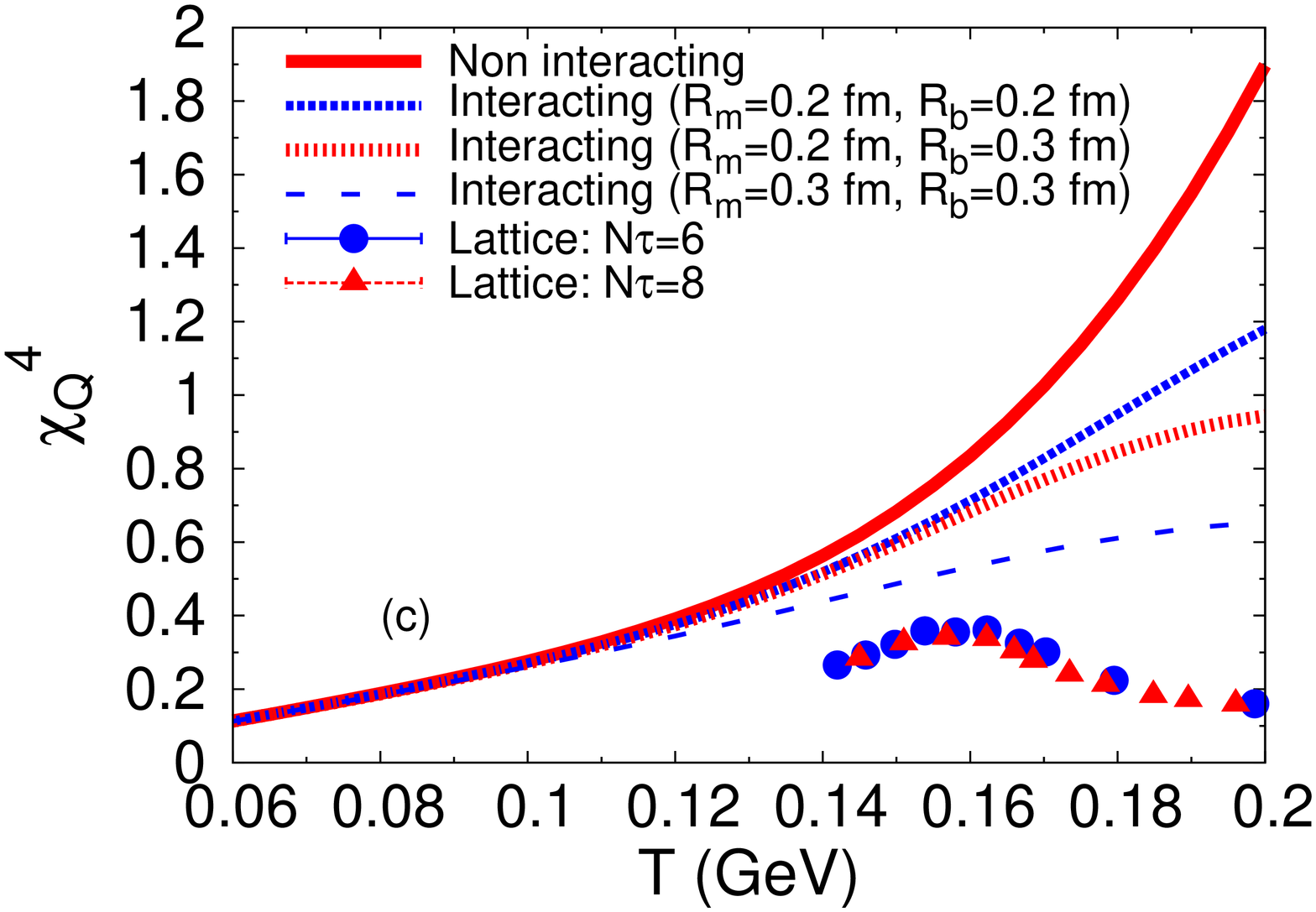}\label {chi_Q4_temp}}
 \caption{(Color online). \label{fig:chi^4_temp}Variation of fourth order susceptibilities $\chi_B^4$, $\chi_S^4$, $\chi_Q^4$
 with temperature ($\mu=0$). Lattice data is taken from Ref.~\cite{Schmidt}.}
\end{figure}

Fluctuations and correlations of conserved charges such as net electric charge, baryon number and strangeness
have been considered
as probes of  hadronization and thermalization of the system created in nuclear collisions  
\cite{Asakawa, Bower, Aziz}. 
Fluctuation are considered to be important signatures for
the existence of CEP on the phase diagram as it would lead to large fluctuations in the thermodynamic quantities. Moreover, fluctuations are expected to show distinctly different 
behaviour in a hadron resonance gas and a QGP.

\subsection{Fluctuations of conserved charges}
Derivatives of the grand canonical partition function ($Z$) with respect to chemical potential define 
susceptibilities which experimentally become accessible through event-by-event analysis of fluctuations in observable quantities
such as baryon number, electric charge, strangeness and others.

The $n^{th}$ order susceptibility is defined as
\begin{equation}\label{eq:chi}
 \chi^n_x=\frac{1}{V T^3}\frac{\partial^n {(ln Z)}}{\partial {(\frac{\mu_x}{T})}^n},
\end{equation}
where $\mu_x$ is the chemical potential for conserved charge $x$. For our purpose $x=B$ (baryon),
$S$ (strangeness) and $Q$ (electric charge).

In Fig. \ref{fig:chi^2_temp} we have shown temperature dependence of pressure and second 
order susceptibilities for various conserved charges at zero chemical potentials ($\mu_B=\mu_S=\mu_Q=0$).
From Fig. \ref{press_temp}, where we plot pressure as a function of temperature, one can see that
the continuum limit lattice data from~\cite{Borsanyi_JHEP_11_77, Bazavov_eos_contm} agrees within error-bar with HRG 
and EVHRG model upto $T \sim 0.17$ GeV. 
The second order susceptibilities are found to increase rapidly with increase of temperature.
Near $T=0.1$ GeV magnitude of $\chi_Q^2$ is almost double compared to $\chi_S^2$ and the 
magnitude of $\chi_B^2$ is almost zero at this temperature.
As at low temperature fluctuations of 
a particular charge are dominated by  lightest hadrons carrying that charge.
The dominant contribution to $\chi_B^2$ at low temperatures comes from protons (lightest baryon), while $\chi_S^2$ 
receives leading 
contribution from kaons (lightest strange hadron) and $\chi_Q^2$ from pions (lightest charged hadron).
Since pion is lighter compared to proton and kaon,
magnitude of $\chi_Q^2$ is more than that of $\chi_B^2$ and $\chi_S^2$.
In EVHRG we have considered different values of $R_b$ and $R_m$ as shown in Fig. \ref{fig:chi^2_temp}.
It can be seen that there is almost no effect of interaction till $T=0.13$ GeV in fluctuations.
The reason for this is that the effective degree of freedom does not increase much up to this temperature
and therefore correction due to excluded volume is small.
Above $T=0.13$ GeV we find quite a substantial change in second order susceptibilities.
It can be seen from this figure that at large $T$, second order fluctuations are reduced by almost $30 \%$ compared to 
the non-interacting hadrons, if we 
take radii of all the hadrons to be $0.2$ fm. 
If we increase radii of hadrons further to $0.3$ fm, the suppression is even more.
We have compared our result with LQCD data~\cite{Bazavov, Borsanyi}. It can be seen that up to $T=0.18$ GeV,
$\chi_B^2$ is in good agreement with LQCD if we consider radii of all hadrons to be $0.2$ fm whereas
$\chi_Q^2$ is in good agreement with LQCD for $R_m=0.2$ fm and $R_b=0.3$ fm.
The meson radius plays an important role for $\chi_S^2$ and $\chi_Q^2$ but not for $\chi_B^2$ which can be
seen from the figure. This is an expected result since in $\chi_S^2$ and $\chi_Q^2$ both the baryons and mesons contribute
whereas in $\chi_B^2$ only baryons contribute. One should, however, note that the dependence of $\chi_B^2$ on $R_m$ is not
completely negligible. We will address the issue later.

In non-interacting HRG model, under Boltzmann approximation, $\chi_B^4 \approx \chi_B^2$ and $\chi_B^1 \approx \chi_B^3$ as only baryons
with baryon number one contribute to
various susceptibilities~\cite{PLB695_Karsch}. In contrast, in case of higher order susceptibilities, electric charge and strangeness is 
expected to show larger values as hadrons with multiple charge/strangeness get larger weight. A similar behaviour is expected for EVRHG 
model as well. In Fig. \ref{fig:chi^4_temp} we have shown variation of fourth order susceptibilities with temperature at $\mu=0$.
The nature of all the fourth order susceptibilities is similar to second order susceptibilities. As expected,
magnitudes of fourth order susceptibilities are larger compared to that of second order susceptibilities 
for strangeness and electric charge. However this is not true for baryon number fluctuations.
Although contributions in  $\chi_B$  are only from baryons, these quantities also depend on size of mesons
as can be seen from Fig. \ref{chi_B2_temp} and Fig. \ref{chi_B4_temp}. This dependence can be understood from Eqs. (\ref{eq:p_new}) 
and (\ref{eq:mu}) which shows the
dependence of chemical potential on hadronic radii through pressure. Hence, even for baryon number susceptibilities there may be
small difference in magnitudes of susceptibilities at different $R_m$ for EVHRG. 
We compare our result with LQCD data ($N_\tau = 6, 8$) \cite{Schmidt}. 
The LQCD data for $\chi_B^4$ and $\chi_S^4$ are in good agreement with HRG model up to $T=0.16$ GeV.
Whereas for $\chi_Q^4$, LQCD data is lower compared to both HRG and EVHRG.

It has been argued in~\cite{PLB718_Andronic} that for $\epsilon$, $P$ and $s$, though lattice QCD results agree at lower temperatures,
they are expected to becomes larger than those in HRG at higher temperatures due to deconfinement. On the other hand, the difference 
between lattice and HRG for the higher susceptibilities may have effects due to the difference in hadronic masses. It has also been 
reported that the thermodynamic quantities calculated in lattice QCD may agree well with the HRG model if the masses of the hadrons 
in the model are tuned appropriately to take into account the discretization errors in the hadron spectrum present in the lattice 
calculations~\cite{huovinen_2010, borsanyi_2011}. In the present study EVHRG is found to show a better agreement with the lattice
data in the continuum limit.

 \subsection{\label{sec:Correlations}Correlations among different conserved charges}

Correlations among different conserved charges may act as probes of the structure of QCD at finite temperature.
In QGP, as baryon number as well as electric charge  are carried by different flavours of quarks, a strong correlation is 
expected between B-Q, Q-S as well as B-S. On the other hand, in the hadronic sector presence of baryons and mesons would generate 
an entirely different type of correlations between these quantities. 
Hence, these correlations are expected to show changes across the phase transition which are characteristics of the changes
in the relevant degrees of freedom.

  \begin{figure}[]
   \subfigure {\includegraphics[scale=0.34,angle=-90]{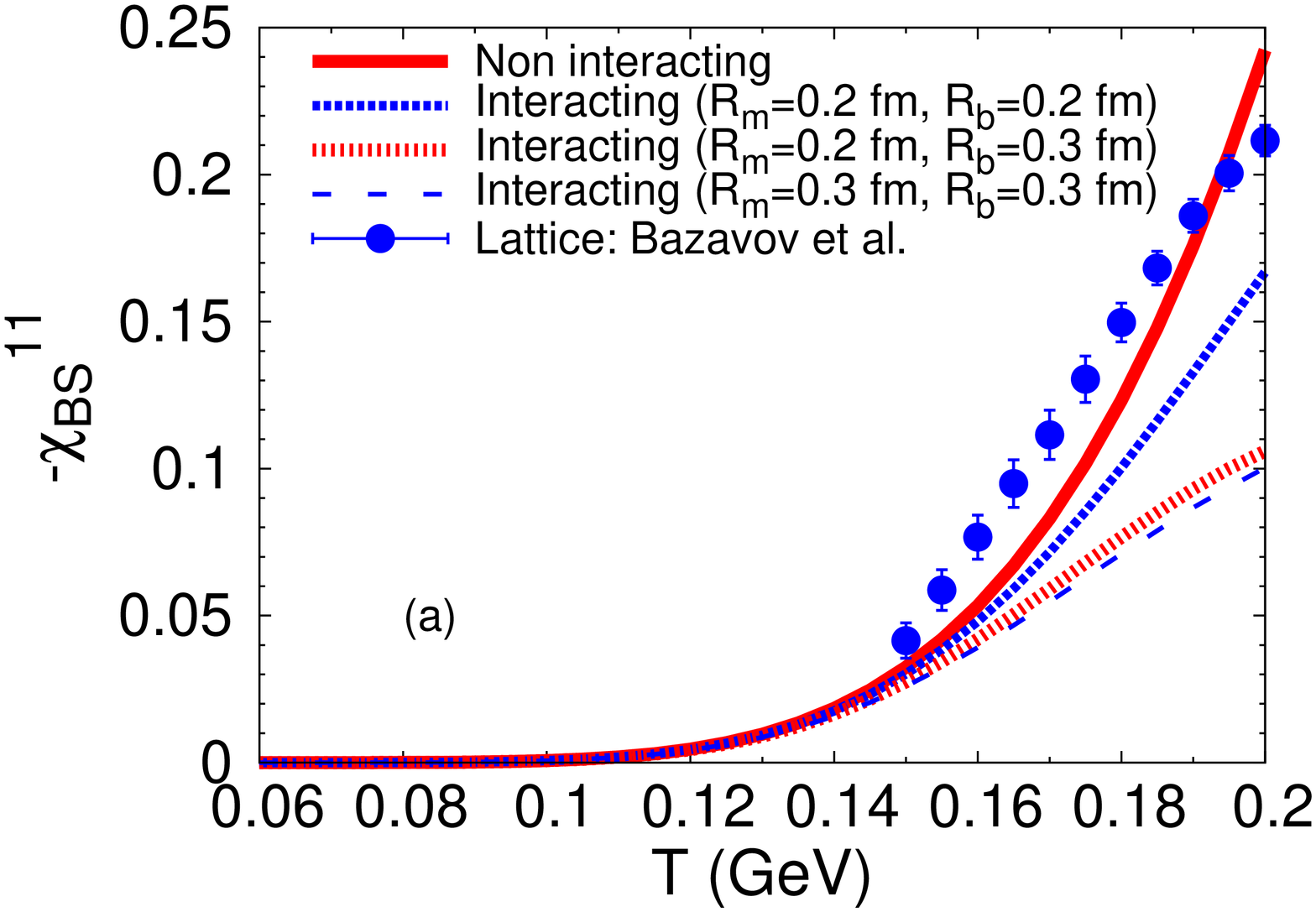}\label {chi_BS11_temp}}
    \subfigure {\includegraphics[scale=0.34,angle=-90]{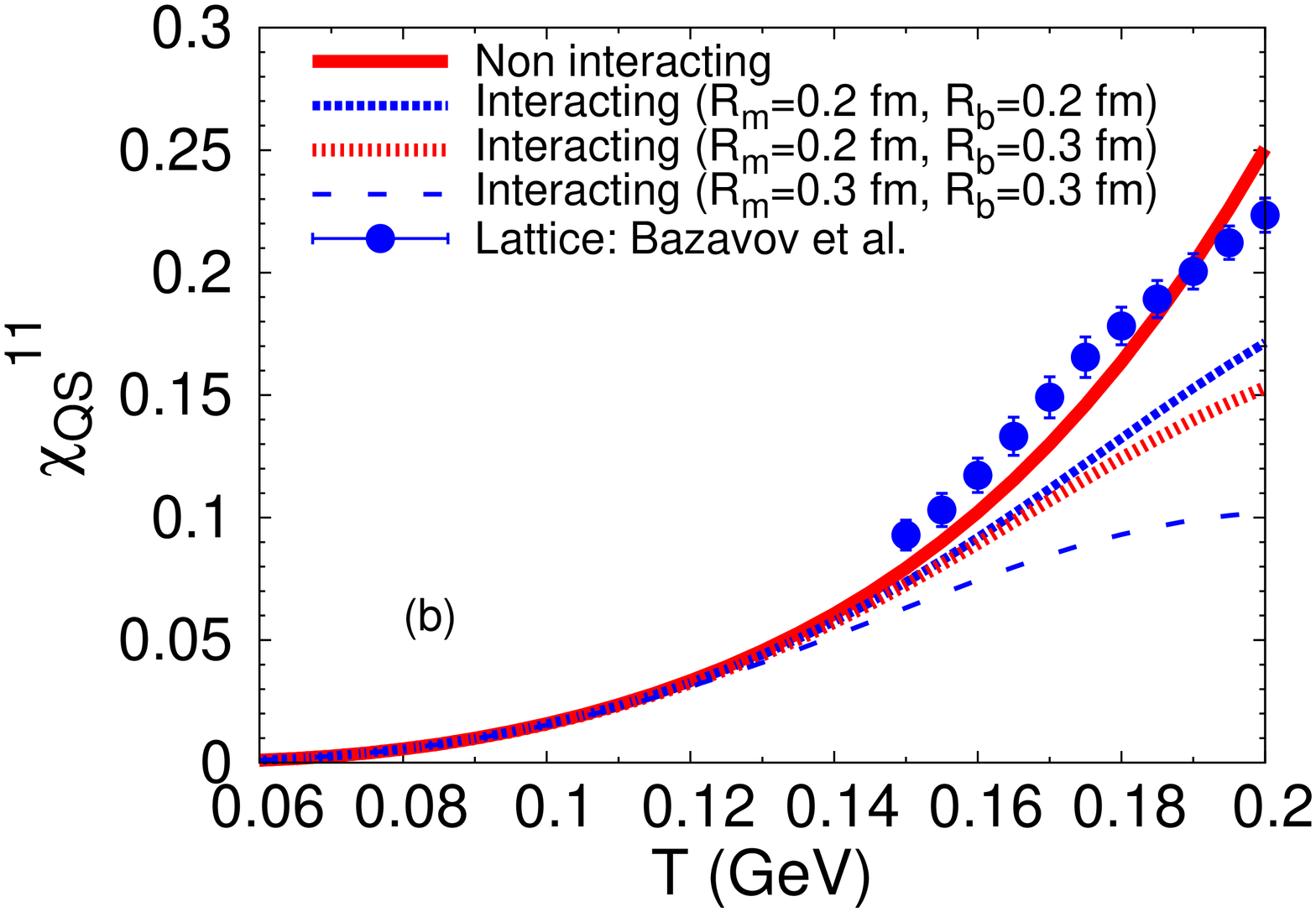}\label {chi_QS11_temp}}
     \subfigure {\includegraphics[scale=0.34,angle=-90]{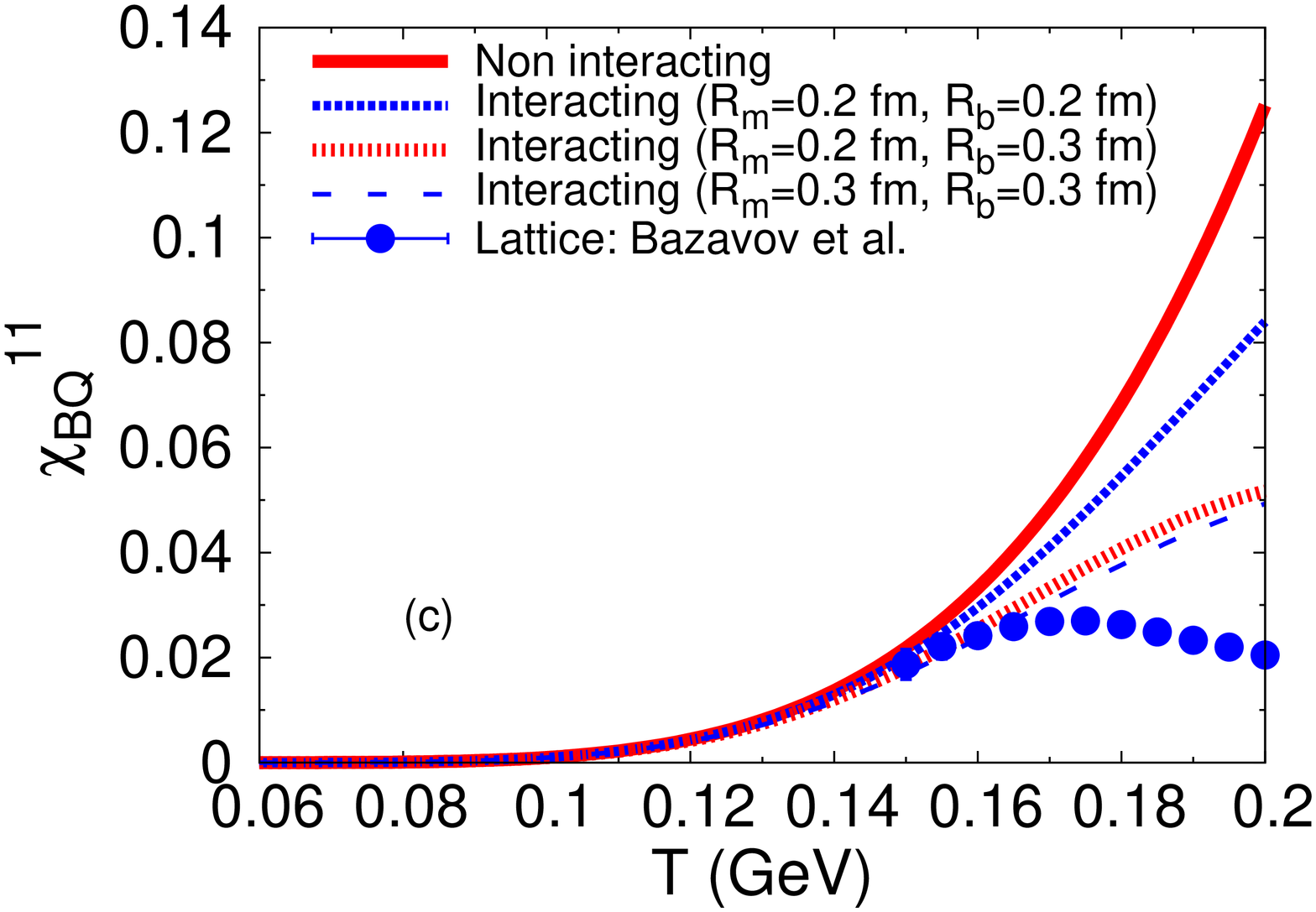}\label {chi_BQ11_temp}}
 \caption{(Color online). \label{fig:chi_BSQ11_temp}Variation of $\chi_{BS}^{11}$, $\chi_{BQ}^{11}$ and $\chi_{QS}^{11}$ 
 with temperature at $\mu=0$. Lattice data for continuum extrapolation is taken from Ref.~\cite{Bazavov}.}
 \end{figure}
\begin{figure}[]
  \subfigure {\includegraphics[scale=0.34,angle=-90]{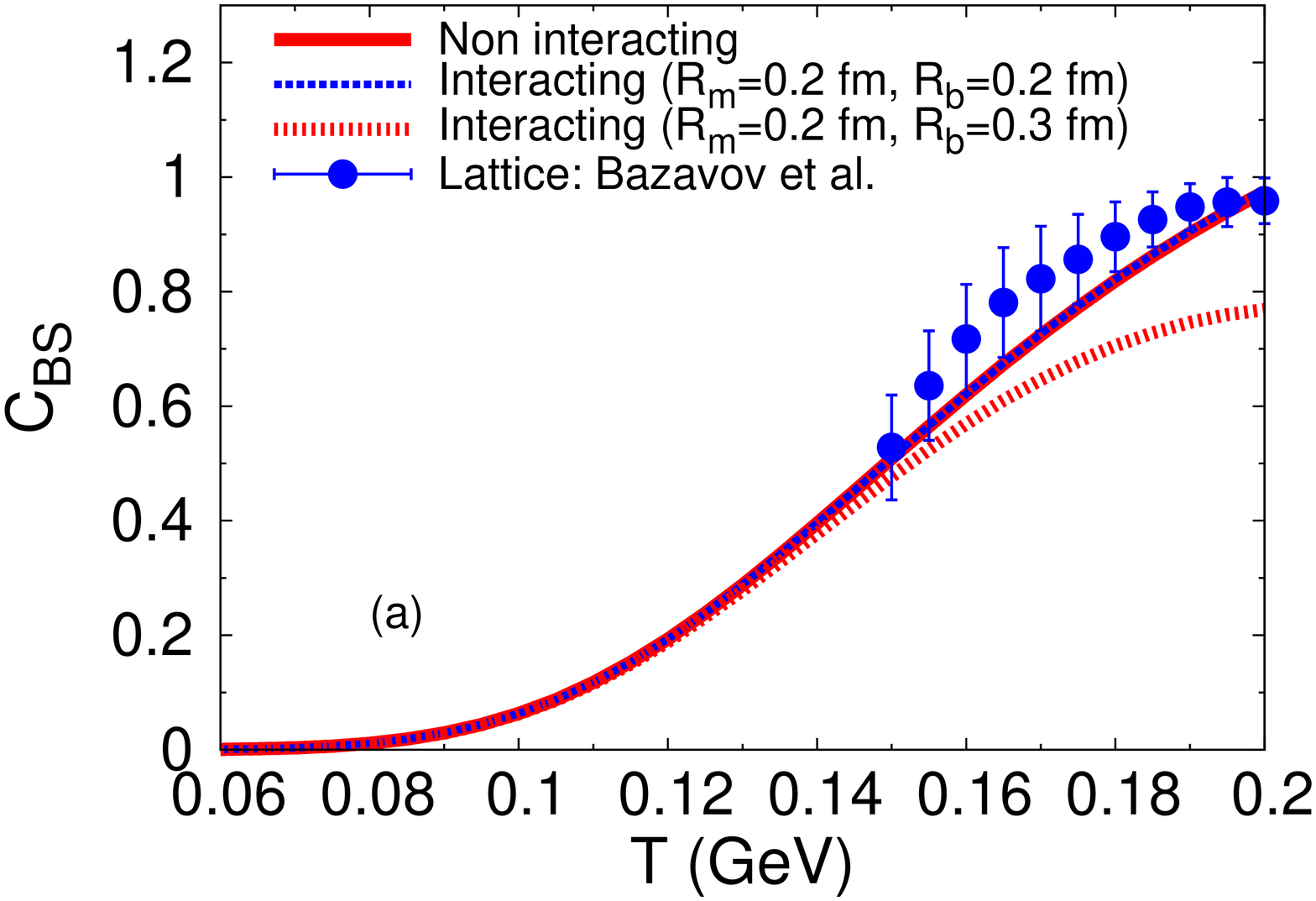}\label {cbs_temp}}
   \subfigure {\includegraphics[scale=0.34,angle=-90]{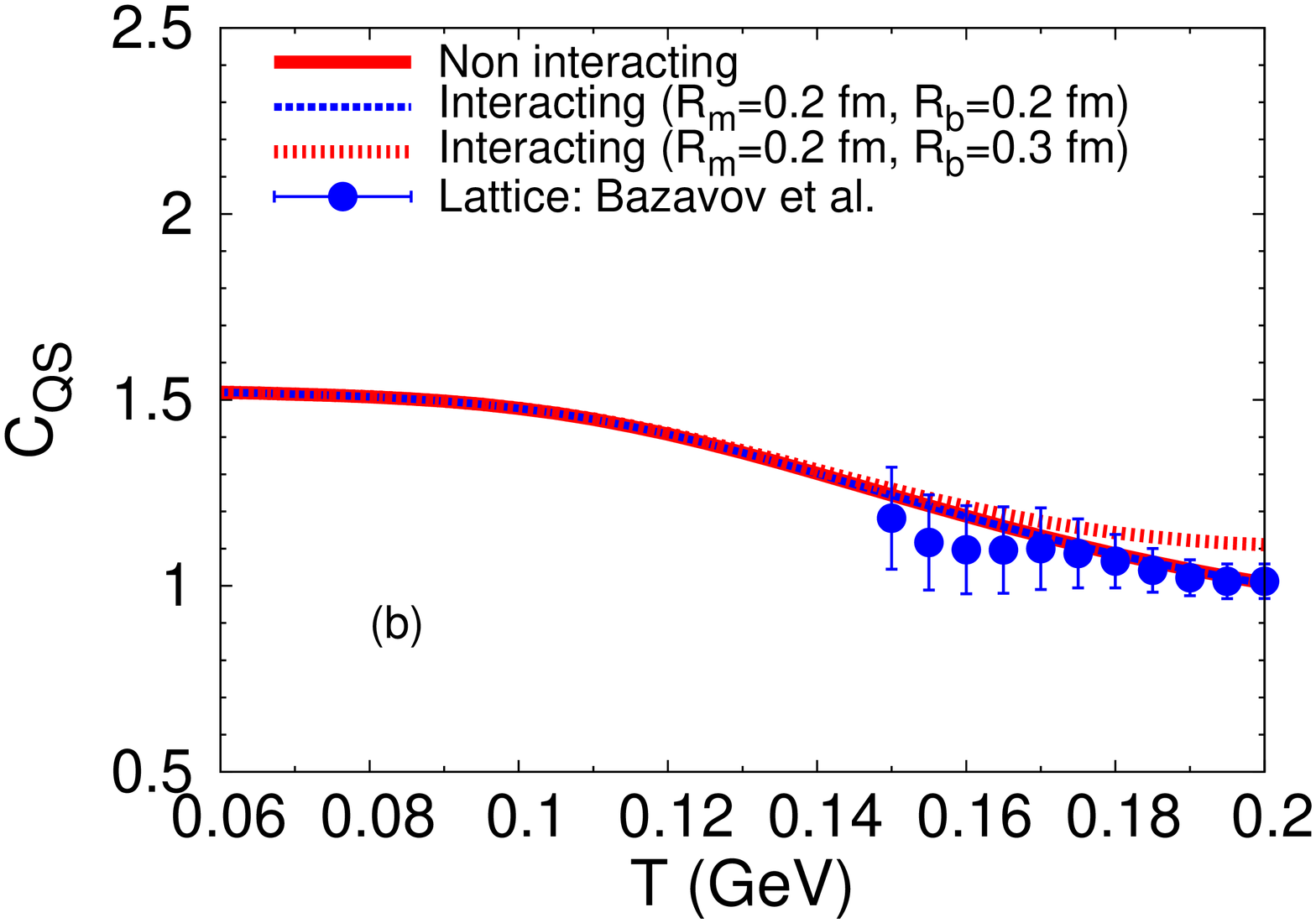}\label {cqs_temp}}
    \subfigure {\includegraphics[scale=0.34,angle=-90]{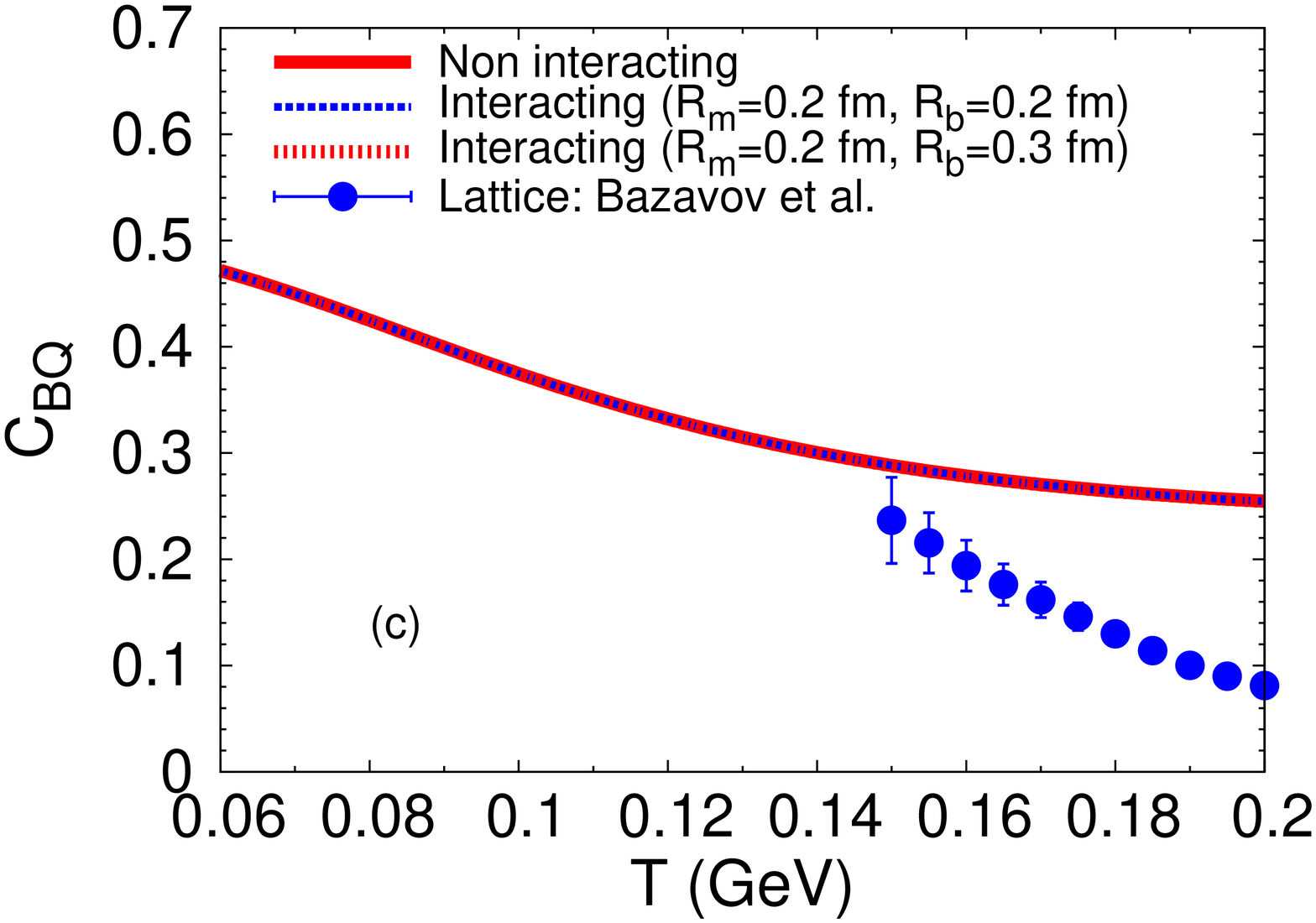}\label {c_bq_temp}}
 \caption{(Color online). \label{fig:C_BS_QS_temp}Variation of $C_{BS}$, $C_{QS}$ and $C_{BQ}$ with temperature at $\mu=0$.
 Lattice data for continuum extrapolation is taken from Ref.~\cite{Bazavov}.}
 \end{figure}

Correlation functions, $\chi^{ij}_{xx'}$, are defined by
 \begin{equation}\label{eq:corr}
 \chi^{ij}_{xx'}=\frac{1}{V T^3}\frac{\partial^{i+j} {(ln Z)}}{(\partial {(\frac{\mu_x}{T})}^i)(\partial {(\frac{\mu_{x'}}{T})}^j)},
\end{equation}
where {$x$} and {$x'$} correspond to conserved charges {\it B,S,Q} and $\mu$'s are chemical potentials of
corresponding conserved charges. 
 In this section we show temperature dependence of various correlation functions at $\mu_B=0$. We also compare our result with LQCD 
data~\cite{Bazavov}.

Fig. \ref{fig:chi_BSQ11_temp} shows the variation of correlations between conserved charges with temperature around $\mu_x=\mu_{x'}=0$.
Since at very low temperature main contribution comes from pions (baryon number as well as strange quantum number $=0$),
all the correlations remain zero.
The next particle to be excited is kaon and as a result $\chi_{QS}^{11}$ becomes non zero around $T=0.075$ GeV. The leading contribution to 
$\chi_{QS}^{11}$ in the hadronic sector is due to charge kaons which has same sign for charge and strange quantum number. So
it remains positive and increases with temperature.
Other correlations pick up non-zero values above $T=0.1$ GeV when baryons
start populating the system.  
 $\chi_{BS}^{11}$ is proportional to the product of baryon number and
strange quantum number. So here most of the contribution is due to lightest baryon $\varLambda^{}$ which
has baryon number $+1$ and strange quantum number $-1$.  Other contributing particles,
such as, $\Sigma$, $\Xi$ etc. also have relative negative sign between baryon number and strange quantum number. 
As a result $\chi_{BS}^{11}$ remains negative. This negative value increases with $T$ due to the increase in the population 
of the strange baryons. Again, HRG results show sharp increase compared to those from EVHRG model. Due to the contribution 
from mesons (mainly kaons)
in $\chi_{QS}^{11}$, meson radius plays a dominating role and an increase in meson radius produces larger suppression in EVHRG model.
On the other hand, baryon radius plays a major role for both $\chi_{BS}^{11}$ as well as $\chi_{BQ}^{11}$. The dominant contribution in 
$\chi_{BQ}^{11}$ at low temperature is due to proton and antiproton and for both of those  baryon 
number as well as charge carry the same sign. So the value of $\chi_{BQ}^{11}$ remains positive and shows a sharp increase
with temperature in HRG, EVHRG being suppressed due to the effect of hard core radius. 
In the quark sector strange quarks (antiquarks) carry $1/3$ (-$1/3$) baryon number and the relative sign between $B$ and $S$ is 
always negative. As a result lattice result for $\chi_{BS}^{11}$ show a similar trend as that of HRG. On the other hand, $\chi_{QS}^{11}$ 
is always positive as charge as well as strangeness for strange quark is negative. Though the lattice values at lower temperature 
($0.15$ GeV) is close to HRG and EVHRG, $\chi_{BS}^{11}$ as well $\chi_{QS}^{11}$ for lattice increase faster than 
those for HRG as strange quark mass is much smaller 
compared to strange baryons.
At high temperature quark masses decrease (which will eventually become zero at Stefan-Boltzmann limit) and lattice results
would start saturating whereas HRG results would keep on increasing. As a result lattice values
are suppressed compared to HRG values  both for $\chi_{QS}^{11}$ and $\chi_{BS}^{11}$.
In the case of $\chi_{BQ}^{11}$, EVHRG ($R_m$=0.2 fm and $R_b$=0.3 fm) results agree well with the lattice data for $T < 0.17$ GeV.
But at higher temperatures, the weighted sum of charges of massless quarks vanishes so the lattice results would start saturating with the 
further increase in temperature whereas HRG results would keep on increasing similar like $\chi_{QS}^{11}$ and $\chi_{BS}^{11}$.
 
The LQCD results of charge correlations has been found to deviate from the ideal gas limit even at twice the transition temperature and
the large contributions of flavour fluctuations are expected to be the cause of these deviations~\cite{Bazavov}.
In order to analyse the experimental results, suitable ratios have been proposed to eliminate these contributions~\cite{PRL95_182301_Majumder}.
Three ratios, namely, baryon-strangeness correlation coefficient $C_{BS}$~\cite{PRL95_182301_Majumder}, electric charge-strangeness 
correlation coefficient $C_{QS}$ and baryon-electric charge correlation coefficient $C_{BQ}$ can be written as~\cite{Bazavov}
\begin{equation}
 C_{BS}=-3\frac{\chi_{BS}^{11}}{\chi_S^2}, ~~~~~     C_{QS}=3\frac{\chi_{QS}^{11}}{\chi_S^2}, ~~~~~
 C_{BQ}=\frac{\chi_{BQ}^{11}}{\chi_B^2}.
\end{equation}

Figure \ref{fig:C_BS_QS_temp} shows variation of $C_{BS}$, $C_{QS}$ and $C_{BQ}$ with temperature at $\mu=0$. 
We compare our result with LQCD data~\cite{Bazavov}.
At low temperature $C_{BS}$ is much lower than one because denominator contains mostly kaon whereas contribution to numerator comes
from baryons, mostly $\varLambda$, which is less populated than kaon. It increases with temperature but remains below 1 due to 
larger contribution from mesons.  If all the hadronic radii are same, the interaction effect gets cancelled. On the other hand, 
for larger baryonic radii, baryon population gets suppressed and $C_{BS}$ value for EVHRG becomes lower than HRG. In case of 
simple QGP, or more generally, for a system where the quark flavours
are uncorrelated, this value will be unity. Since lattice $\chi_{BS}^{11}$ is higher, 
lattice $C_{BS}$ values remain above the HRG (and EVHRG) and saturates at high temperature.

At very low temperature, contribution to numerator of $C_{QS}$ come from  charged kaons only while the contributions
to denominator
come from all the kaons. So $C_{QS}$ comes out to be around $1.5$.
With increase of temperature lambda starts populating the system and $C_{QS}$ becomes close to $1$.
There is no effect of interaction in most of the temperature range except at very high temperatures. Here, a higher baryonic radii 
(compared to meson radii) induces a small increase in $C_{QS}$. Interaction effects get cancelled if radii are same for all hadrons. 
$C_{BS}$ and $C_{QS}$ are related through the equation $C_{QS}=0.5(3-C_{BS})$~\cite{Bazavov}. So in lattice studies~\cite{Bazavov} 
at low temperature, where $C_{BS}$ is very small $C_{QS}$ is near 1.5, same compared to HRG (and EVHRG) 
and at high temperature tends to one.

At low temperatures, contribution to numerator of $C_{BQ}$ come from  protons only while the contributions to denominator
come from both protons and neutrons. As a result $C_{BQ}$ is close to $0.5$.
It decreases slowly with increase in $T$ as at high $T$ heavier 
neutral baryons like $\varLambda$ start contributing to the denominator whereas only
charged baryons contribute to numerator. For both HRG and EVHRG $C_{BQ}$ is very close to each other due to the cancellation 
of interaction effects. In lattice, $C_{BQ}$ approaches zero at high $T$ as the system is in QGP phase. However at lower
temperature it approaches the HRG/EVHRG values.

\subsection{\label{sec:mub}High density scenario}
  \begin{figure}[]
  \centering
  \subfigure {\includegraphics[scale=0.34,angle=-90]{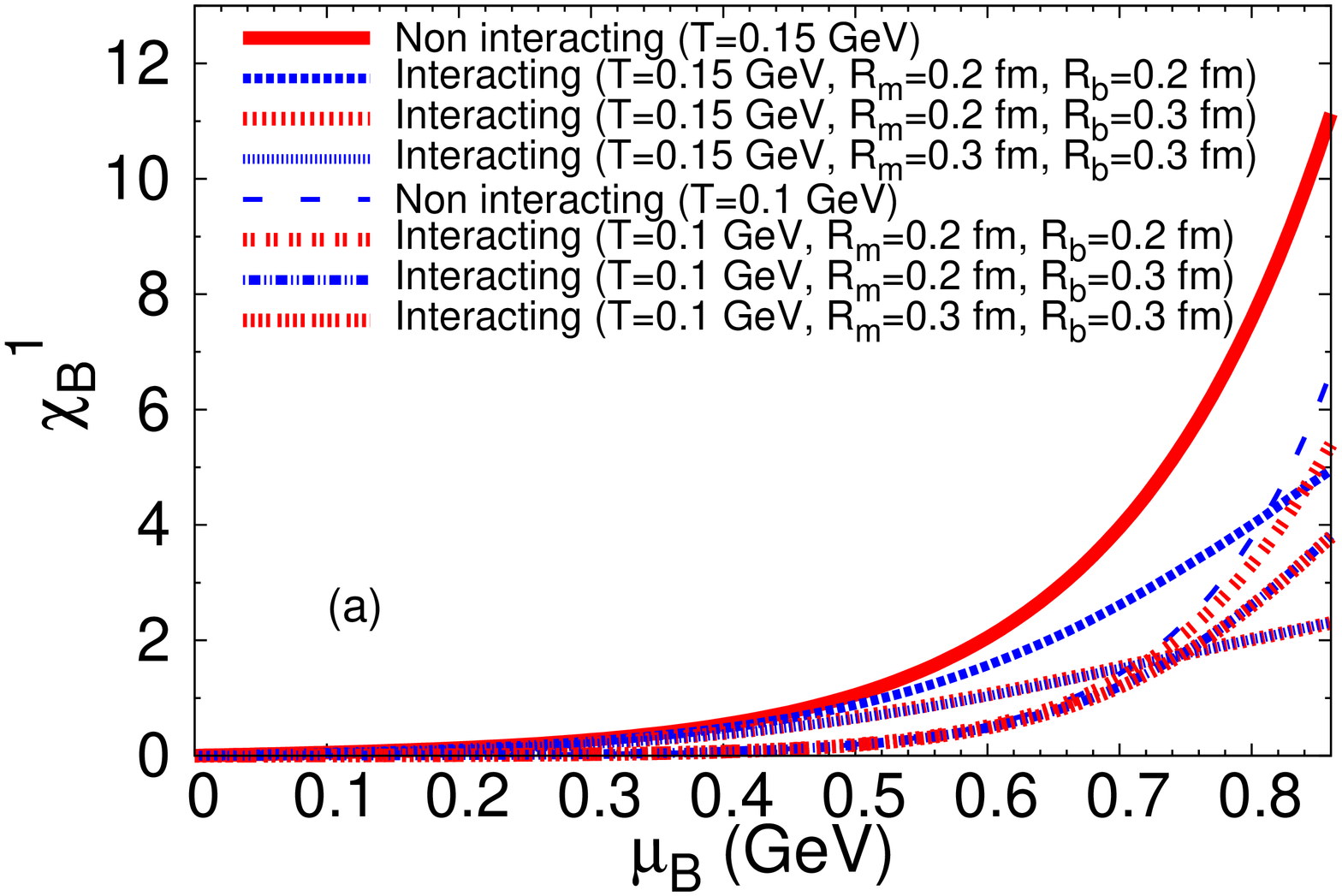}\label {chi_B1_mub}}
  \subfigure {\includegraphics[scale=0.34,angle=-90]{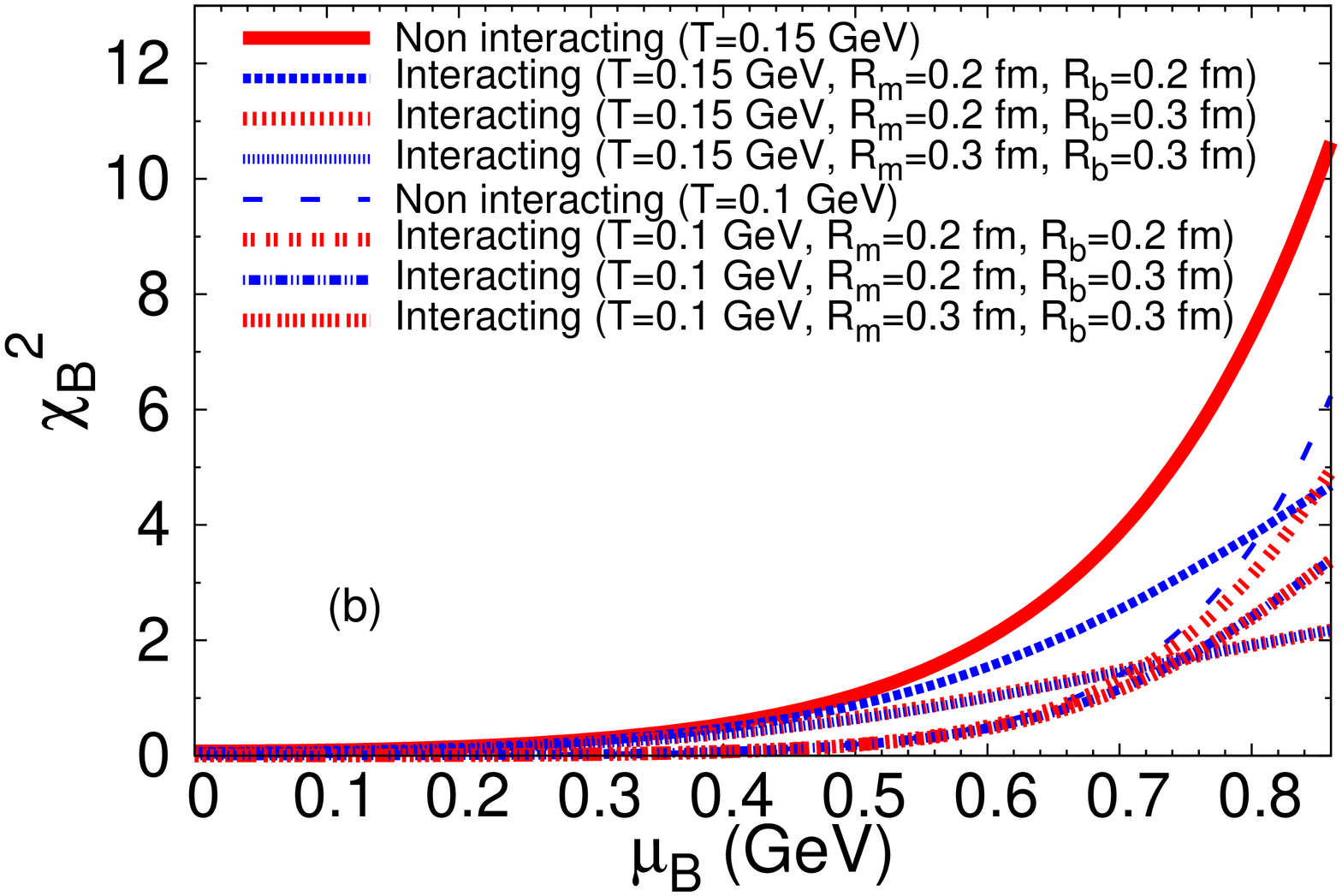}\label {chi_B2_mub}}
   \subfigure {\includegraphics[scale=0.34,angle=-90]{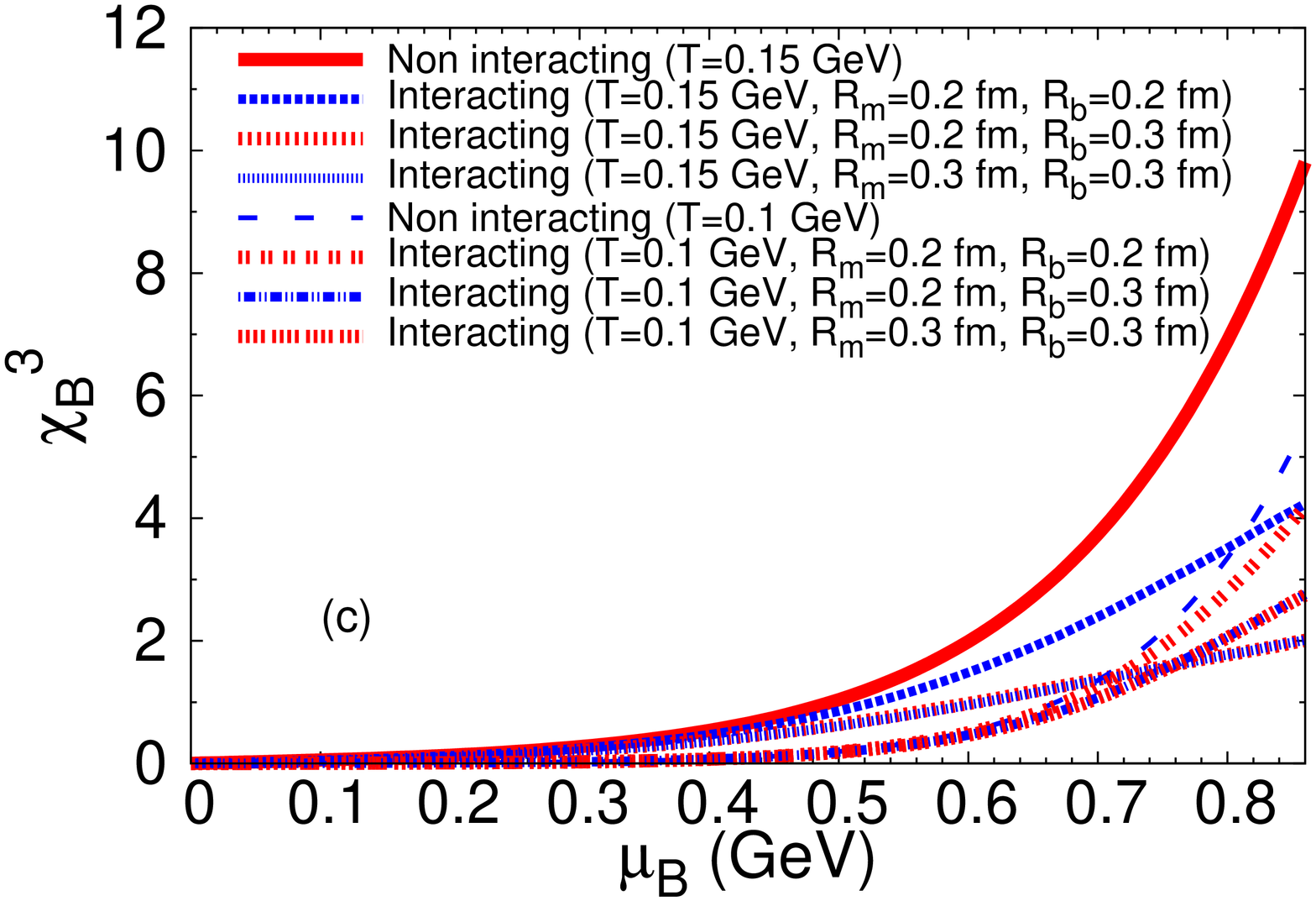}\label {chi_B3_mub}}
    \subfigure {\includegraphics[scale=0.34,angle=-90]{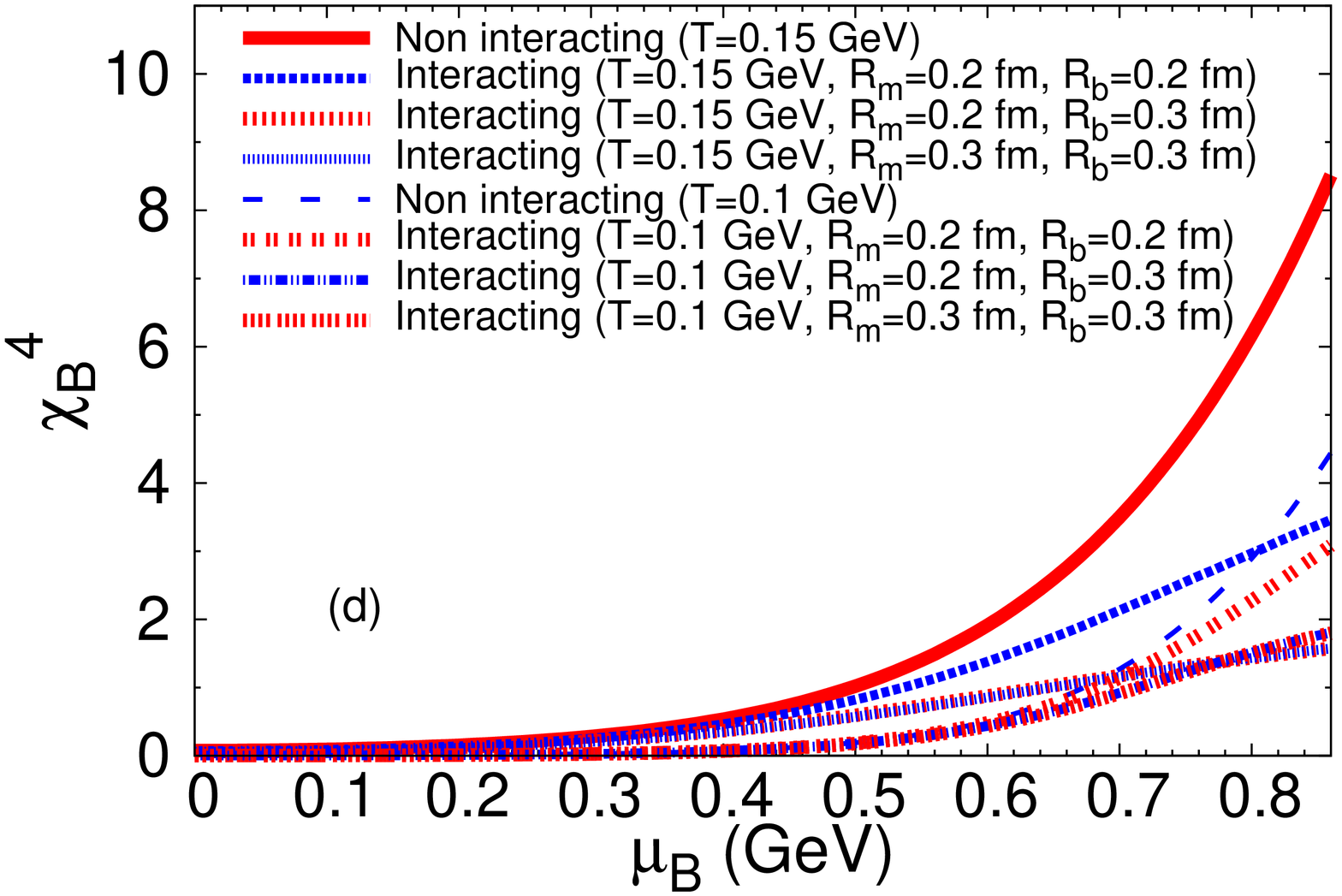}\label {chi_B4_mub}}
 \caption{(Color online). \label{fig:chi_B_mub}Susceptibilities $\chi_B^1$, $\chi_B^2$, $\chi_B^3$, $\chi_B^4$ as a function of $\mu_B$ keeping $T$ fixed and $\mu_S=\mu_Q=0$.}
 \end{figure}
 \begin{figure}
 \centering
   \subfigure {\includegraphics[scale=0.34,angle=-90]{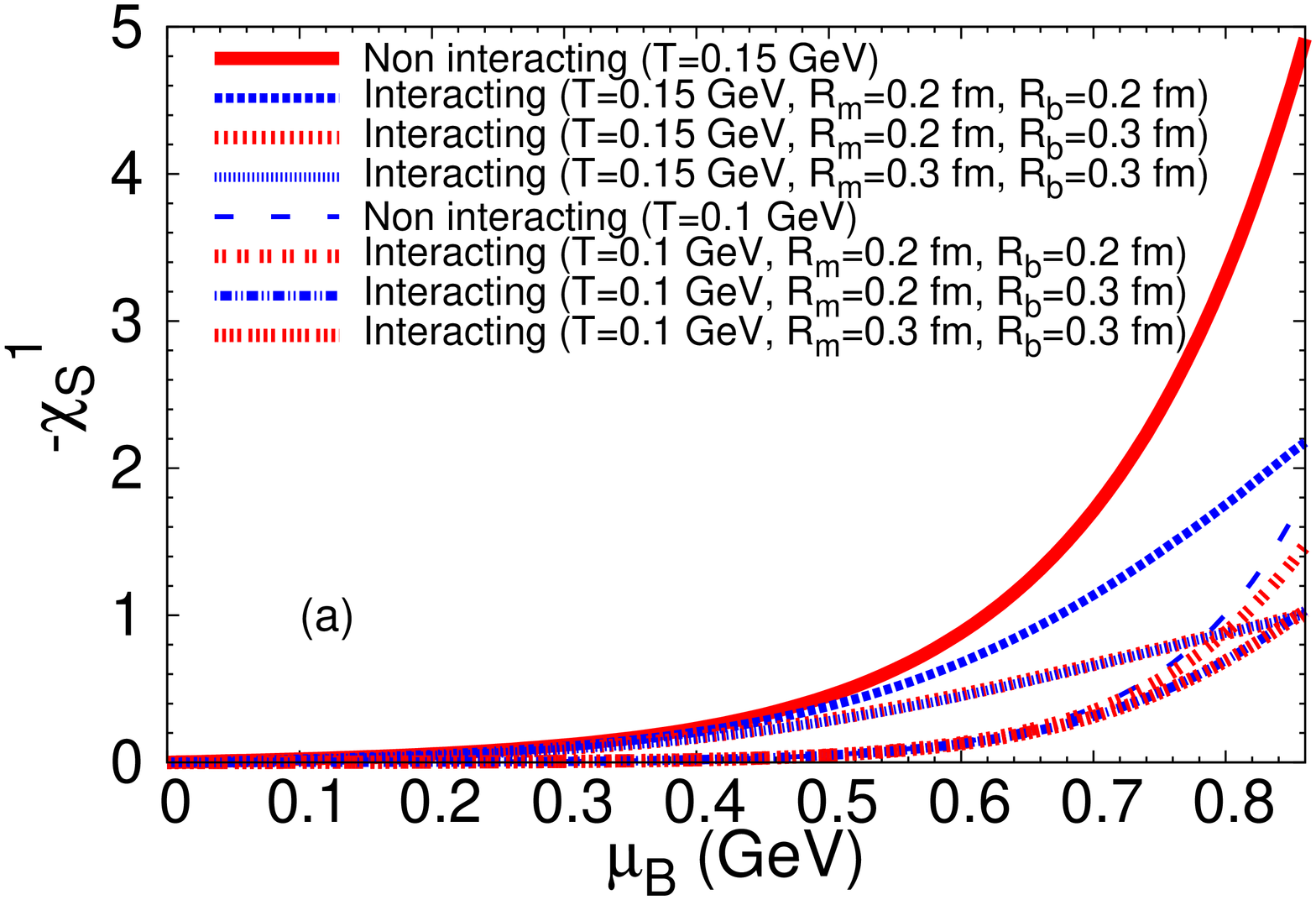}\label {chi_S1_mub}}
    \subfigure {\includegraphics[scale=0.34,angle=-90]{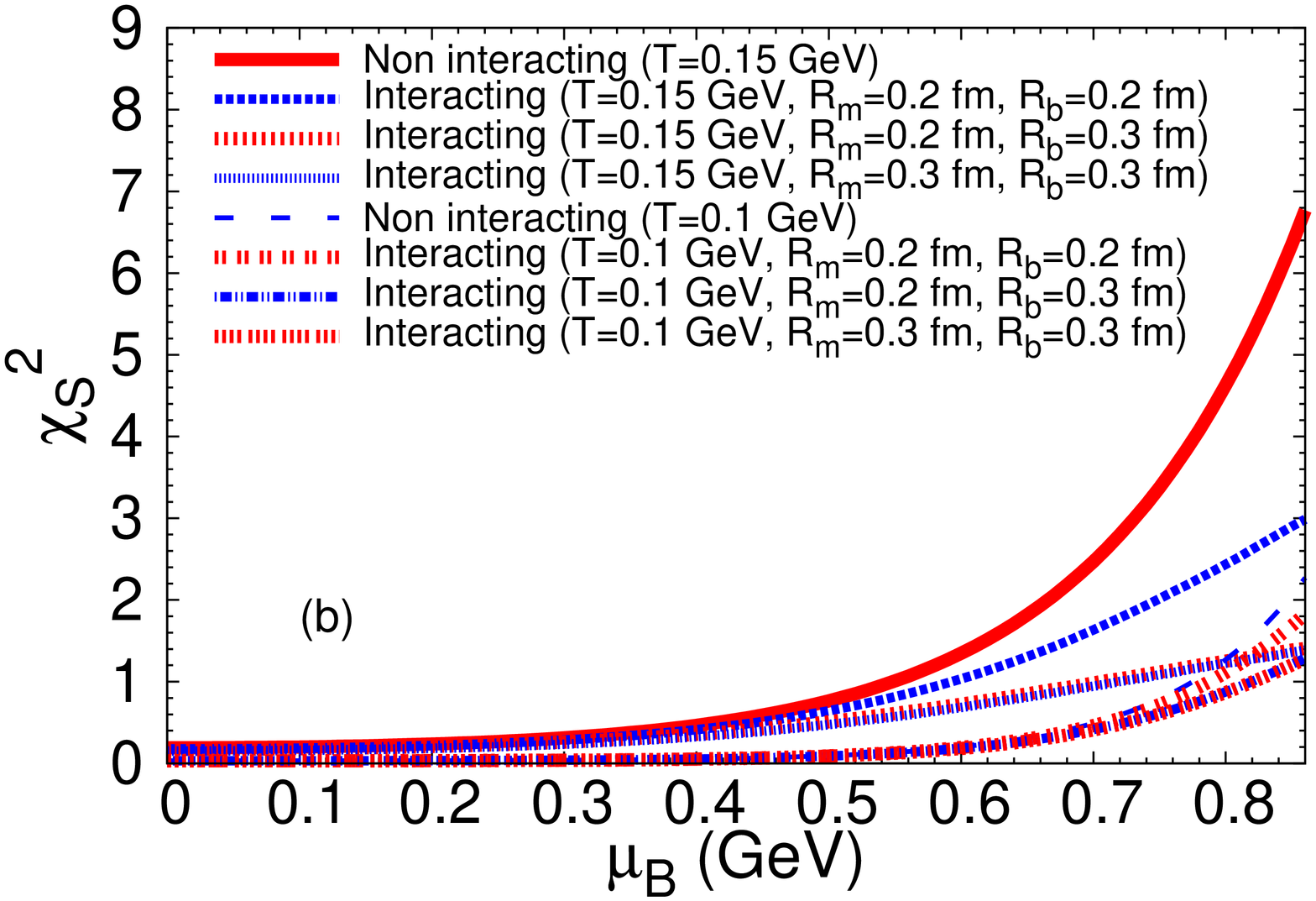}\label {chi_S2_mub}}
     \subfigure {\includegraphics[scale=0.34,angle=-90]{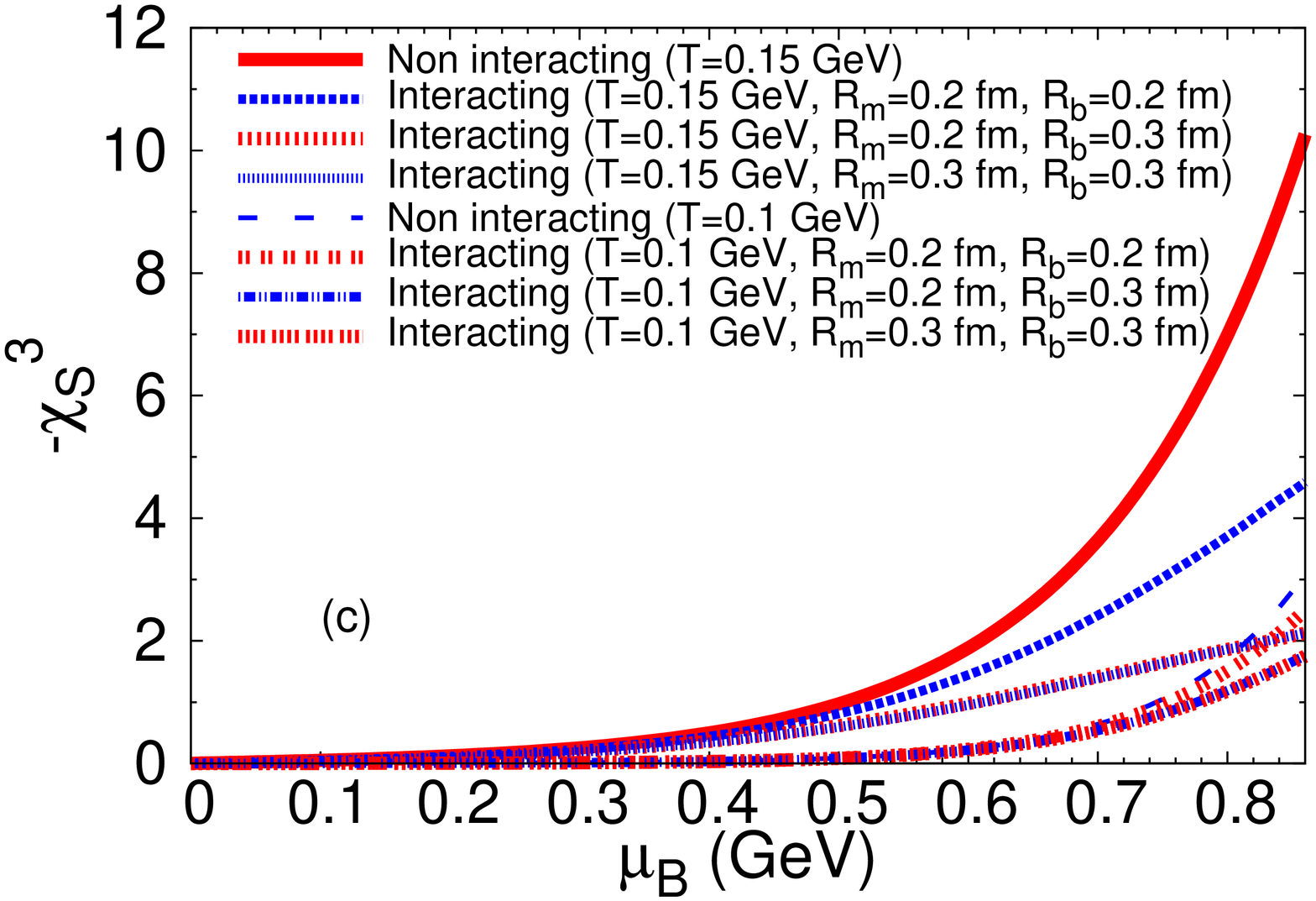}\label {chi_S3_mub}}
      \subfigure {\includegraphics[scale=0.34,angle=-90]{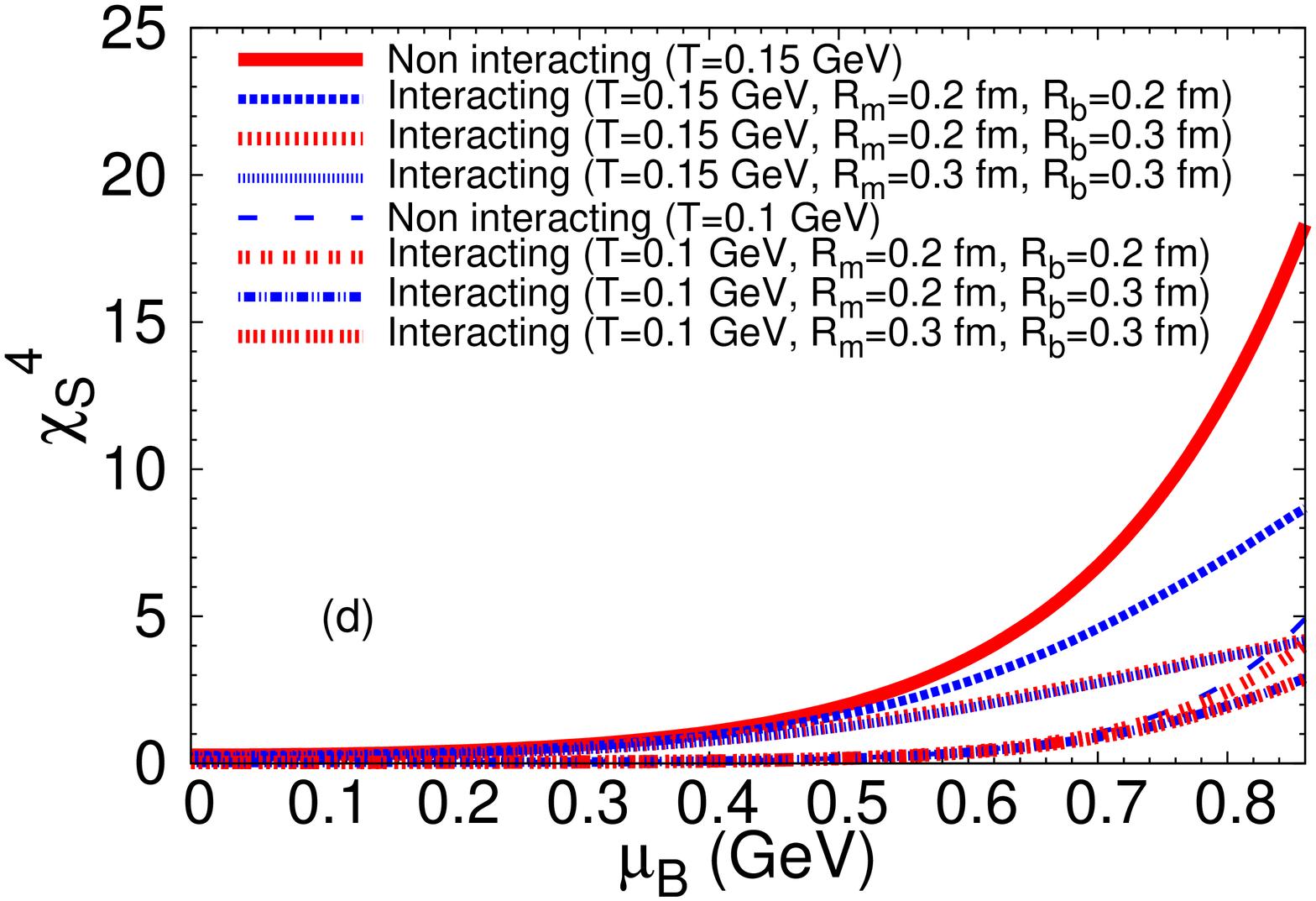}\label {chi_S4_mub}}
 \caption{(Color online). \label{fig:chi_S_mub}Susceptibilities $\chi_S^1$, $\chi_S^2$, $\chi_S^3$, $\chi_S^4$ as a 
 function of $\mu_B$ keeping $T$ fixed and $\mu_S=\mu_Q=0$.}
 \end{figure}
 \begin{figure}
 \centering
   \subfigure {\includegraphics[scale=0.34,angle=-90]{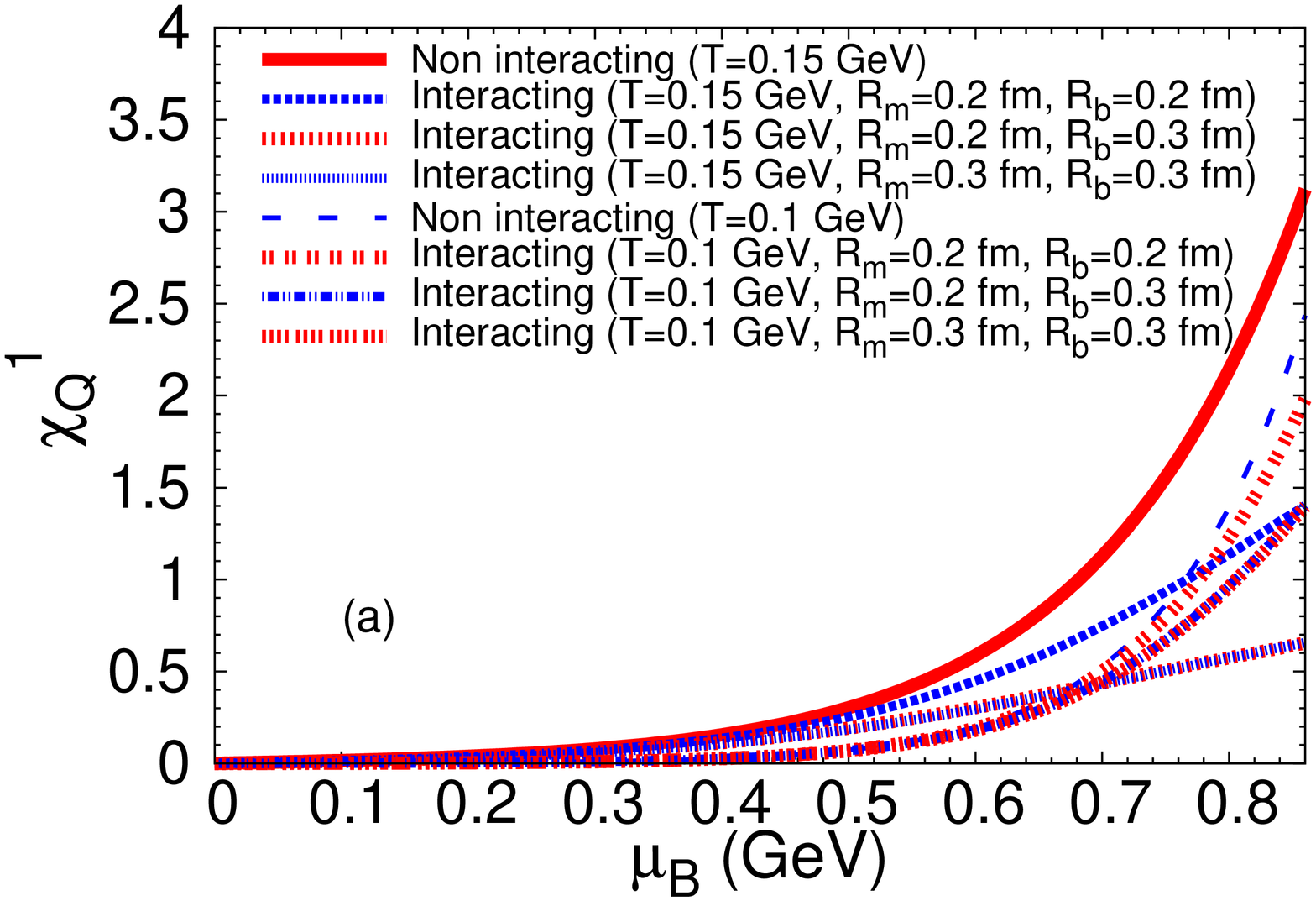}\label {chi_Q1_mub}}
    \subfigure {\includegraphics[scale=0.34,angle=-90]{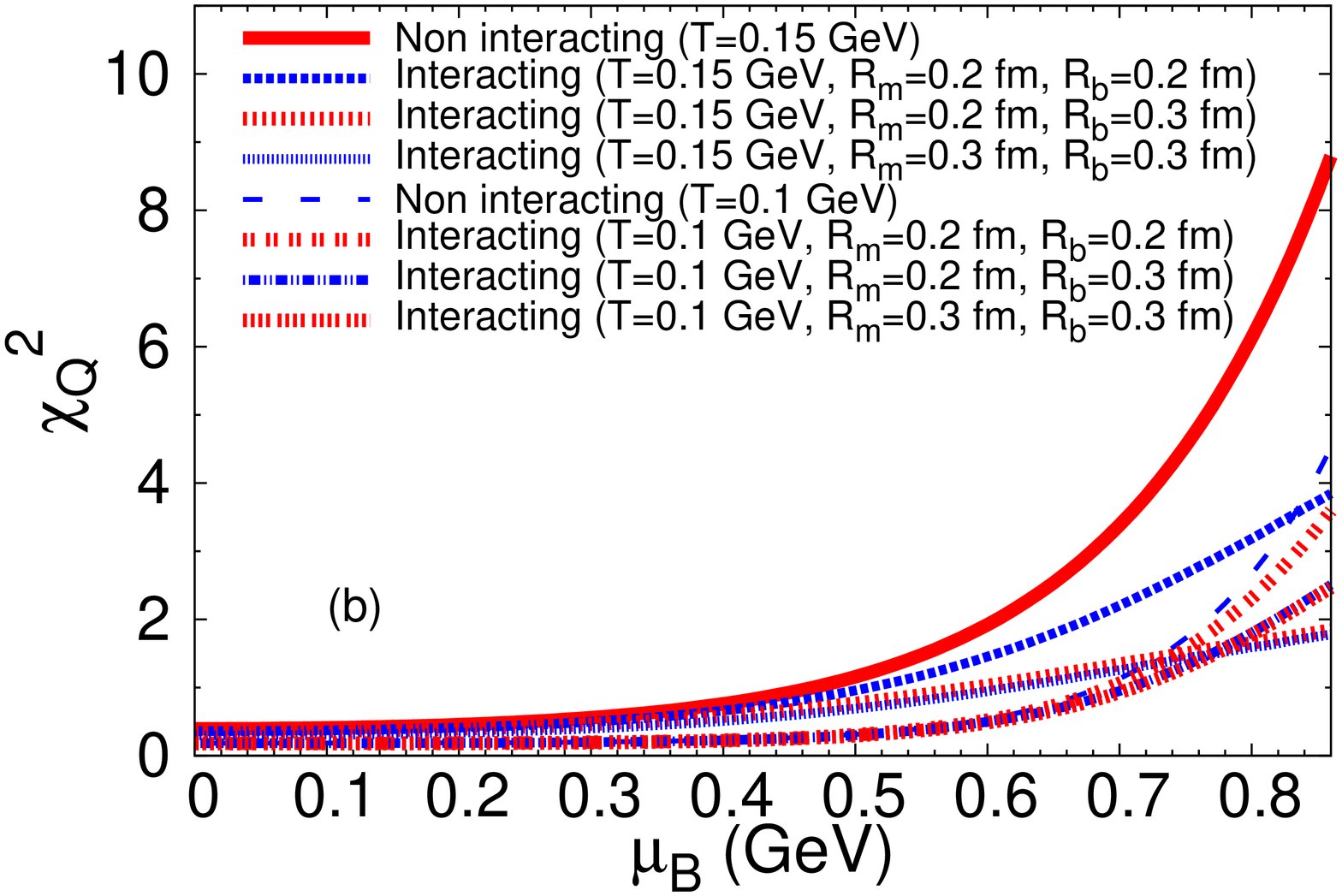}\label {chi_Q2_mub}}
     \subfigure {\includegraphics[scale=0.34,angle=-90]{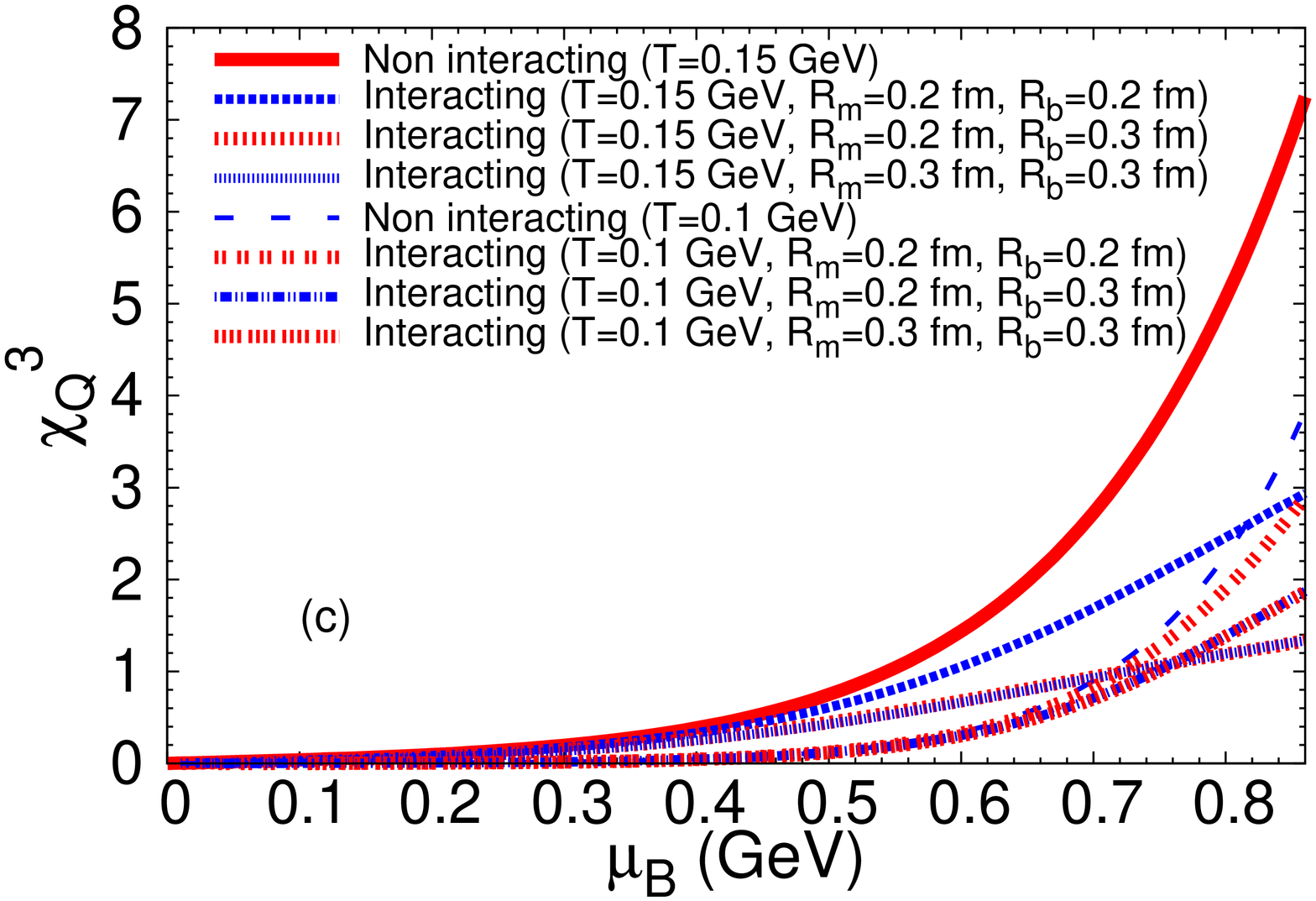}\label {chi_Q3_mub}}
      \subfigure {\includegraphics[scale=0.34,angle=-90]{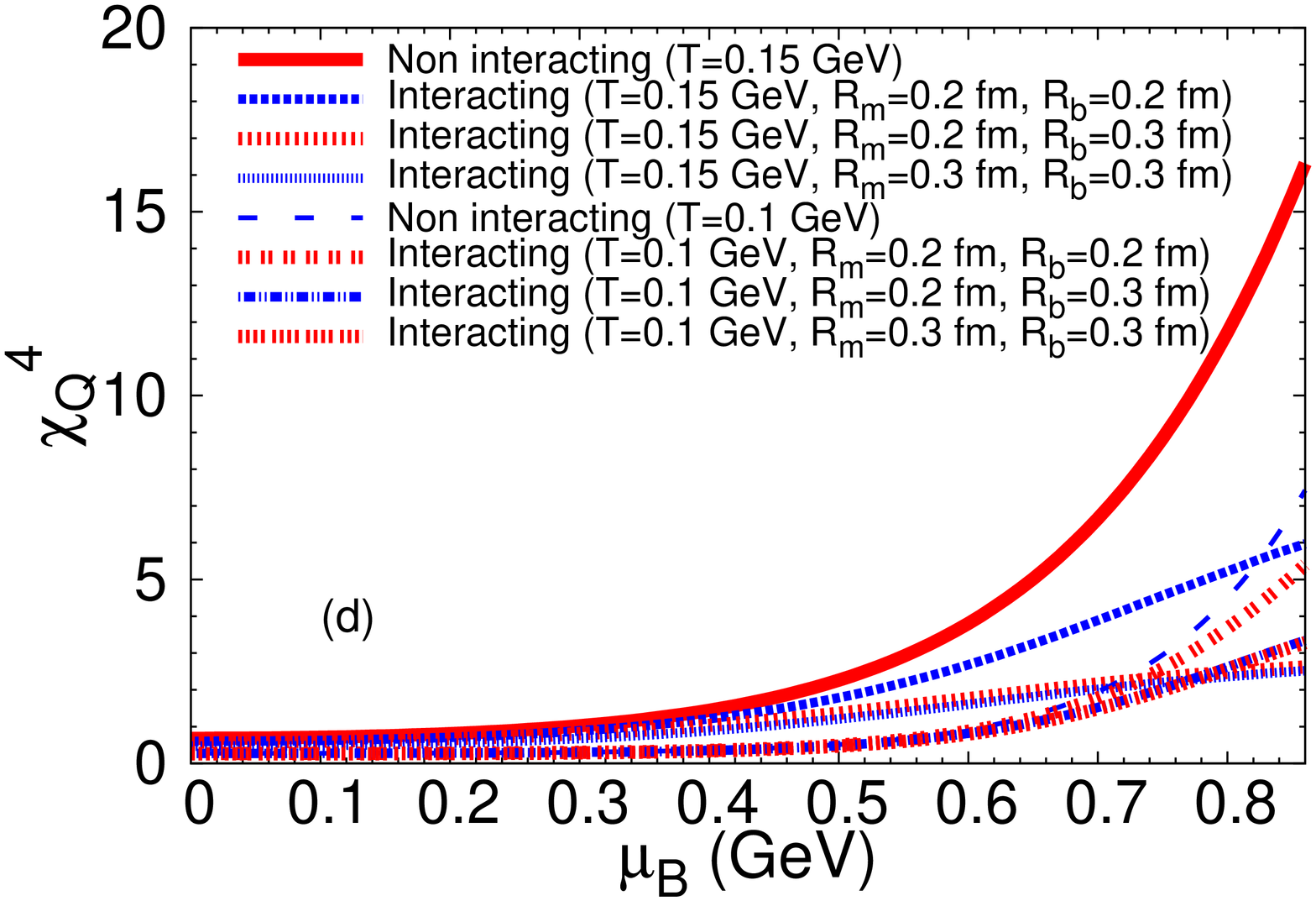}\label {chi_Q4_mub}}
 \caption{(Color online). \label{fig:chi_Q_mub}Susceptibilities $\chi_Q^1$, $\chi_Q^2$, $\chi_Q^3$, $\chi_Q^4$ as a function of $\mu_B$ keeping $T$
 fixed and $\mu_S=\mu_Q=0$.}
 \end{figure}

At very high densities and low temperatures, strongly interacting matter is expected to undergo a first order phase transition 
which ends at CEP as one moves towards higher $T$ and lower $\mu_B$ along the phase boundary. At finite $\mu_B$ due to the 
expected stronger divergence at the phase boundary and at CEP, study of higher order moments is extremely important at finite $\mu_B$. 
In fact, not only the magnitude but the signs of third moments have also been proposed as a signature for CEP~\cite{asakawa_prl}.

In this section we have discussed the results of fluctuations and correlations at large $\mu_B$. 
In Fig. \ref{fig:chi_B_mub} - \ref{fig:chi_Q_mub}. we show variation of $\chi_x^1$, $\chi_x^2$, $\chi_x^3$, $\chi_x^4$ 
with $\mu_B$ for a fixed temperature and at $\mu_S=\mu_Q=0$ where $x=B, S, Q$. Here we have considered
 $T=0.1$ GeV and $0.15$ GeV for the purpose of illustration. For each temperatures we show variation of the quantities with $\mu_B$ in 
HRG and EVHRG model with different baryon ($R_b$) and meson ($R_m$) radii. 
The effect of interaction is found to be more pronounced at 
high temperatures. It can be seen from these figures that at high $T$ and large
$\mu_B$ the magnitude of fluctuations are suppressed by a factor of $2$ or more, compared to HRG, if we take the radii of all 
the hadrons to be $0.2$ fm.
At high $\mu_B$  fluctuations are very sensitive to $R_b$ but not to $R_m$ as in this region system is dominated 
by baryons. Similar behaviour was obtained in Fig. \ref{fig:EOS_mub} as well.
$\chi^1_S$ and $\chi^3_S$ are proportional to the odd power of strange quantum number of the particles.
Dominant contribution to $\chi_S$ comes from $\varLambda$, the lightest strange baryon  
which has strange quantum number $-1$ (at $\mu_S=0$ strange mesons don't contribute to $\chi^1_S$ and $\chi^3_S$
because particle and anti-particle terms are equal in magnitude but opposite in sign).
As a result, $\chi^1_S$ and $\chi^3_S$ remain negative (Fig. \ref{chi_S1_mub} and Fig. \ref{chi_S3_mub}).

As mentioned earlier, there are contributions from multiple charged hadrons in electrical charge and strangeness sector only, 
baryon number being always one. Hence, at high $\mu_B$ region magnitudes of $\chi_x^3$ and $\chi_x^4$ are found to be almost double compared 
to that of $\chi_x^1$ and $\chi_x^2$ respectively for $x=S,Q$ (Fig. \ref{fig:chi_S_mub}
and Fig. \ref{fig:chi_Q_mub}), though the change in $\chi_B$ remains small as shown in Fig. \ref{fig:chi_B_mub}. At very small $\mu_B$ and given $T$ (0.1 or 0.15 GeV) system
is dominated by mesons and $\chi_B$ is small. With increase in $\mu_B$ system will be populated by proton, neutron as well as hyperons and $\chi_B$ will increase. This increase will be sharper for HRG due to the absence of repulsive interaction.

\begin{figure}
  \subfigure {\includegraphics[scale=0.34,angle=-90]{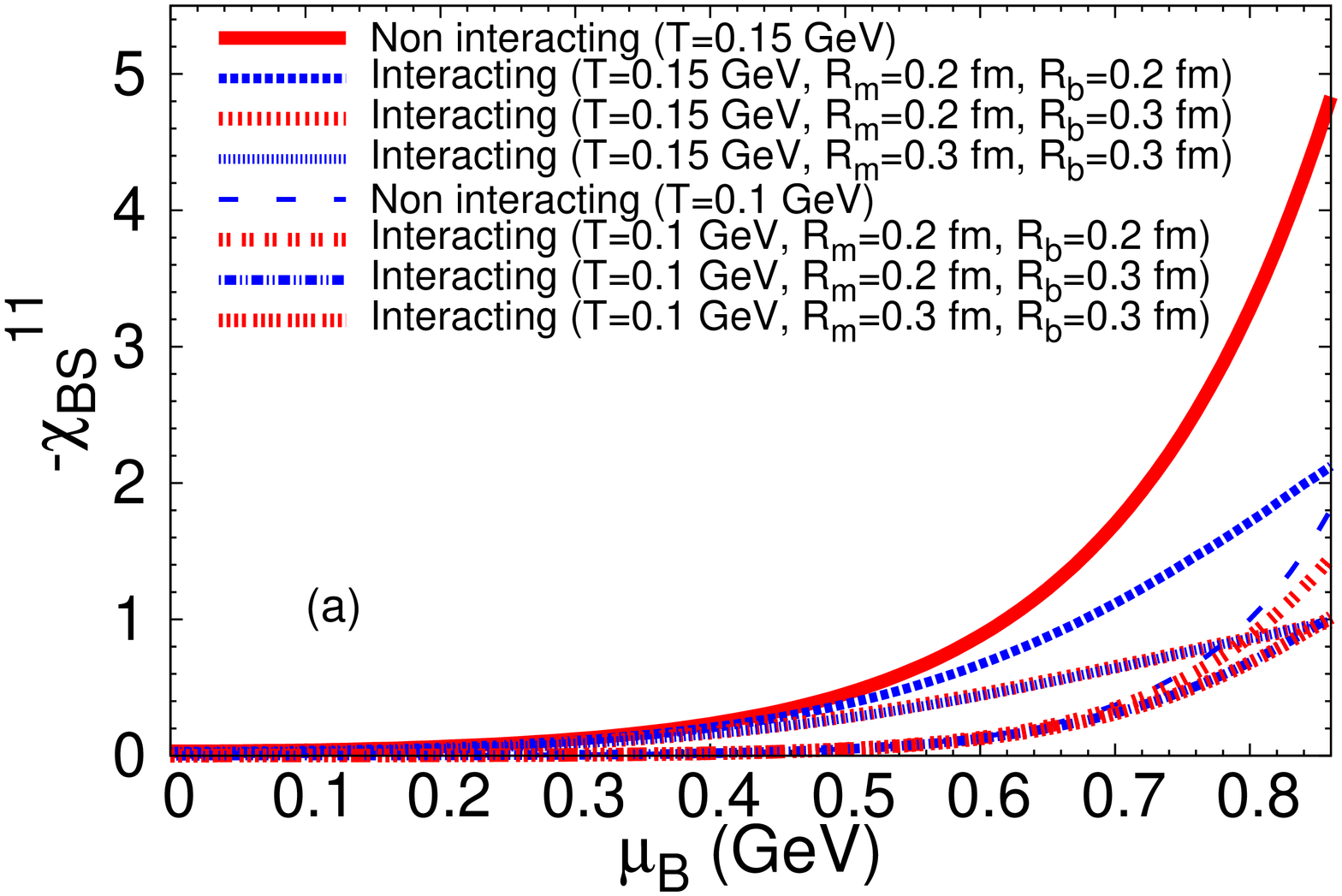}\label {chi_BS11_mub}}
   \subfigure {\includegraphics[scale=0.34,angle=-90]{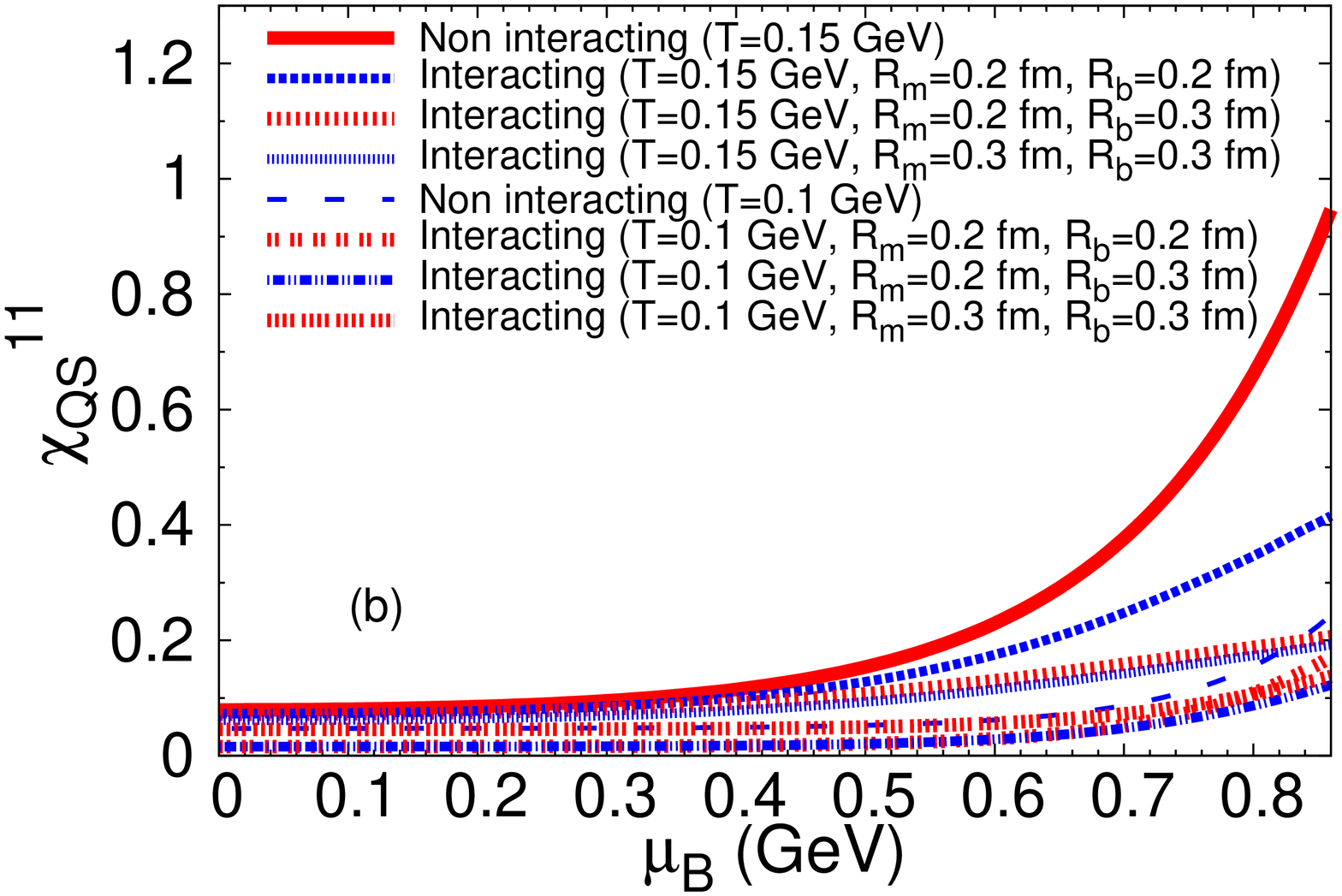}\label {chi_QS11_mub}}
    \subfigure {\includegraphics[scale=0.34,angle=-90]{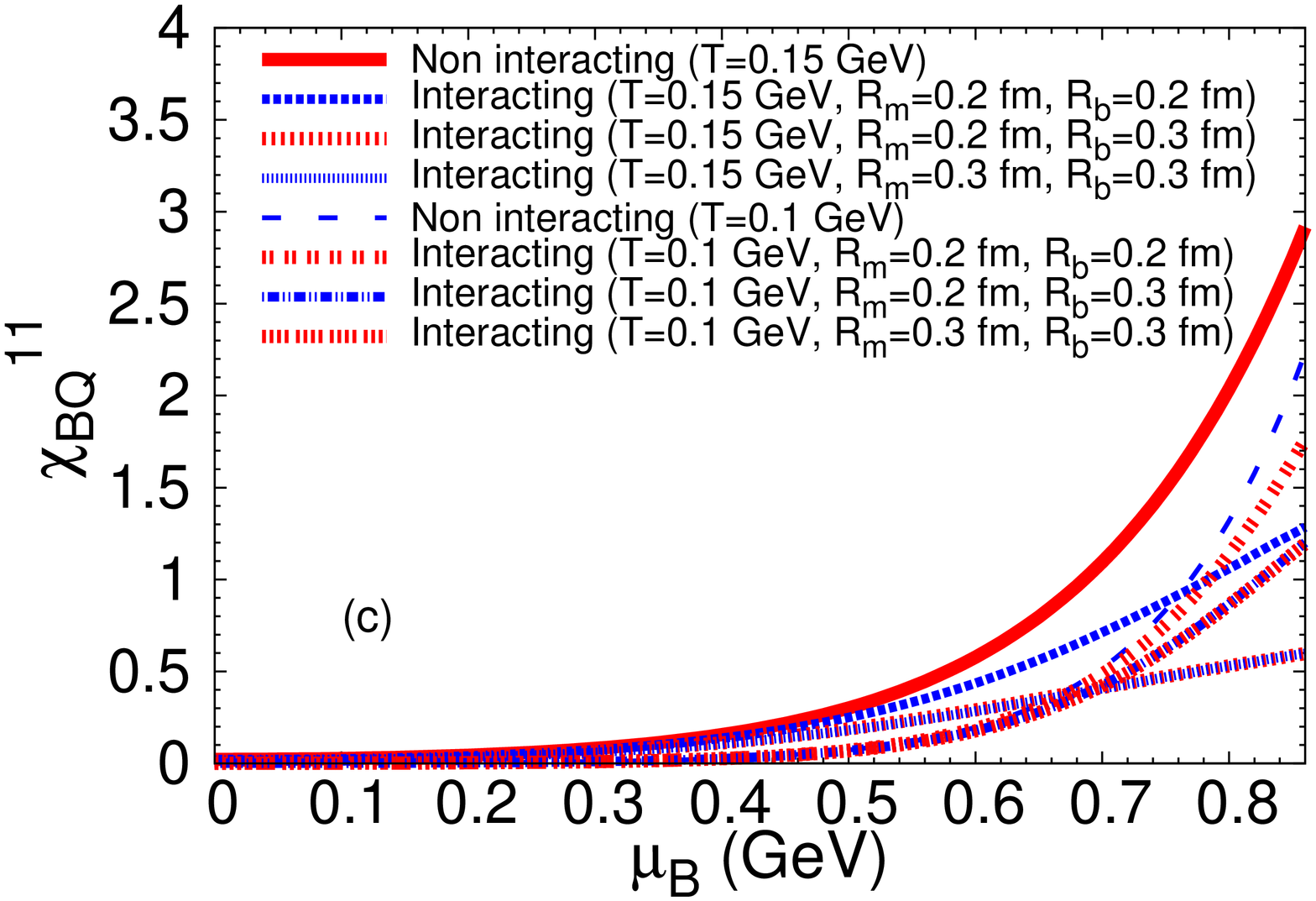}\label {chi_BQ11_mub}}
 \caption{(Color online).  \label{fig:chi_BSQ11_mub}$\chi_{BS}^{11}$, $\chi_{BQ}^{11}$, $\chi_{QS}^{11}$ as a function of $\mu_B$ keeping $\mu_S=\mu_Q=0$.}
\end{figure}
 \begin{figure}
   \subfigure {\includegraphics[scale=0.34,angle=-90]{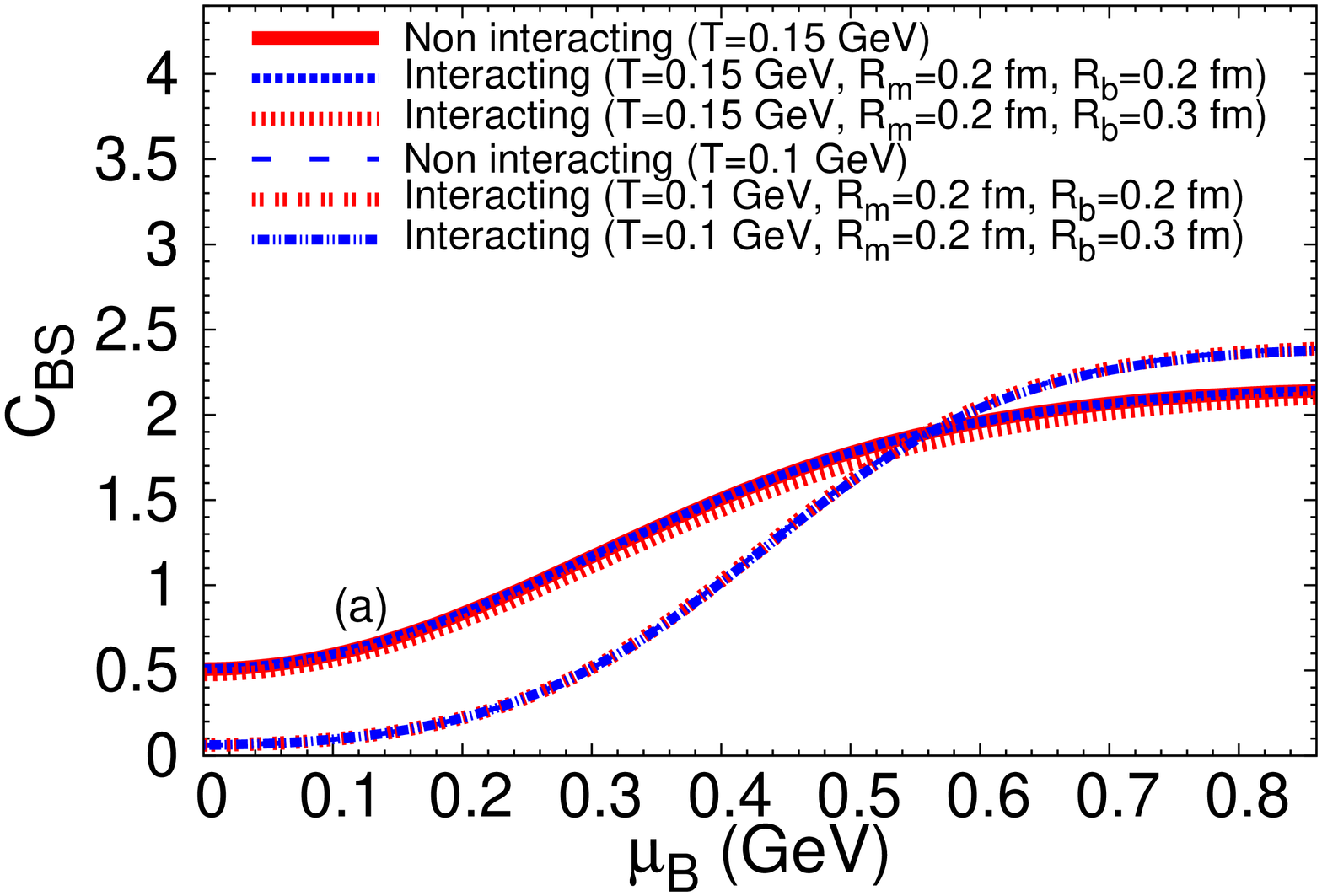}\label {C_BS_mub}}
   \subfigure {\includegraphics[scale=0.34,angle=-90]{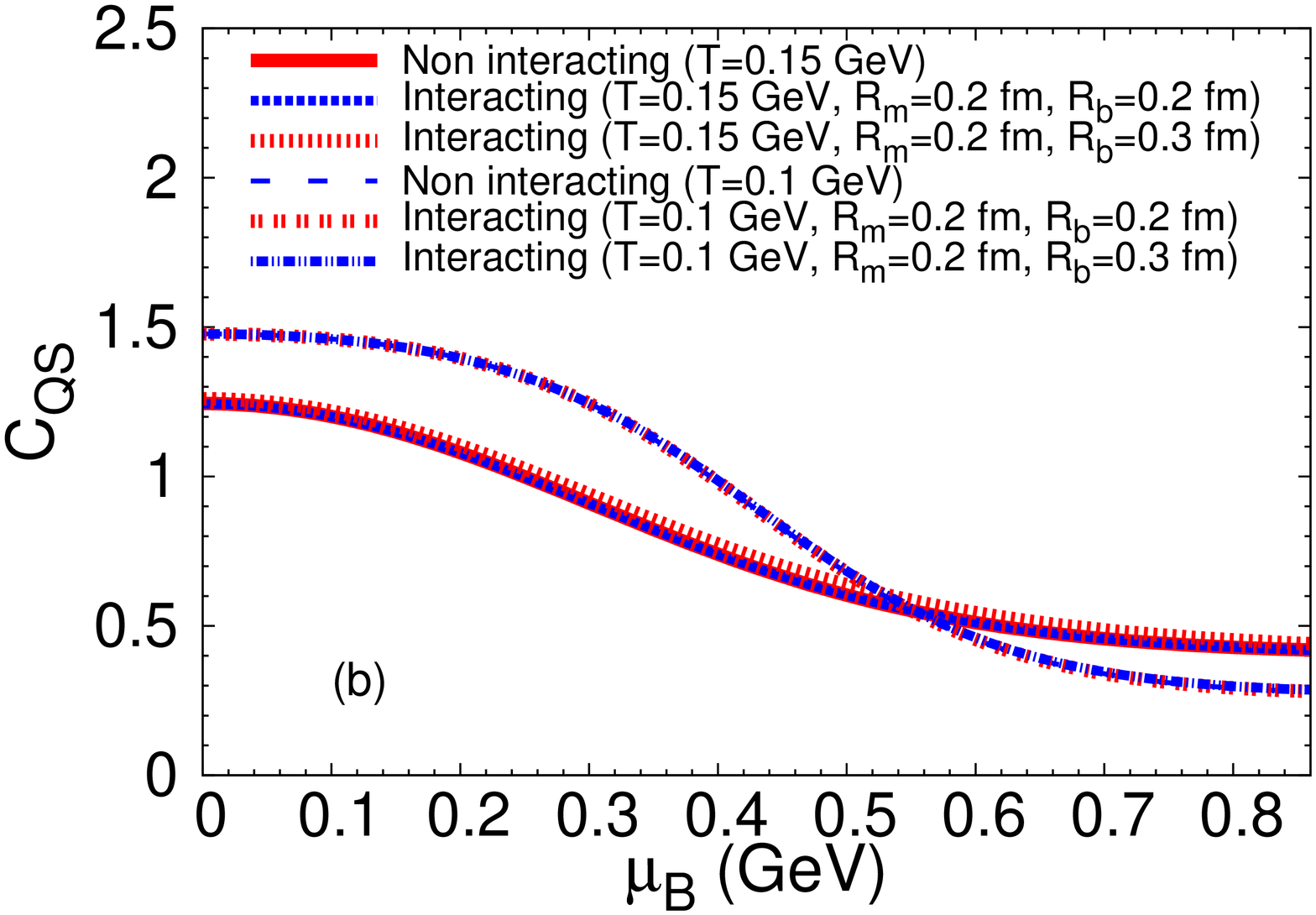}\label {C_QS_mub}}
    \subfigure {\includegraphics[scale=0.34,angle=-90]{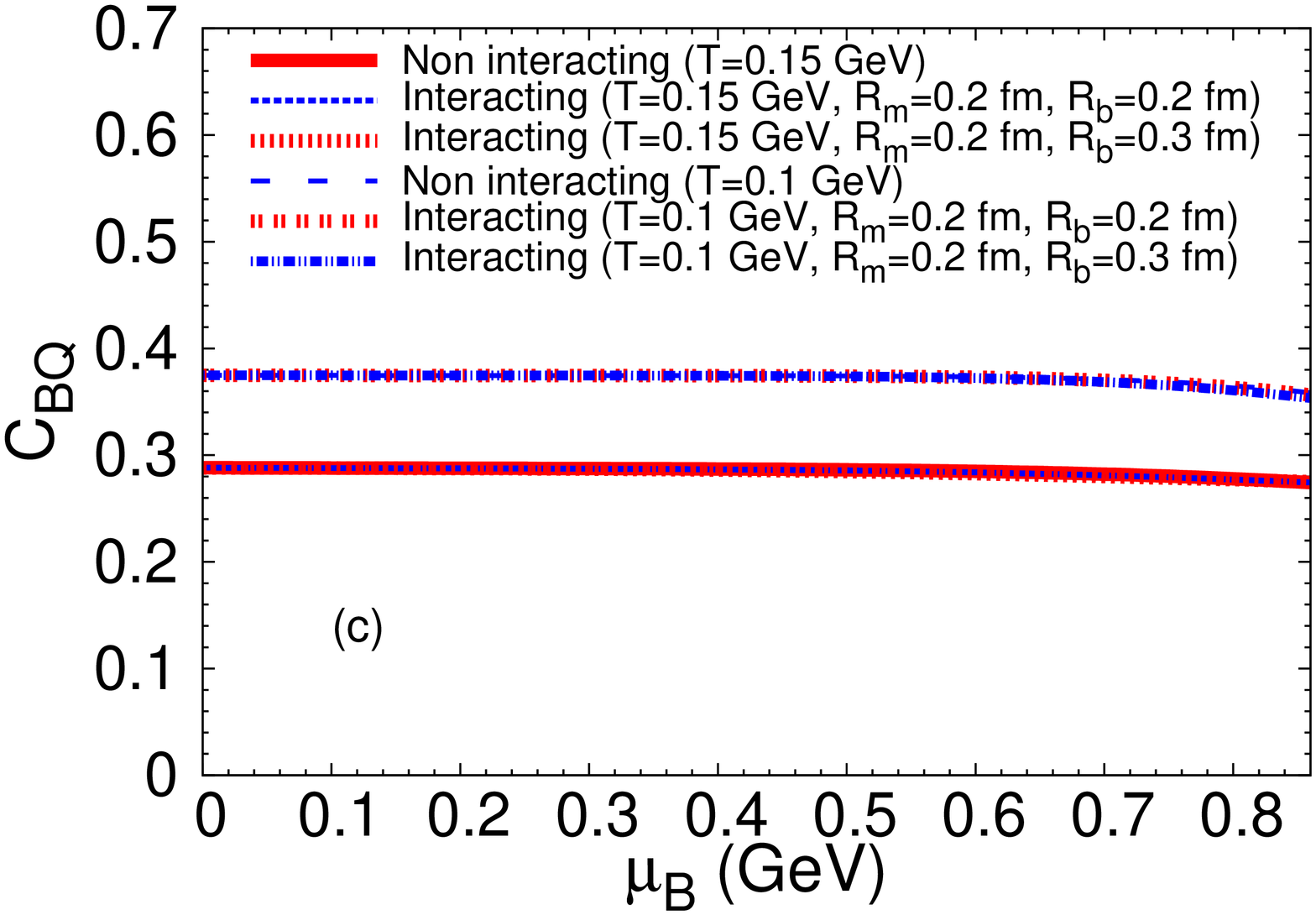}\label {c_bq_mub}}
 \caption{(Color online). \label{fig:chi_C_BS_QS_mub}Variation of $C_{BS}$, $C_{QS}$ and $C_{BQ}$ with $\mu_B$ keeping $\mu_S=\mu_Q=0$.}
\end{figure}
At finite $\mu_B$ (fixed $T$) due to larger contribution from effective chemical potentials, the effect of interaction, compared to 
zero $\mu_B$ case, is more 
prominent on correlations as shown in Fig. \ref{fig:chi_BSQ11_mub}.
The plots are given for different $R_m$ and $R_b$.
It can be seen that at low $\mu_B$, $\chi_{BS}^{11}$  and $\chi_{BQ}^{11}$ are zero whereas $\chi_{QS}^{11}$
is non-zero as pions and kaons are main contributors in this range of parameters. At large $\mu_B$, baryons start populating the system 
and both B-S and B-Q correlations increase more sharply compared to $\chi_{QS}$ as kaons remain the main contributor in $\chi_{QS}$ 
and charge hyperon population is much smaller. One can also see that there is no effect of interaction till $\mu_B=0.3$ GeV in
$\chi_{BS}^{11}$ and $\chi_{BQ}^{11}$. However at high $T$, $\chi_{QS}^{11}$ is affected by interaction even at low $\mu_B$.

In Fig. \ref{fig:chi_C_BS_QS_mub} we show variation of $C_{BS}$, $C_{QS}$ and $C_{BQ}$  with $\mu_B$ at a fixed
 $T$ and $\mu_S= \mu_Q =0$. At low $\mu_B$, kaon population is dominating in the denominator of $C_{BS}$ 
 and it is less than unity. On the other hand increase in $\mu_B$ facilitates the strange baryon population 
 and $C_{BS}$ increases with increasing $\mu_B$. Moreover, the effect of large $\mu_B$ becomes more 
 pronounced for lower $T$. So that $C_{BS}$ at $T$=0.1 GeV is larger than that at $T=0.15$ GeV for 
large $\mu_B$ ($> 0.6$ GeV). Similarly at low $\mu_B$, $C_{QS}$ is around $1.5$ as it gets contribution from
mainly charged kaons in the numerator and both charged and neutral kaons in denominator.  With 
increase in $\mu_B$, $\varLambda$ starts populating the system so that $C_{QS}$ keeps decreasing. 
Since this effect is more pronounced for larger $T$, $C_{QS}$ is lower for larger $T$ at large  $\mu_B$ ($> 0.7$ GeV). 

At low $\mu_B$, for small $T$, $C_{BQ}$ gets contribution from protons and neutrons in the denominator and only protons
in the numerator. So it is
close to $0.5$ as shown earlier in Fig. ~\ref{c_bq_temp}. At larger $T$ ($0.1$ or $0.15$ GeV), $\varLambda$ starts contributing in the
denominator and $C_{BQ}$ becomes
less than $0.5$. 
Similarly as $\mu_B$ increases, population of protons, neutrons and lambda increases. As a result $C_{BQ}$ becomes less than $0.5$.
At high $\mu_B$, $C_{BQ}$ saturates as rate of increase becomes same for different species. For HRG and EVHRG
$C_{BQ}$ is close to each other due to cancellation of interaction effect.

\subsection{\label{sec:Experiment}Experimental scenario}
Experimentally measured moments such as mean ($M$), standard deviation ($\sigma$), skewness ($S$) and kurtosis ($\kappa$) of 
conserved charges are used to characterize the shape of charge distribution. 
The standard deviation corresponds to the width 
of the distribution, the skewness is a measure of the asymmetry of the distribution and kurtosis describes 
peakness of the distribution.
Products of moments are related to susceptibilities ($\chi_x$) by the following relations,
\begin{equation}\label{moment_product}
 \frac{\chi_x^2}{\chi_x^1}=\frac{\sigma_x^2}{M_x}, ~~~~ \frac{\chi_x^3}{\chi_x^2}=S_x\sigma_x, ~~~~ 
 \frac{\chi_x^4}{\chi_x^2}=\kappa_x\sigma_x^2.
\end{equation}

The advantage of using the above mentioned products of moments is that they are expected to be independent of the volume 
of the system. On the other hand, our discussion on the variation of $\chi_x$ with $T$ (or $\mu_B$) shows that the cancellation of the volume effect, depending on the degrees of freedom involved, will become less as one increases the order of susceptibility 
in the numerator (or denominator). In nucleus-nucleus collision experiments, beam energy ($\sqrt{s_{NN}}$) is varied to scan the
phase plane. At a particular $\sqrt{s_{NN}}$, if CEP is reached, the signature of CEP could survive during the
time evolution of the system. Hence along the freeze-out line one would expect a non-monotonic behaviour of the 
quantities given in Eq. \ref{moment_product}~\cite{PRL105_022302_Aggarwal, AHEP_Tawfik}.
To compare the products of moments with experiment one has to know freeze-out $T$ and $\mu$ 
at a particular $\sqrt{s_{NN}}$.
Freeze-out curve $T(\mu_B)$ can be parametrized by~\cite{PRC73_Cleymans}
\begin{equation}
 T(\mu_B)=a-b\mu_B^2-c\mu_B^4,
 \end{equation}
where $a = 0.166 \pm 0.002$ GeV, $b =0.139 \pm 0.016$ GeV$^{-1}$, $c=0.053 \pm 0.021$ GeV$^{-3}$.
The energy dependence of the $\mu_q$ can be parametrized as~\cite{PLB695_Karsch}
\begin{equation}
 \mu_x(\sqrt{s_{NN}})=\frac{d_x}{1+e_x\sqrt{s_{NN}}},
\end{equation}
where $d_x$ and $e_x$ are listed in table \ref{tableLabel_muq}.

\begin{table}[h]
\centering
\begin{tabular}{|c|c|c|}
\hline
\it x&$d_x$ (\it GeV)&$e_x$ (\it GeV$^{-1}$)\\
\hline
\it B &$1.308 \pm 0.028$ &$0.273 \pm 0.008$ \\
\it S &$0.214$&$0.161 $\\
\it Q &$-0.0211$ &$0.106$\\
\hline
\end{tabular}
\caption{Parametrization of chemical potentials $\mu_x$ along the freeze-out curve.}
\label{tableLabel_muq}
\end{table}

\begin{figure}[]
   \subfigure {\includegraphics[scale=0.34,angle=-90]{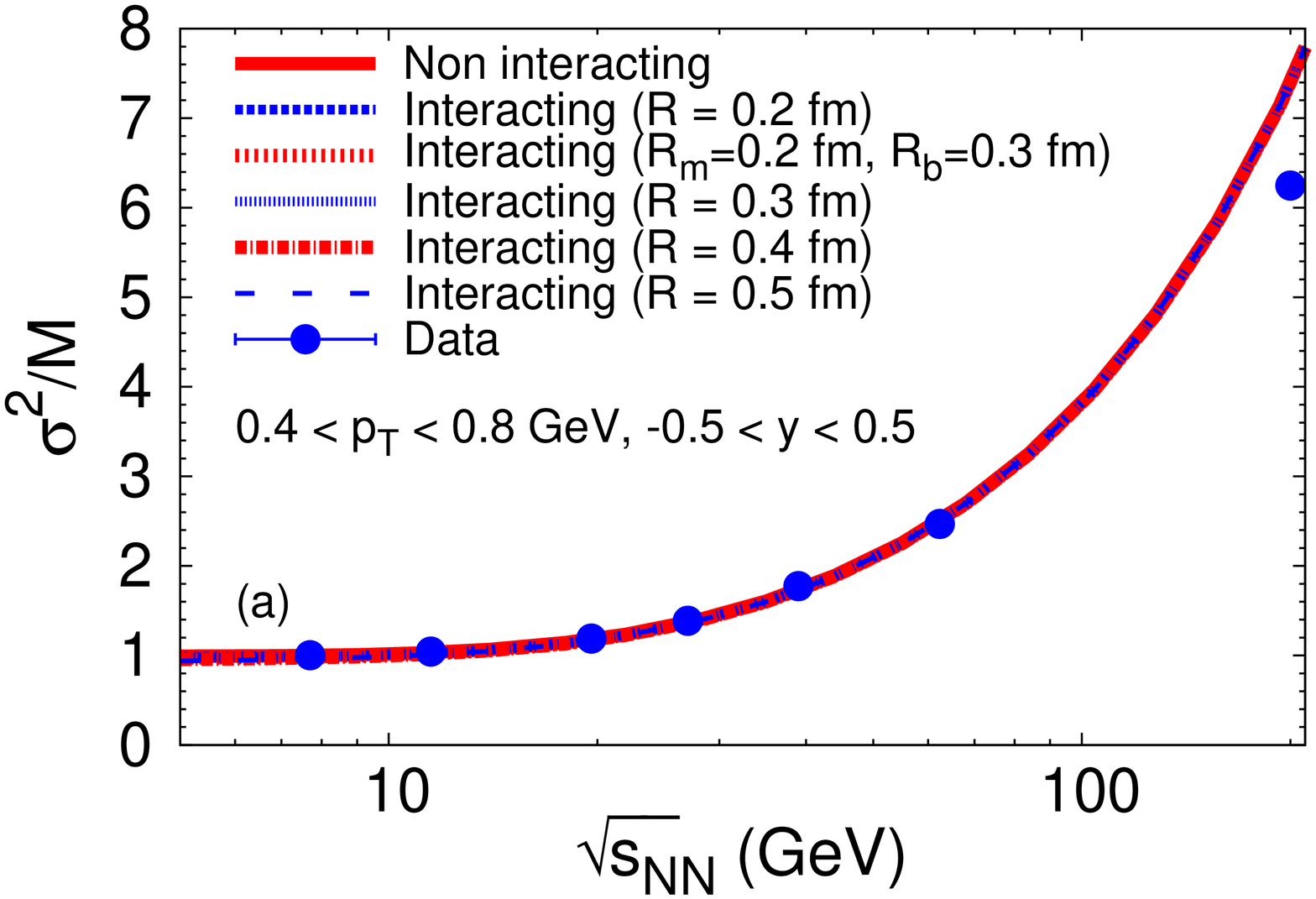}\label {chi_2_1_np_roots}}
    \subfigure {\includegraphics[scale=0.34,angle=-90]{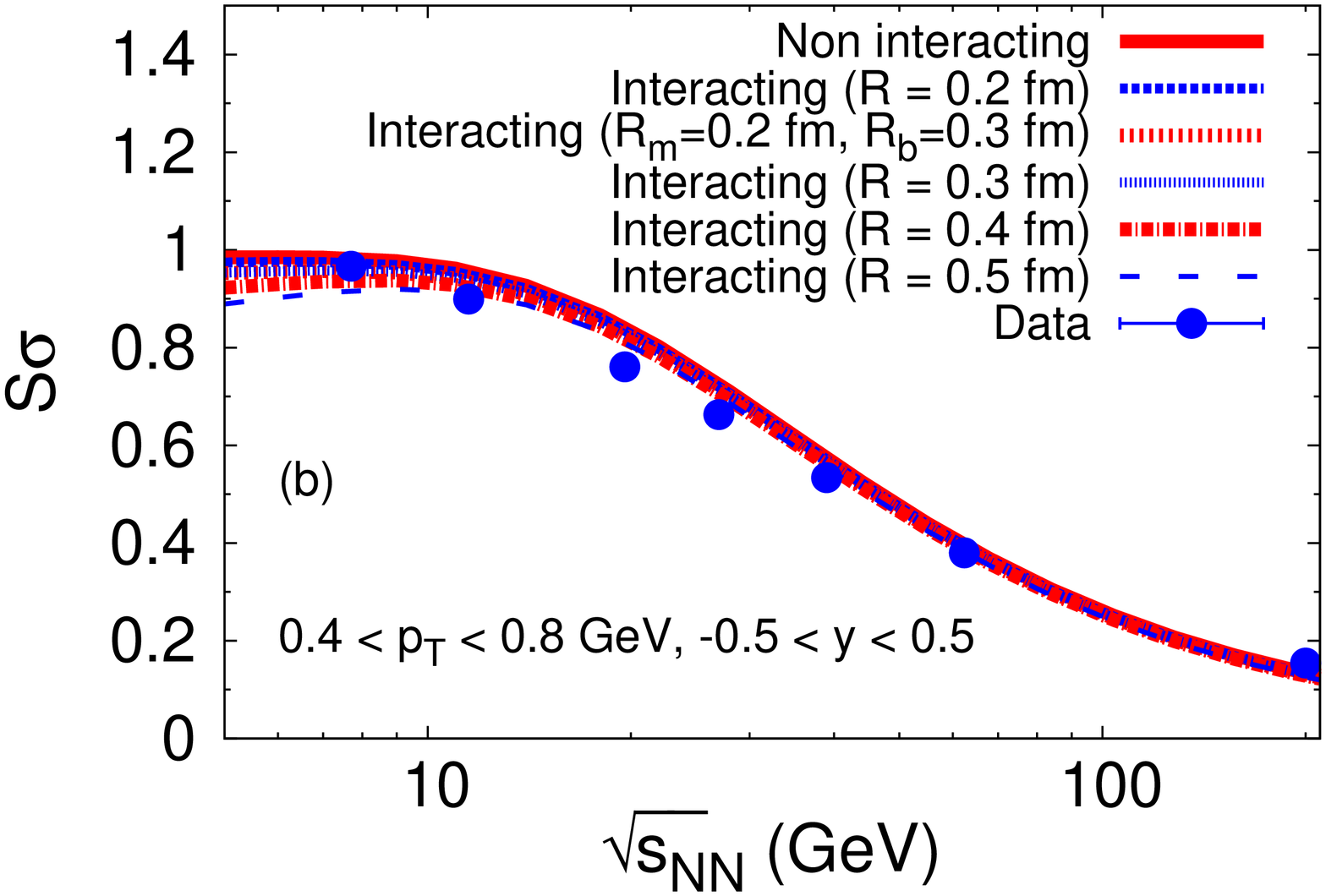}\label {chi_3_2_np_roots}}
     \subfigure {\includegraphics[scale=0.34,angle=-90]{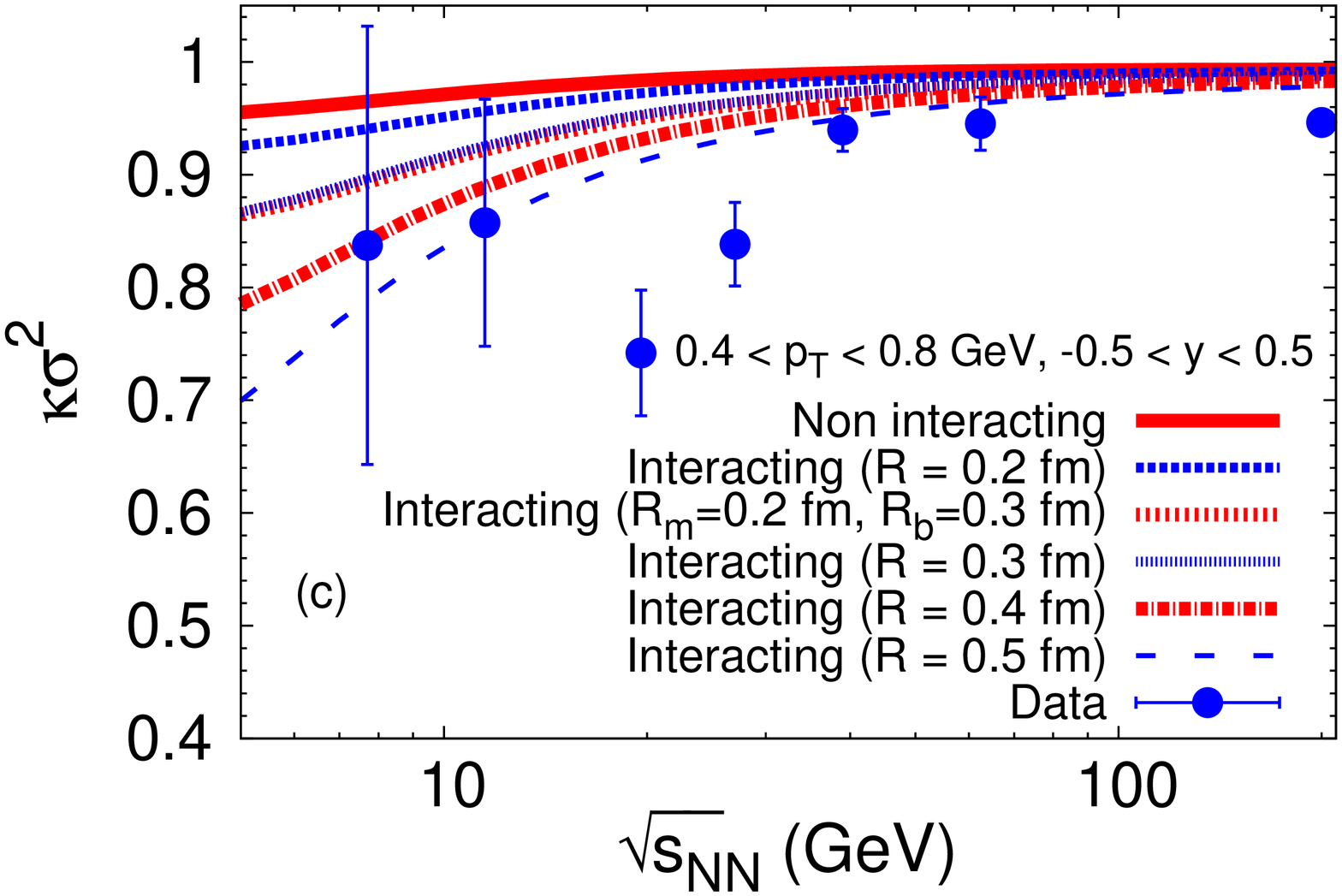}\label {chi_4_2_np_roots}}
 \caption{(Color online). \label{fig:np_roots}Energy dependence of $\sigma^2/M$, $S\sigma$ and $\kappa\sigma^2$ 
 for net-proton. Experimental data is taken from Ref. ~\cite{STAR_Luo}.}
 \end{figure}

The $T$ - $\mu_B$ parametrisation with $\sqrt{s_{NN}}$ is obtained by fitting the different particle ratios obtained 
experimentally at different
$\sqrt{s_{NN}}$. So for EVHRG model, in general, the parameters are expected to be different for different hadronic radii. 
In our study, considering all the particle ratios, we found the parameters to be same as 
\cite{PRC73_Cleymans} for $R_b=R_m$. On the other hand, for $R_B \ne R_m$ case, for the radii used here, the parameter are 
found to be within $\pm 0.005- 0.008$ GeV of the above fit.
Moreover, phenomenological freeze-out condition of fixed energy per nucleon about $1$ GeV, calculated in our model,
matches quite well with the above mentioned freeze-out parametrization.
Hence we have used the above mentioned freeze-out parametrization for comparison
with the experimental results. It may also be noted here that as $\sqrt{s_{NN}}$ increases from 7.7 GeV to 200 GeV, $T$ increases from 
around $0.140 $ GeV to $0.166 $ GeV and corresponding $\mu_B$ varies between $0.421$ GeV to $0.023$ GeV. Corresponding $\mu_S$ and $\mu_Q$ are expected
to vary between $0.095$ GeV to $0.006$ GeV and $-0.011$ GeV to $-0.001$ GeV~\cite{PLB695_Karsch}.

In terms of transverse momentum ($p_T$) and rapidity ($y$) and azimuthal angle ($\phi$)
 $d^3p$ and $E_i$ can be written as
$d^3p=p_T \, m_{Ti} \, \cosh y \, dp_T \, dy \, d\phi$
and $E_i=m_{Ti} \, \cosh y$, where $m_{Ti}=\sqrt{(p_T^2+m_i^2)}$. One can write a similar expression in terms of $p_T$ and
pseudo-rapidity $\eta$ as well. These prescriptions have been used to set the momentum and rapidity acceptance range to
compare the present results with the experimental data.

In Fig. \ref{fig:np_roots} we have shown energy dependence of $\sigma^2/M$, $S\sigma$ and $\kappa\sigma^2$ 
for net-proton. We compare our result with experimental data of net-proton fluctuations for $(0-5) \%$  central 
Au-Au collisions measured at STAR~\cite{STAR_Luo}. Experimental data is measured at mid rapidity $(|y|<0.5)$ and within 
the transverse momentum range $0.4<p_T<0.8$ GeV.
Same acceptances range has been used for different hadronic radii in the present manuscript. 
At low energy, $\sigma^2/M$ is almost unity and its value
increases with increase of $\sqrt{s_{NN}}$ as can be seen from the Fig. \ref{chi_2_1_np_roots}.
Both HRG and EVHRG model give almost same result irrespective of the value of radii.
At low energy $S\sigma$ for HRG is almost unity and its value
decreases with increase of $\sqrt{s_{NN}}$ as can be seen from the Fig. \ref{chi_3_2_np_roots}.
At low $\sqrt{s_{NN}}$, $S\sigma$ for EVHRG is less than that
of HRG model and suppression increases with increase of radii. However, at high $\sqrt{s_{NN}}$, both HRG and EVHRG model give almost same
result. Experimental data of $S\sigma$ can be described well with EVHRG model for radii of hadrons between $0.3 - 0.4$ fm.
At low energy $\kappa\sigma^2$ (Fig. \ref{chi_4_2_np_roots}) in HRG model is slightly less than unity and its value
reaches to unity as we move to high energy. There is prominent suppression of $\kappa\sigma^2$ in EVHRG model at low $\sqrt{s_{NN}}$.
The $\kappa\sigma^2$ of net-proton matches within error-bar with EVHRG model at $\sqrt{s_{NN}} \geq 39$ GeV and  $\sqrt{s_{NN}} \leq 11.5$ 
GeV but at intermediate energies EVHRG model over estimates the experimental data. Deviation of experimental data of $S\sigma$ and 
$\kappa\sigma^2$ for net-proton at intermediate energies ($\sqrt{s_{NN}}= 19$ GeV and $27$ GeV) 
may be an indication of existence of quark degrees of freedom or different physics process which is not included
in HRG/EVHRG model.

In Ref.~\cite{PLB722_Fu}, STAR data for net-proton fluctuations is already compared with EVHRG model in full space.  
Energy dependence of fluctuations for net-proton using LQCD can be found in Ref.~\cite{Mohanty_science, Karsch_PRD84_R}.
Though at high energies, where $\mu_B$ is small, LQCD agrees well with the experimental data, at low energies {\it i.e.} for
higher $\mu_B$ it fails.

\begin{figure}[]
   \subfigure {\includegraphics[scale=0.34,angle=-90]{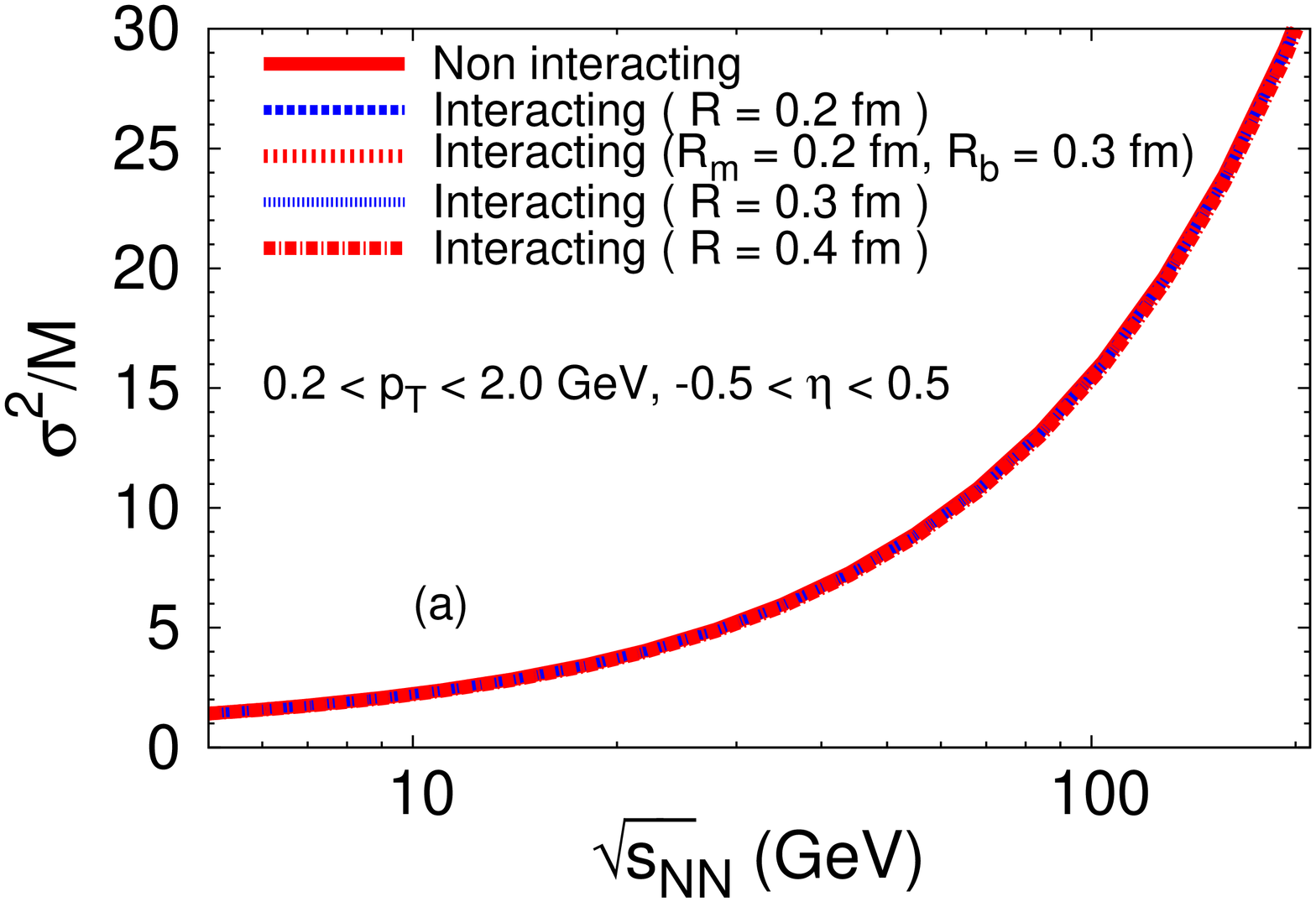}\label{chi_2_1_nk_roots}}
    \subfigure {\includegraphics[scale=0.34,angle=-90]{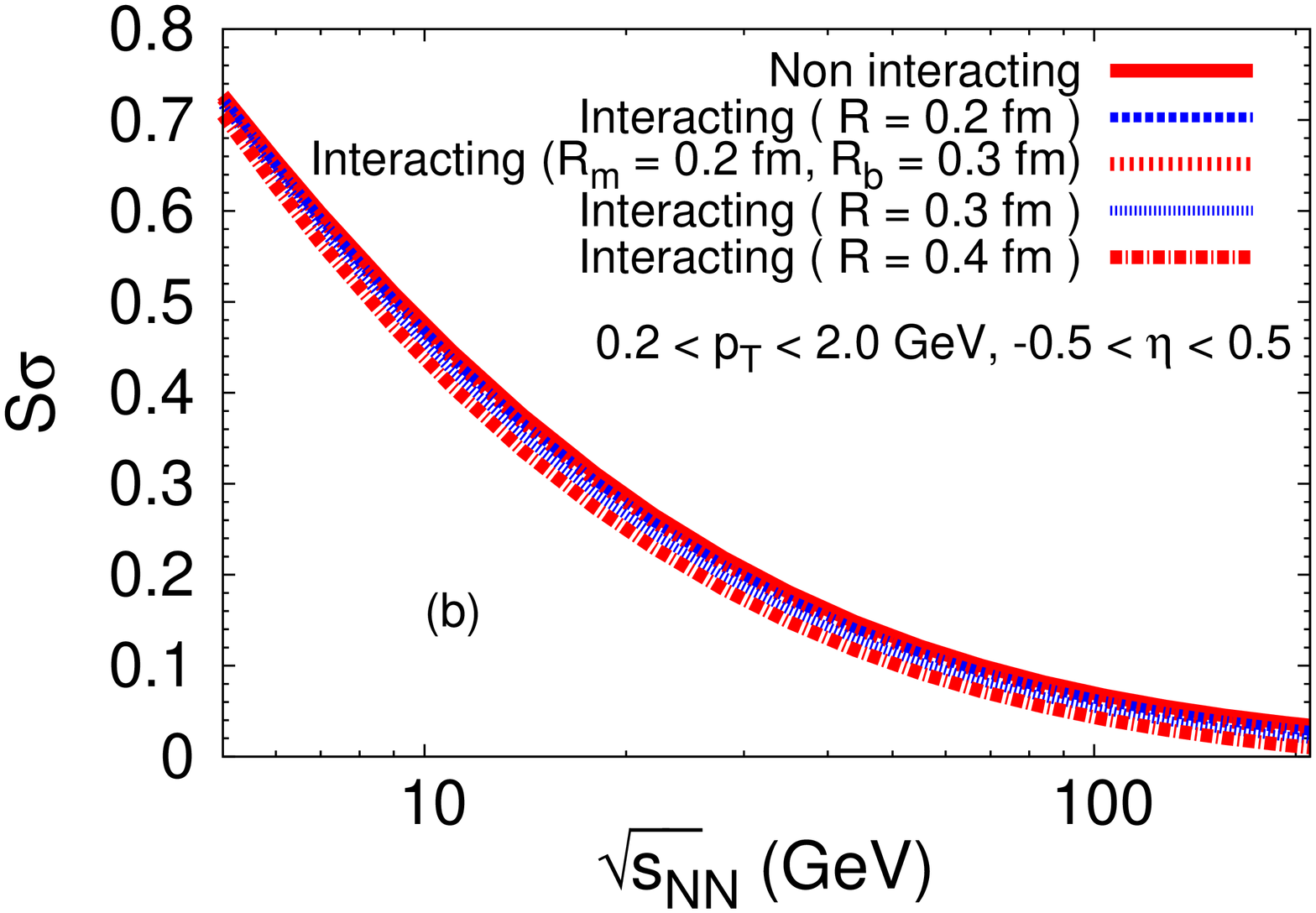}\label{chi_3_2_nk_roots}}
     \subfigure {\includegraphics[scale=0.34,angle=-90]{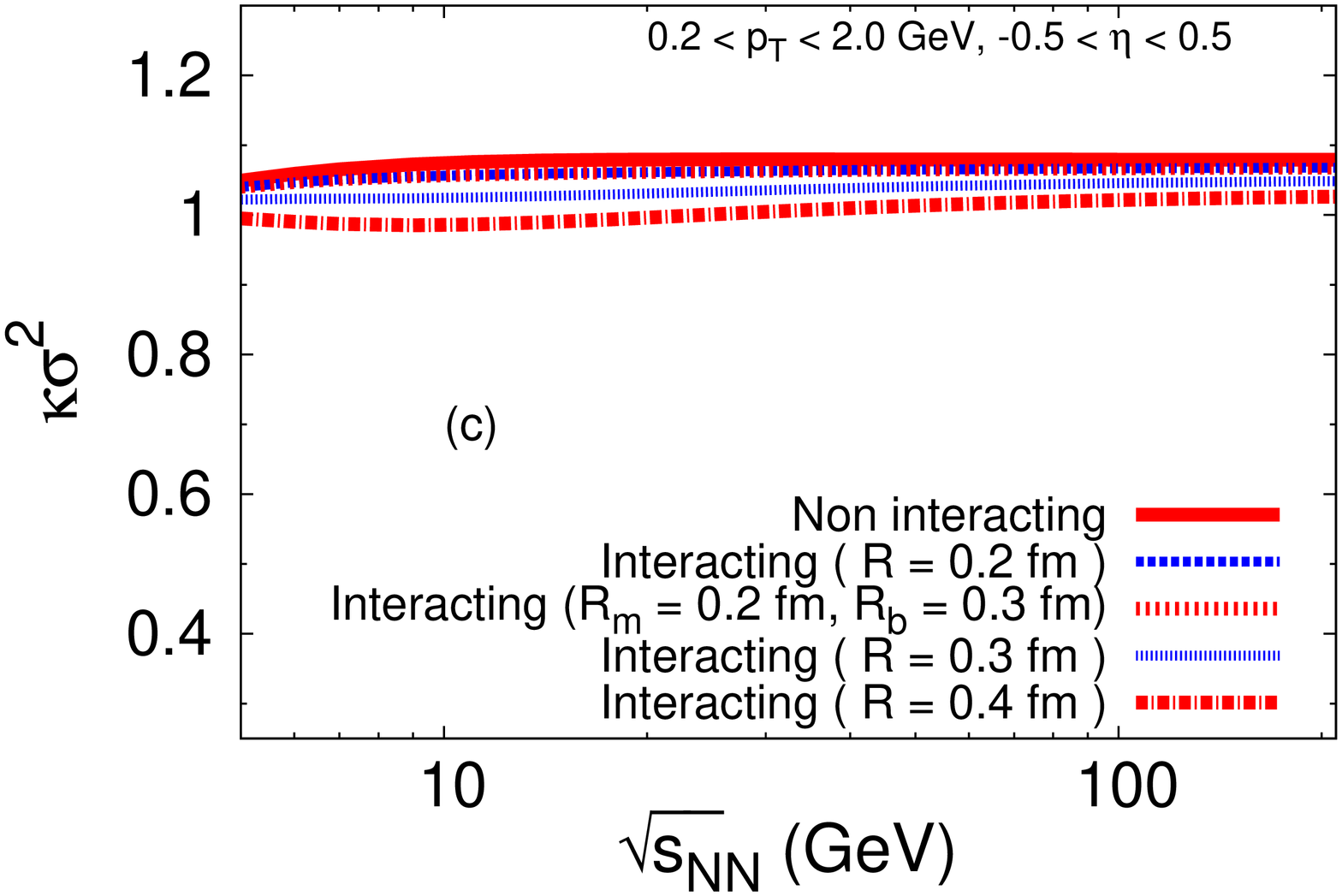}\label{chi_4_2_nk_roots}}
 \caption{(Color online). \label{fig:nk_roots}Energy dependence of $\sigma^2/M$, $S\sigma$ and $\kappa\sigma^2$ 
 for net-kaon.}
 \end{figure}

In Fig. \ref{fig:nk_roots} we have shown energy dependence of $\sigma^2/M$, $S\sigma$ and $\kappa\sigma^2$ 
for net-kaon. We have chosen transverse momentum and pseudo-rapidity within the range $0.2 < p_T < 2.0$ GeV and $-0.5 < \eta < 0.5$
in our calculation. 
At low energy $\sigma^2/M$ for net-kaon is slightly more than unity and its value increases rapidly with increase
of $\sqrt{s_{NN}}$ and result is almost same for both HRG and EVHRG model. Radii of hadrons
practically have no effect on the results as  shown in Fig. \ref{chi_2_1_nk_roots}.
$S\sigma$ for net-kaon decreases with increase of $\sqrt{s_{NN}}$ as can be seen from Fig. \ref{chi_3_2_nk_roots}. 
Value of $S\sigma$ in EVHRG model is suppressed compared to HRG model and suppression increases with 
increase of radii of hadrons.
In Fig. \ref{chi_4_2_nk_roots} energy dependence of $\kappa\sigma^2$ for net-kaon is shown. In this case, volume effect is more 
prominent, as discussed earlier. $\kappa\sigma^2$ for net-kaon increases with increase of $\sqrt{s_{NN}}$ and then 
saturates. Value of $\kappa\sigma^2$ is
suppressed in EVHRG model compared to HRG model and suppression increases with increase of radii of hadrons.
At all energies $\kappa\sigma^2$ for net-kaon lie between $0.9$ to $1.1$ in both HRG and EVHRG.

\begin{figure}[]
   \subfigure {\includegraphics[scale=0.34,angle=-90]{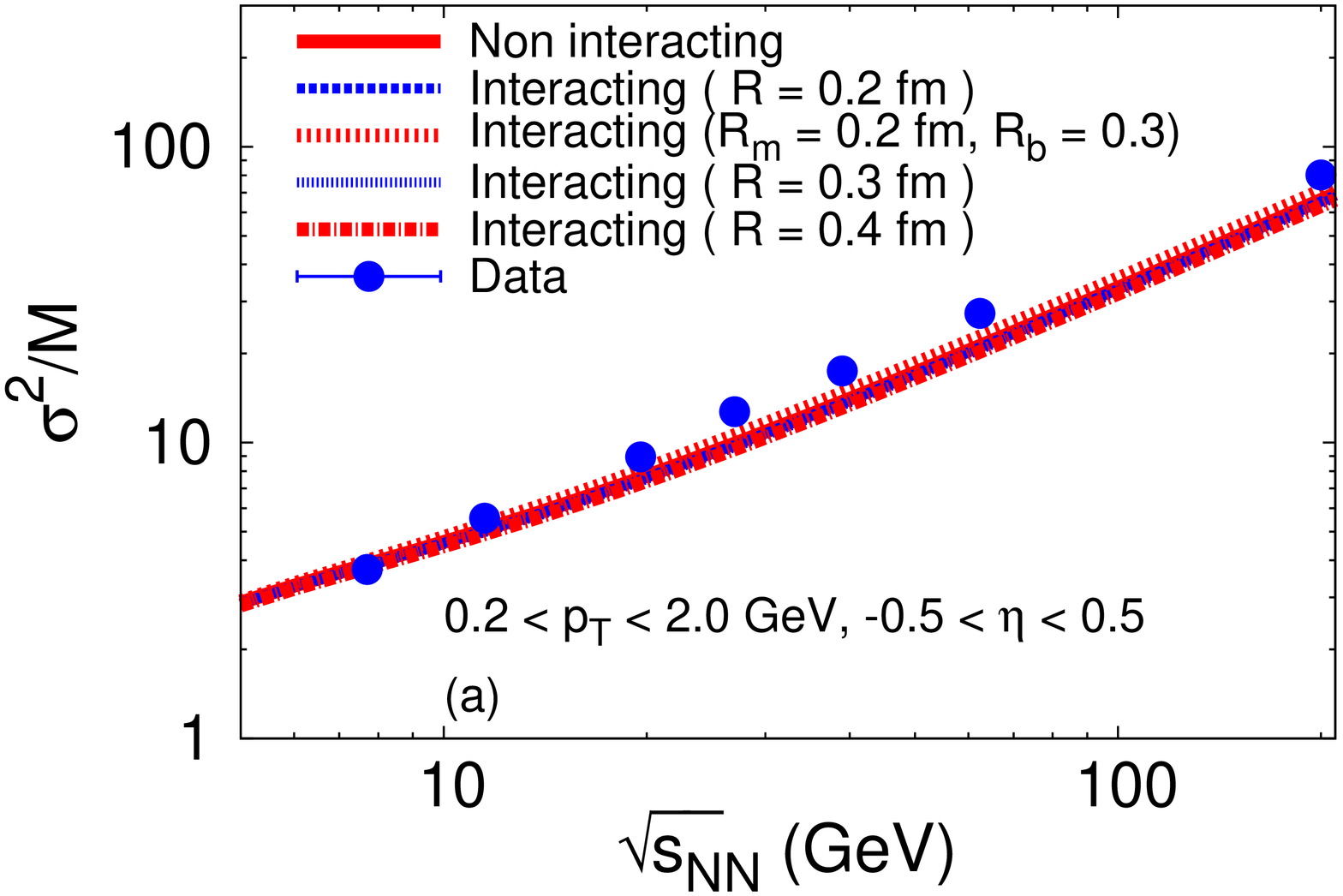}\label{chi_2_1_nc_roots}}
    \subfigure {\includegraphics[scale=0.34,angle=-90]{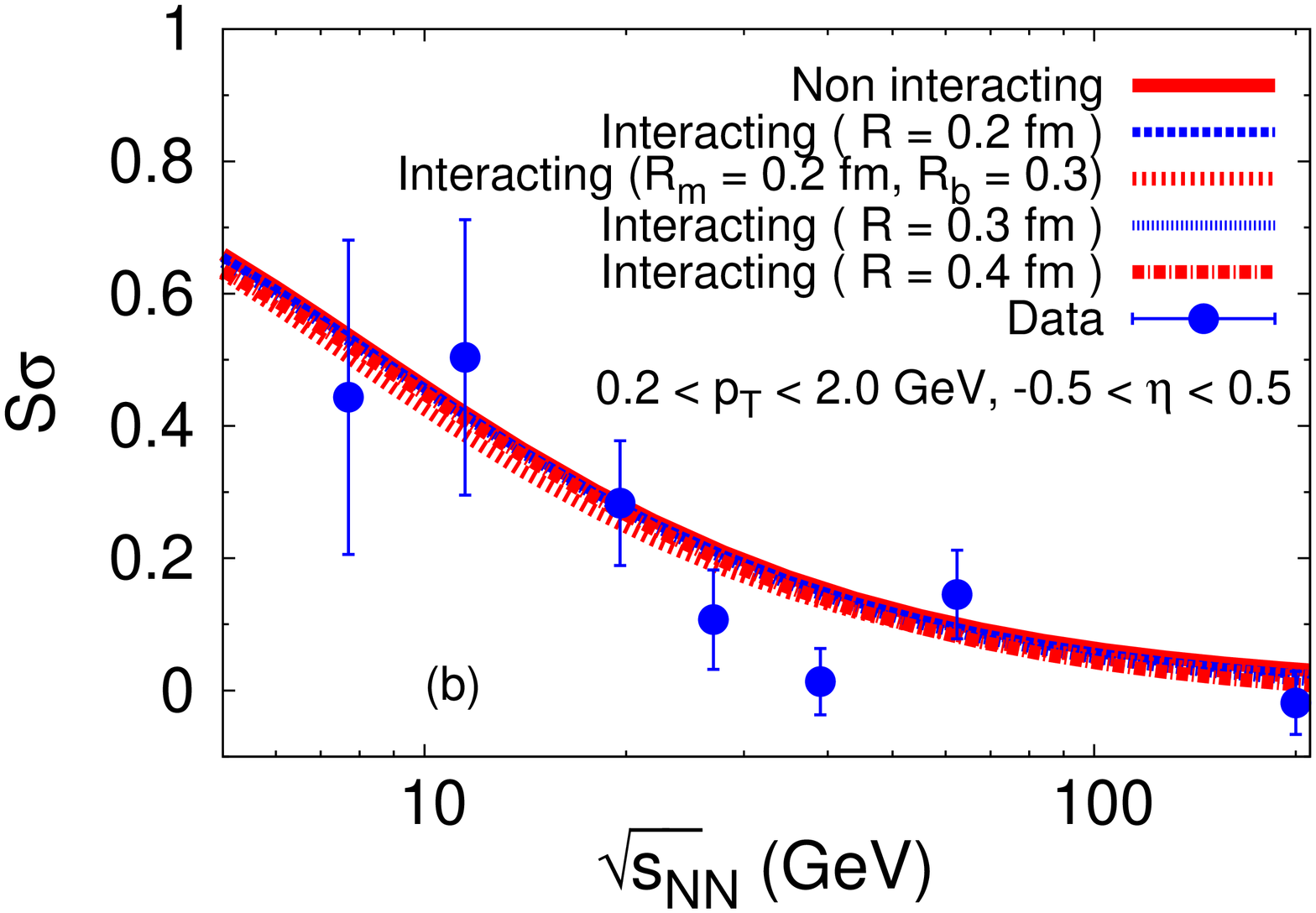}\label{chi_3_2_nc_roots}}
     \subfigure {\includegraphics[scale=0.34,angle=-90]{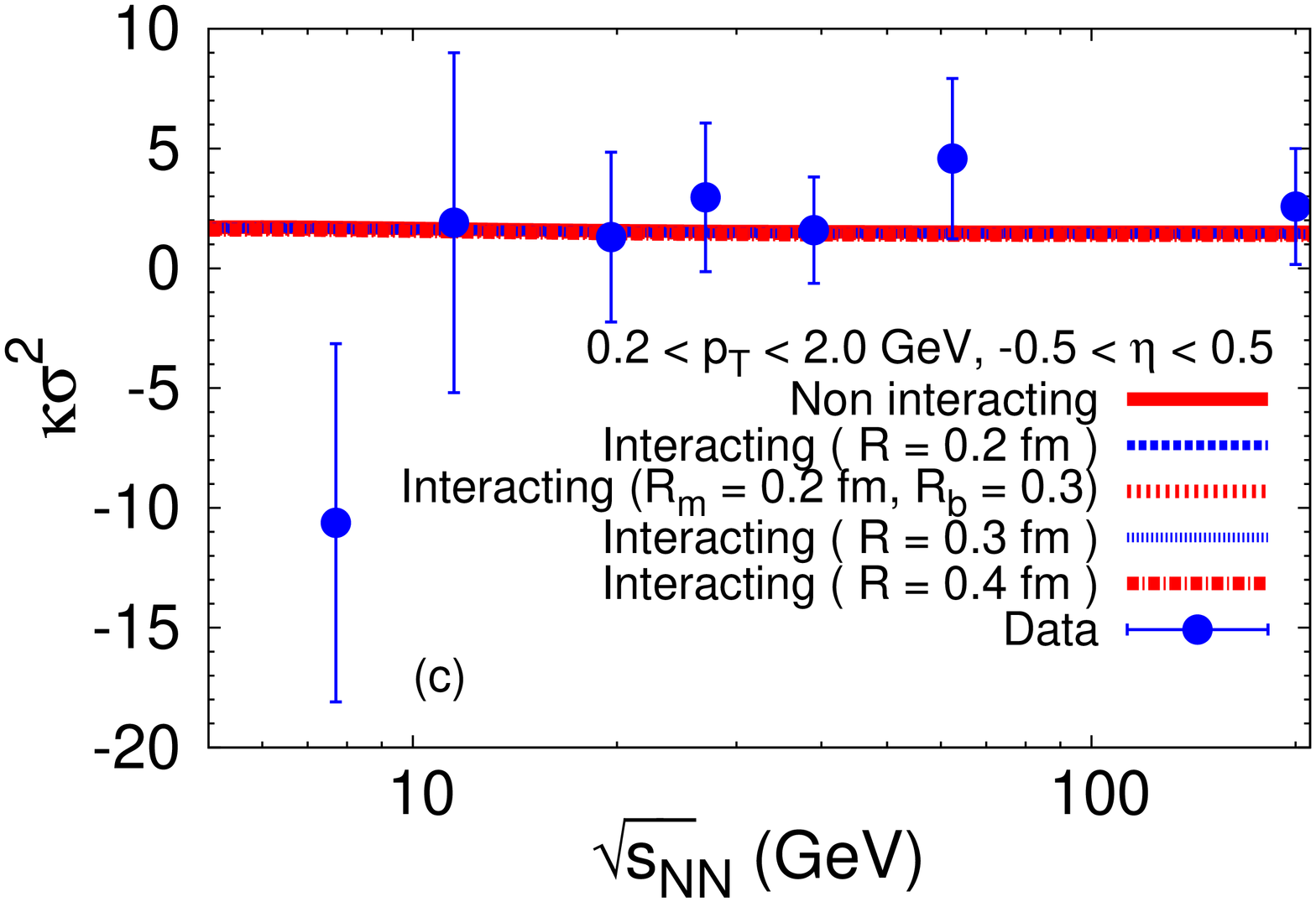}\label{chi_4_2_nc_roots}}
 \caption{(Color online). \label{fig:nc_roots_2}Energy dependence of $\sigma^2/M$, $S\sigma$ and $\kappa\sigma^2$ 
 for net-charge. Experimental data is taken from Ref. ~\cite{STAR_nihar}.}
 \end{figure}

In Fig. \ref{fig:nc_roots_2} we have shown energy dependence of $\sigma^2/M$, $S\sigma$ and $\kappa\sigma^2$ 
for net-charge. 
We have chosen transverse momentum and pseudo-rapidity within the range $0.2 < p_T < 2.0$ GeV and $-0.5 < \eta < 0.5$
in our model calculation.
We have compared our results with experimental data of net-charge fluctuations for $(0-5) \%$ central 
Au-Au collisions measured at STAR~\cite{STAR_nihar}.
Figure \ref{chi_2_1_nc_roots} shows $\sigma^2/M$, for net-charge, increases rapidly with increase of 
$\sqrt{s_{NN}}$ and compared to experimental values our results are suppressed at higher energies. 
$S\sigma$ for net-charge decreases with increase of $\sqrt{s_{NN}}$ as can be seen from Fig. \ref{chi_3_2_nc_roots}.
Experimental data of $S\sigma$ for net-charge matches within error-bar 
with HRG/EVHRG model
at $\sqrt{s_{NN}} \leq 19.6 $ GeV 
and $\sqrt{s_{NN}} \geq 62.4$ GeV but at other intermediate energies HRG/EVHRG model over estimate the experimental data.
Figure \ref{chi_4_2_nc_roots} shows $\kappa\sigma^2$ for net-charge slowly decreases with increase of $\sqrt{s_{NN}}$ and then saturates.
At all  $\sqrt{s_{NN}}$, $\kappa\sigma^2$ for net-charge lie between $1.4$ to $1.8$ in both HRG and 
EVHRG model which is in agreement with experiment within error-bar at $\sqrt{s_{NN}} \geq 11.5$ GeV. Calculations 
for ratios of higher order fluctuations of electric charge using LQCD can be found in Ref.~\cite{PRL_111_Borsanyi}.

\section{\label{sec:Conclusion} Discussion and Conclusion}
We have shown fluctuations ($\chi^i_x$; $i=1-4; x=B, S, Q$) of various conserved charges like net baryon number, net strangeness 
and net charge at finite temperatures and chemical potentials using interacting and non-interacting hadron resonance gas model and compared them with the lattice as well as experimental data.

We can draw following inferences from the present study:
\begin{itemize}

\item In general HRG results are larger compared to EVHRG, difference being higher for higher temperatures and densities.

\item At high temperatures and $\mu_B = 0$, compared to HRG, second order fluctuations from EVHRG fits better to LQCD continuum data for radius of hadrons between $0.2-0.3$ fm. 

\item The LQCD data ($N_\tau = 6, 8$) for $\chi^4_B$ and  $\chi^4_S$ are in good agreement with HRG/EVHRG model up to $T = 0.16$ GeV.
Whereas, both HRG and EVHRG model overestimates LQCD data for  $\chi^4_Q$.

\item Correlations at $T\ne 0$ and $\mu_B=0$ shows much stronger dependence on the degrees of freedom involved. 
Lattice continuum data for both $\chi_{BS}^{11}$ and $\chi_{QS}^{11}$ are closer to HRG results (higher than EVHRG) at 
lower $T$ but rises less sharply and becomes less than HRG for higher $T$.  
On the other hand, $\chi_{BQ}^{11}$ calculated in lattice is close to EVHRG results at lower $T$. Initially it increases with $T$ but then 
decreases beyond $T=0.165$GeV as discussed earlier. 

\item The correlation ratios from both HRG and EVHRG disagree with LQCD continuum data though the trends are qualitatively similar.
This difference could be attributed to the fact that LQCD incorporates transition from hadronic to quark degrees of freedom, whereas 
only hadronic degrees of freedom are present in HRG or EVHRG model. At high $\mu_B$, the magnitudes of susceptibilities as well
as correlations are higher correspond to $\mu=0$ case.

\item The susceptibilities as well as correlations are found to increase with $\mu_B$ at fixed temperatures for both HRG and EVHRG,
the effect of interaction being larger for higher $T$.


\item Since low energy collisions usually correspond to larger $\mu_B$, the effect of repulsive interaction, as present in EVHRG, 
is distinguishable for lower energies. Moreover this difference is more pronounced for higher order susceptibilities.

\item With increase of $\sqrt{s_{NN}}$, chemical potential decreases sharply whereas temperature increases
slowly. Therefore, $\chi^1$ and $\chi^3$  decrease with increase
of $\sqrt{s_{NN}}$ whereas $\chi^2$ and $\chi^4$ decrease slowly and then saturate. 
As a result $\sigma^2/M$ increases with increase of $\sqrt{s_{NN}}$
, $S\sigma$ decreases with increase of $\sqrt{s_{NN}}$ whereas $\kappa\sigma^2$ remains almost constant in all energies.

\item Although the variations of $\sigma^2/M$ and $S\sigma$ with $\sqrt{s_{NN}}$ seem to describe the experimental data well,
higher moment $\kappa\sigma^2$ shows large deviations.
It may be an indication of quark degrees of freedom.
It would be
interesting to look at the other higher moments as they are supposed to have stronger dependence on correlation length and hence 
will be more sensitive to the critical fluctuations.

\end{itemize}
In general, EVHRG seems to have a better agreement with the LQCD continuum data at low $T$ and $\mu_B = 0$. But the 
comparison with the experimental data does not provide us with a clear preference between HRG and EVHRG. 

It may be noted that in figs.\ref{chi_S2_temp} and \ref{chi_BS11_temp} LQCD results are larger than all the HRG results in the
crossover temperature range of $0.14-0.15$ GeV. The LQCD may become closer to the HRG by the inclusion of non-PDG additional strange beryons
which, though not observed experimentally, are predicted by the quark-model and also observed in the LQCD spectrum \cite{14046511}.

Here we would like to mention that the, grand canonical (GC) description, as used for the present study is usually 
considered as the relevant
description for the systems in heavy ion collision~\cite{jeon&koch-Hwa}. But if the conserved quanta, under 
consideration, is small then one need to take care of the charge conservations and the present description will 
not be appropriate. More specifically, it has been found that the results for scaled variances vary for different 
ensembles though the mean multiplicities from different ensembles tend to agree with each other in the infinite 
volume limit~\cite{begunprc76}. Moreover, micro canonical ensemble seems to give a better
agreement with the data~\cite{begunprc76}. These findings definitely suggest the importance of the conservation laws.
We plan to study these aspects in our future work.

\section *{Acknowledgement}
The work is funded by CSIR, UGC (DRS) and DST of the Government of India.
The authors thank STAR collaboration for providing the experimental data.
We would like to thank Anirban Lahiri, Sarbani Majumder, Xiaofeng Luo, Nihar Ranjan Sahoo and Swagato
Mukherjee for useful discussion.


\begin{thebibliography}{99}

\bibitem{nature05120_Aoki} Y. Aoki, G. Endrodi, Z. Fodor, S. D. Katz and K. K. Szabo, 
\href{http://www.nature.com/nature/journal/v443/n7112/full/nature05120.html}{Nature {\bf 443}, 675 (2006)}.

\bibitem{PRL65_2491_Brown}F. R. Brown {\it et al.}, \href{http://prl.aps.org/abstract/PRL/v65/i20/p2491_1}
{Phys. Rev. Lett. {\bf 65}, 2491 (1990)}.

\bibitem{PRD78_074507_Ejiri}S. Ejiri, \href{http://prd.aps.org/abstract/PRD/v78/i7/e074507}{Phys. Rev. {\bf D 78}, 074507 (2008)}.

\bibitem{PRD29_Pisarski}R. D. Pisarski and F. Wilczek, 
\href{http://prd.aps.org/abstract/PRD/v29/i2/p338_1]}{Phys. Rev. {\bf D 29}, 338 (1984)}.

\bibitem{NPA504_Asakawa}M. Asakawa and K. Yazaki, 
\href{http://www.sciencedirect.com/science/article/pii/037594748990002X}{Nucl. Phys. {\bf A 504}, 668 (1989)}. 

\bibitem{PRD58_096007_Halasz}M. A. Halasz, A. D. Jackson, R. E. Shrock, M. A. Stephanov and J. J. M. Verbaarschot, 
\href{http://prd.aps.org/abstract/PRD/v58/i9/e096007}{Phys. Rev. {\bf D 58},  096007 (1998)}.

\bibitem{PRD67_014028_Hatta}Y. Hatta and T. Ikeda, 
\href{http://prd.aps.org/abstract/PRD/v67/i1/e014028}{Phys. Rev. {\bf D 67}, 014028 (2003)}.

\bibitem{PRC79_Bowman}E. S. Bowman and J. I. Kapusta, 
\href{http://prc.aps.org/pdf/PRC/v79/i1/e015202}{Phys. Rev. {\bf C 79}, 015202 (2009)}.

\bibitem{JHEP_Fodor}Z. Fodor and S. D. Katz, 
\href{http://iopscience.iop.org/1126-6708/2004/04/050/pdf/1126-6708_2004_04_050.pdf}{J. High Energy Phys. {\bf 04}, 050 (2004)}.

\bibitem{PTPS_Stephanov}M. A. Stephanov, Prog. Theor. Phys. Suppl. {\bf 153}, 139 (2004).

\bibitem{PRD71_114014_Gavai}R. V. Gavai and S. Gupta, 
\href{http://prd.aps.org/abstract/PRD/v71/i11/e114014}{Phys. Rev. {\bf D 71}, 114014 (2005)}.

\bibitem{PRD75_Schaefer}B.-J. Schaefer and J. Wambach, 
\href{http://prd.aps.org/abstract/PRD/v75/i8/e085015}{Phys. Rev. {\bf D 75}, 085015 (2007)}.

\bibitem{PRD78_114503_Gavai}R. V. Gavai and S. Gupta,
\href{http://prd.aps.org/abstract/PRD/v78/i11/e114503}{Phys. Rev. {\bf D 78}, 114503 (2008)}.

\bibitem{PRD79_074505_Cheng}M. Cheng {\it et al.}, \href{http://prd.aps.org/pdf/PRD/v79/i7/e074505}{Phys. Rev. {\bf D 79}, 074505 (2009)}.


\bibitem{boyd} G. Boyd, J. Engels, F. Karsch, E. Laermann, C. Legeland,
M. Lugermeier and B. Peterson, Nucl. Phys. {\bf B 469}, 419 (1996).

\bibitem{engels} J. Engels, O. Kaczmarek, F. Karsch, and E. Laermann, Nucl.
Phys. {\bf B 558}, 307 (1999).

\bibitem{fodor1} Z. Fodor and S. D. Katz, Phys. Lett. {\bf B 534}, 87 (2002).

\bibitem{fodor2} Z. Fodor, S. D. Katz and K. K. Szabo, Phys. Lett. {\bf B 568}, 73 (2003).

\bibitem{allton1} C. R. Allton, S. Ejiri, S. J. Hands, O. Kaczmarek, F. Karsch,
E. Laermann, Ch. Schmidt, and L. Scorzato, \href{http://prd.aps.org/abstract/PRD/v66/i7/e074507}{Phys. Rev. {\bf D 66}, 074507 (2002)}.

\bibitem{allton2} C. R. Allton, S. Ejiri, S. J. Hands, O. Kaczmarek, F. Karsch,
E. Laermann, and Ch. Schmidt, \href{http://prd.aps.org/abstract/PRD/v68/i1/e014507}{Phys. Rev. {\bf D 68}, 014507 (2003)}.

\bibitem{allton3} C. R. Allton, M. Doring, S. Ejiri, S. J. Hands, O. Kaczmarek,
F. Karsch, E. Laermann, and K. Redlich, \href{http://prd.aps.org/abstract/PRD/v71/i5/e054508}{Phys. Rev. {\bf D 71}, 054508 (2005)}.

\bibitem{forcrand} P. de Forcrand, and O. Philipsen, Nucl. Phys. {\bf B 642},
290 (2002); {\bf B 673}, 170 (2003).

\bibitem{aoki1} Y. Aoki, Z. Fodor, S. D. Katz, and K. K. Szabo, Phys. Lett. 
{\bf B 643}, 46 (2006).


\bibitem{PRD77_014511_Cheng}M. Cheng {\it et al.}, 
\href{}{Phys. Rev. {\bf D 77}, 014511 (2008)}.

\bibitem{Borsanyi_JHEP_11_77} S. Bors\'{a}nyi {\it et al.}, J. High Energy Phys. {\bf 11}, 077 (2010).



\bibitem{Bazavov}A. Bazavov {\it et al.}, Phys. Rev. {\bf D 86}, 034509 (2012).

\bibitem{Borsanyi} S. Bors\'{a}nyi {\it et al.}, J. High Energy Phys. {\bf 1201}, 138 (2012).


\bibitem{Ray}P. N. Meisinger and M. C. Ogilvie, Phys. Lett. {\bf B 379}, 163
(1996); Nucl. Phys. {\bf B} (Proc. Suppl.) {\bf 47}, 519 (1996); K. Fukushima, Phys. Lett. {\bf B 591}, 277 (2004); 
E. Megias, E. R. Arriola and L. L. Salcedo, Phys. Rev. {\bf D 74}, 065005 (2006); {\bf 74}, 114014 (2006);
J. High Energy Phys. {\bf 01} 073 (2006); C. Ratti, M. A. Thaler and W. Weise, Phys. Rev. {\bf D 73}, 014019 (2006);
S. K. Ghosh, T. K. Mukherjee, M. G. Mustafa and R. Ray, 
\href{http://prd.aps.org/abstract/PRD/v73/i11/e114007}{Phys. Rev. {\bf D 73}, 114007 (2006)}.

\bibitem{Bhattacharyya}S. Mukherjee, M. G. Mustafa and R. Ray, 
\href{http://prd.aps.org/abstract/PRD/v75/i9/e094015}{Phys. Rev. {\bf D 75},094015 (2007)};
S. K. Ghosh, T. K. Mukherjee, M. G. Mustafa and R. Ray, 
\href{http://prd.aps.org/abstract/PRD/v77/i9/e094024}{Phys. Rev. {\bf D 77}, 094024 (2008)}; 
A. Bhattacharyya, P. Deb, A. Lahiri and R. Ray, 
\href{http://prd.aps.org/abstract/PRD/v82/i11/e114028}{Phys. Rev. {\bf D 82}, 114028 (2010)}; 
\href{http://prd.aps.org/abstract/PRD/v83/i1/e014011}{{\bf D 83}, 014011 (2011)}.

\bibitem{PRL102_032301_Stephanov}M. A. Stephanov, 
\href{http://prl.aps.org/abstract/PRL/v102/i3/e032301}{Phys. Rev. Lett. {\bf 102}, 032301 (2009)}.


\bibitem{HRG_Braun-Munzinger}P. Braun-Munzinger, K. Redlich and J. Stachel, in {\it Quark Gluon Plasma 3}, edited by 
R.C. Hwa and X.N. Wang, (World Scientific Publishing, 2004).



\bibitem{PLB344_Braun-Munzinge}P. Braun-Munzinger, J. Stachel, J. P. Wessels and N. Xu, 
\href{http://www.sciencedirect.com/science/article/pii/037026939401534J}{Phys. Lett. {\bf B 344}, 43 (1995)}.

\bibitem{arXiv:nucl-th/9603004_Cleymans}J. Cleymans, D. Elliott, H. Satz and R. L. Thews, Z. Phys. {\bf C 74}, 319 (1997);
\href{http://arxiv.org/pdf/nucl-th/9603004.pdf}{[nucl-th/9603004]}.

\bibitem{PLB465_Braun-Munzinge}P. Braun-Munzinger, I. Heppe and J. Stachel, 
\href{http://www.sciencedirect.com/science/article/pii/S037026939901076X}{Phys. Lett. {\bf B 465}, 15 (1999)}.

\bibitem{PRC60_054908_Cleymans}J. Cleymans and K. Redlich, 
\href{http://prc.aps.org/abstract/PRC/v60/i5/e054908}{Phys. Rev. {\bf C 60}, 054908 (1999)}.

\bibitem{PRC73_Becattini}F. Becattini, J. Manninen and M. Gaździcki,
\href{http://prc.aps.org/abstract/PRC/v73/i4/e044905}{Phys. Rev. {\bf C 73}, 044905 (2006)}.

\bibitem{PLB518_Braun-Munzinger}P. Braun-Munzinger, D. Magestro, K. Redlich and J. Stachel, 
\href{http://www.sciencedirect.com/science/article/pii/S0370269301010693}{Phys. Lett. {\bf B 518}, 41 (2001)}.

\bibitem{NPA772_Andronic}A. Andronic, P. Braun-Munzinger and J. Stachel, 
\href{http://www.sciencedirect.com/science/article/pii/S0375947406001485}{Nucl. Phys. {\bf A 772}, 167 (2006)}.

\bibitem{PLB673_Andronic}A. Andronic, P. Braun-Munzinger and J. Stachel, 
\href{http://www.sciencedirect.com/science/article/pii/S0370269309001609}{Phys. Lett. {\bf B 673}, 142 (2009)}.

\bibitem{gottlieb} S. A. Gottlieb {\it et al.}, \href{http://prl.aps.org/abstract/PRL/v59/i20/p2247_1}{Phys. Rev Lett. {\bf 59}, 2247 (1987)}.

\bibitem{gavai} R. V. Gavai, S. Gupta and P. Majumdar, 
\href{http://prd.aps.org/abstract/PRD/v65/i5/e054506}{Phys. Rev. {\bf  D 65}, 054506 (2002)}.

\bibitem{bernard1} C. Bernard {\it et al.}, 
\href{http://prd.aps.org/abstract/PRD/v71/i3/e034504}{Phys. Rev. {\bf D 71}, 034504 (2005)}.

\bibitem{bernard2} C. Bernard {\it et al.}, 
\href{http://prd.aps.org/abstract/PRD/v77/i1/e014503}{Phys. Rev. {\bf D 77}, 014503 (2008)}.


\bibitem{PLB97_Hagedorn}R. Hagedorn and J. Rafelski, Phys. Lett. {\bf B 97}, 136 (1980).

\bibitem{ZPC51_Rischke}D.H. Rischke, M. I. Gorenstein, H. St$\ddot{o}$cker and W. Greiner, Z. Phys. {\bf C 51}, 485 (1991).

\bibitem{PS48_277_Cleymans}J. Cleymans, M. I. Gorenstein, J. Stalnacke and E. Suhonen, 
\href{http://iopscience.iop.org/1402-4896/48/3/004/}{Phys. Scripta {\bf 48}, 277 (1993)}.

\bibitem{Singh}C. P. Singh, B. K. Patra and K. K. Singh, 
\href{http://www.sciencedirect.com/science/article/pii/0370269396011173}{Phys. Lett. {\bf B 387}, 680 (1996)}.

\bibitem{PRC56_Yen}G. D. Yen, M. I. Gorenstein, W. Greiner and S. N. Nang, 
\href{http://prc.aps.org/abstract/PRC/v56/i4/p2210_1}{Phys. Rev. {\bf C 56}, 2210 (1997)}.

\bibitem{PRC77_Gorenstein}M. I. Gorenstein, M. Hauer, and O. N. Moroz, 
\href{http://journals.aps.org/prc/abstract/10.1103/PhysRevC.77.024911}{Phys. Rev. {\bf C 77}, 024911 (2008)}.

\bibitem{PLB718_Andronic}A. Andronic, P. Braun-Munzinger, J. Stachel and M. Winn, 
\href{http://www.sciencedirect.com/science/article/pii/S037026931201043X}{Phys. Lett. {\bf B 718}, 80, (2012)}.

\bibitem{PRC85_Fu}J. Fu, 
\href{http://journals.aps.org/prc/abstract/10.1103/PhysRevC.85.064905}{Phys. Rev. {\bf C 85}, 064905 (2012)}.

\bibitem{begunprc88}V. V. Begun, M. Ga\'{z}dzicki, and M. I. Gorenstein, 
\href{https://journals.aps.org/prc/abstract/10.1103/PhysRevC.88.024902}{Phys. Rev. {\bf C 88}, 024902 (2013)}.

\bibitem{PLB722_Fu}J. Fu,
\href{http://www.sciencedirect.com/science/article/pii/S0370269313002979}{Phys. Lett. {\bf B 722}, 144 (2013)}.

\bibitem{PRC88_Tawfik}A. Tawfik, 
\href{http://prc.aps.org/abstract/PRC/v88/i3/e035203}{Phys. Rev. {\bf  C 88}, 035203 (2013)}.

\bibitem{dashen} R. Dashen and S. K. Ma, 
\href{https://journals.aps.org/pra/abstract/10.1103/PhysRevA.4.700}{Phys. Rev. {\bf A 4}, 700 (1971}.

\bibitem{hama} Y. Hama, T. Kodama, and O. Socolowski, Braz. J. Phys. {\bf 35}, 24 (2005).  

\bibitem{werner} K. Werner, Iu. Karpenko, T. Pierog, M. Bleicher and K. Mikhailov, 
\href{http://journals.aps.org/prc/abstract/10.1103/PhysRevC.82.044904}{Phys. Rev. {\bf C 82} , 044904 (2010)}.

\bibitem{satarov}A. V. Merdeev, L. M. Satarov, and I. N. Mishustin,
\href{http://journals.aps.org/prc/abstract/10.1103/PhysRevC.84.014907}{Phys. Rev. {\bf C 84}, 014907 (2011)}.


\bibitem{Mohanty}P. Garg, D. K. Mishra, P. K. Netrakanti, B. Mohanty, A. K. Mohanty, B. K. Singh and N. Xu,
Phys. Lett. {\bf B 726}, 691 (2013).



\bibitem{PLB633_275_Ejiri}S. Ejiri, F. Karsch and K. Redlich, 
\href{http://www.sciencedirect.com/science/article/pii/S0370269305017521}{Phys. Lett. {\bf B 633}, 275 (2006)}.

\bibitem{PLB695_Karsch} F. Karsch and K. Redlich,  
\href{http://www.sciencedirect.com/science/article/pii/S0370269310012578}{Phys. Lett. {\bf B 695}, 136 (2011)}.

\bibitem{PRL105_022302_Aggarwal}M. M. Aggarwal {\it et al.} (STAR Collaboration), 
\href{http://prl.aps.org/pdf/PRL/v105/i2/e022302}{Phys. Rev. Lett. {\bf 105}, 022302 (2010)}.

\bibitem{PDG}K Nakamura {\it et al.} (Particle Data Group) 
\href{http://iopscience.iop.org/0954-3899/37/7A/075021}{J. Phys. G: Nucl. Part. Phys. {\bf 37}, 075021 (2010)}.


\bibitem{amnedolia} S. R. Amendolia {\it et al.}, Nucl Phys. {\bf B 277}, 168 (1986); Phys. Lett. {\bf B 178}, 435 (1986);
G. Simon, C. Schmitt, F. Borkowski and V. Walther, Nucl. Phys. {\bf A 333}, 381 (1980); T. Udem {\it et al.}, Phys. Rev. Lett. {\bf 79},
2646 (1997); P. Mergell, U. G. Meissner and D. Drechsel, Nucl. Phys. {\bf A 596}, 367 (1996).

\bibitem{povh} B. Povh and J. H$\ddot{u}$ffner, Phys. Rev. Lett. {\bf 58}, 1612 (1987); J. H$\ddot{u}$ffner and B. Povh, 
Phys. Lett. {\bf B 215}, 772 (1988).


\bibitem{bohr} A. Bohr and B. Mottelson, {\it Nucl. Structure}, (Benjamin, New York, 1969), vol. 1,  p. 266.

\bibitem{giltinan}  D. A. Giltinan and R. M. Thaler, Phys. Rev. {\bf 131}, 805 (1963).







\bibitem{Asakawa}M. Asakawa, U. Heinz, and B. M$\ddot{u}$ller, \href{http://prl.aps.org/abstract/PRL/v85/i10/p2072_1}{Phys. Rev. Lett. {\bf 85}, 
 2072 (2000)}.
 
\bibitem{Bower}D. Bower and S. Gavin, \href{http://prc.aps.org/abstract/PRC/v64/i5/e051902}{Phys. Rev. {\bf C 64}, 051902 (2001)}. 
 
 
\bibitem{Aziz}M. A. Aziz and S. Gavin, \href{http://prc.aps.org/abstract/PRC/v70/i3/e034905}{Phys. Rev. {\bf C 70}, 034905 (2004)}.

\bibitem{Bazavov_eos_contm}A. Bazavov {\it et al.},
\href{http://arxiv.org/abs/1407.6387}{arxiv: 1407.6387v1 [hep-lat]}.

\bibitem{Schmidt}A. Bazavov {\it et al.}, Phys. Rev. Lett. {\bf 109}, 192302 (2012);
C. Schmidt (for the BNL-Bielefeld Collaboration), arXiv:1301.6526; 
A. Bazavov {\it et al.}, Phys. Rev. Lett. {\bf 111}, 082301 (2013).

\bibitem{huovinen_2010} P. Huovinen and P. Petreczky, J. Phys. :Conf. Series {\bf 230}, 012012 (2010).

\bibitem{borsanyi_2011} S. Bors\'{a}nyi {\it et al.}, Nucl. Phys. {\bf A855}, 253 (2011).

\bibitem{PRL95_182301_Majumder}V. Koch, A. Majumder and J. Randrup, 
\href{http://journals.aps.org/prl/abstract/10.1103/PhysRevLett.95.182301}{Phys. Rev. Lett. {\bf 95}, 182301 (2005)}.

\bibitem{asakawa_prl}M. Asakawa, S. Ejiri and M. Kitazawa,  
\href{http://journals.aps.org/prl/abstract/10.1103/PhysRevLett.103.262301}{Phys. Rev. Lett. {\bf 103}, 262301 (2009)}.


\bibitem{AHEP_Tawfik}A. Tawfik, 
\href{http://www.hindawi.com/journals/ahep/2013/574871/}{Adv.High Energy Phys. {\bf 2013}, 574871 (2013)}.



\bibitem{PRC73_Cleymans}J. Cleymans, H. Oeschler, K. Redlich and S. Wheaton, 
\href{http://prc.aps.org/abstract/PRC/v73/i3/e034905}{Phys. Rev. {\bf C 73}, 034905 (2006)}.

\bibitem{STAR_Luo}X. Luo (for the STAR collaboration), Nuclear Physics {\bf A 904-905}, 911c, (2013).

\bibitem{Mohanty_science}S. Gupta, X. Luo, B. Mohanty, H.-G. Ritter and Nu Xu, \href{http://www.sciencemag.org/content/332/6037/1525}
{Science {\bf 332}, 1525 (2011)}.

\bibitem{Karsch_PRD84_R}F. Karsch and K. Redlich, \href{http://prd.aps.org/abstract/PRD/v84/i5/e051504}
{Phys. Rev. {\bf D 84}, 051504(R) (2011)}.


\bibitem{STAR_nihar}L. Adamczyk {\it et al.} (STAR Collaboration), Phys. Rev. Lett. {\bf 113}, 092301 (2014).


\bibitem{PRL_111_Borsanyi} S. Bors\'{a}nyi, Z. Fodor, S.D. Katz, S. Krieg, C. Ratti and K. K. Szabo,
\href{http://prl.aps.org/abstract/PRL/v111/i6/e062005}{Phys. Rev. Lett. {\bf 111}, 062005 (2013)}.

\bibitem{jeon&koch-Hwa} S. Jeon and V. Koch, in {\it Quark gluon plasma 3}, edited by R. C. Hwa and X. N. Wang (World Scientific,
Singapore, 2004), p. 430.


\bibitem{begunprc76} V. V. Begun, M. Gazdzicki, M. I. Gorenstein, M. Hauer, V. P. Konchakovski and V. Lugwitz, Phys. Rev. {\bf C 76}, 
024902 (2007) and references therein.



\bibitem{14046511} A. Bazavov {\it et al.}, Phys. Rev. Lett. {\bf 113}, 072001 (2014).


\end{thebibliography}
\end{document}